\documentclass[aps,prc,twocolumn,nofootinbib,superscriptaddress,amsmath,amssymb]{revtex4-1}

 \newcommand{\Ratio}{$G_E^p/G_M^p$}	%

\usepackage{epsfig}
\usepackage{graphicx}
\usepackage{dcolumn}
\usepackage{bm}
\usepackage{subfigure}
\usepackage{epstopdf}
\usepackage{hyperref}
\usepackage[utf8]{inputenc}

\begin{document}
\title{Polarization Transfer Observables in Elastic Electron-Proton Scattering at $\mathbf{Q^2= 2.5}$, 5.2, 6.8 and 8.5 GeV$\mathbf{^2}$}


\author{A. J. R. Puckett} \email[Corresponding author:]{andrew.puckett@uconn.edu}
\affiliation{University of Connecticut, Storrs, CT 06269}
\author{E. J. Brash}
\affiliation{Christopher Newport University, Newport News, VA 23606}
\affiliation{Thomas Jefferson National Accelerator Facility, Newport News, VA 23606}
\author{M. K. Jones}
\affiliation{Thomas Jefferson National Accelerator Facility, Newport News, VA 23606}
\author{W. Luo}
\affiliation{Lanzhou University, Lanzhou 730000, Gansu, Peoples Republic of China}
\author{M. Meziane}
\affiliation{College of William and Mary, Williamsburg, VA 23187}
\author{L. Pentchev}
\affiliation{Thomas Jefferson National Accelerator Facility, Newport News, VA 23606}
\author{C. F. Perdrisat}
\affiliation{College of William and Mary, Williamsburg, VA 23187}
\author{V. Punjabi}
\author{F. R. Wesselmann}
\affiliation{Norfolk State University, Norfolk, VA 23504}
\author{A. Afanasev}
\affiliation{The George Washington University, Washington, DC 20052}
% ...

\begin{abstract}
  \begin{description}
    \item[Background] Interest in the behavior of nucleon electromagnetic form factors at large momentum transfers has steadily increased since the discovery, using polarization observables, of the rapid decrease of the ratio $G_E^p/G_M^p$ of the proton's electric and magnetic form factors for momentum transfers $Q^2 \gtrsim 1$ GeV$^2$, in strong disagreement with previous extractions of this ratio using the traditional Rosenbluth separation technique. 
    \item[Purpose] The GEp-III and GEp-2$\gamma$ experiments were carried out in Jefferson Lab's (JLab's) Hall C from 2007-2008, to extend the knowledge of $G_E^p/G_M^p$ to the highest practically achievable $Q^2$ given the maximum beam energy of 6~GeV, and to search for effects beyond the Born approximation in polarization transfer observables of elastic $\vec{e}p$ scattering. This article provides an expanded description of the common experimental apparatus and data analysis procedures, and reports the results of a final reanalysis of the data from both experiments, including the previously unpublished results of the full-acceptance dataset of the GEp-2$\gamma$ experiment.
    \item[Methods] Polarization transfer observables in elastic $\vec{e}p\rightarrow e\vec{p}$ scattering were measured at central $Q^2$ values of 2.5, 5.2, 6.8, and 8.54 GeV$^2$. At $Q^2 = 2.5$ GeV$^2$, data were obtained for central values of the virtual photon polarization parameter $\epsilon$ of 0.149, 0.632, and 0.783. 
The Hall C High Momentum Spectrometer detected and measured the polarization of protons recoiling elastically from collisions of JLab's polarized electron beam with a liquid hydrogen target. A large-acceptance electromagnetic calorimeter detected the elastically scattered electrons in coincidence to suppress inelastic backgrounds.
    \item[Results] The final GEp-III data are largely unchanged relative to the originally published results. The statistical uncertainties of the final GEp-2$\gamma$ data are significantly reduced at $\epsilon = 0.632$ and $0.783$ relative to the original publication.
\item[Conclusions] The final GEp-III results show that the decrease with $Q^2$ of $G_E^p/G_M^p$ continues to $Q^2 = 8.5$ GeV$^2$, but at a slowing rate relative to the approximately linear decrease observed in earlier Hall A measurements. At $Q^2 = 8.5$ GeV$^2$, $G_E^p/G_M^p$ remains positive, but is consistent with zero. 
At $Q^2 = 2.5$ GeV$^2$, 
 $G_E^p/G_M^p$ derived from the polarization component ratio $R \propto P_t/P_\ell$ shows no statistically significant $\epsilon$-dependence, as expected in the Born approximation. On the other hand, the ratio $P_\ell/P_\ell^{Born}$ of the longitudinal polarization transfer component to its Born value shows an enhancement of roughly 1.4\% at $\epsilon = 0.783$ relative to $\epsilon = 0.149$, with $\approx 1.9\sigma$ significance based on the total uncertainty, implying a similar effect in the transverse component $P_t$ that cancels in the ratio $R$. 
     \end{description}
\end{abstract}

\date{\today}
\maketitle
\section{Introduction}
\label{sec:INTRO}
Electron scattering is of central importance to the characterization
of nucleon and nuclear structure, because of the relative weakness of
the electromagnetic interaction (compared to a strongly interacting probe), the structureless character of the leptonic probe, and the availability of electron beams of high
intensity, duty cycle, energy, and polarization.
The field of elastic electron-nucleus scattering started with the
availability of electron beams with energies up to 550 MeV at the High
Energy Physics Laboratory (HEPL) in Stanford in the mid-1950s. One
notable result of these early experiments was the first determination of a proton radius \cite{Hofstadter:1956qs}, which, together with the
anomalous magnetic moment of the proton, discovered in 1933 by Otto Stern \cite{stern}, completed the picture of 
the proton as a finite-size object with an internal structure. 

The utility of electron-nucleon scattering as a probe of nucleon
structure derives from the validity of the single virtual photon
exchange (Born) approximation, up to radiative corrections that are modest in size compared to the leading (Born) term, and precisely calculable in low-order QED perturbation theory, due to the small value of the fine structure constant $\alpha = \frac{e^2}{4\pi \epsilon_0 \hbar c} \approx 1/137.036$\cite{Olive:2016xmw}. This allows for a theoretically ``clean'' extraction of the electromagnetic structure of the target from the measured scattering observables such as cross sections and polarization asymmetries.
In the Born approximation, the effect of the
proton's internal structure on the Lorentz-invariant elastic $ep
\rightarrow ep$ scattering amplitude is completely specified by two
form factors (FFs), which encode the interaction of the pointlike electromagnetic current of the electron with the proton's charge and magnetic moment distributions. The ``Dirac'' form factor $F_1$ describes the charge and Dirac magnetic moment interactions, while the ``Pauli'' form factor $F_2$ describes the anomalous magnetic moment interaction. $F_1$ and $F_2$ are real-valued functions of the Lorentz-invariant four-momentum transfer squared between the electron and the nucleon, defined as $Q^2 \equiv -q^2 = -(k-k')^2$, with $k$ and $k'$ the four-momenta of the incident and scattered electron. In fixed-target electron scattering, $q^2$ is a spacelike invariant that is always negative. The reaction kinematics and physical observables are thus typically discussed in terms of the positive-definite quantity $Q^2$. A detailed overview of the theoretical formalism of the 
Born approximation for elastic $ep$ scattering is given in
Ref.~\cite{Punjabi:2015bba}.

An equivalent description of the nucleon electromagnetic form factors (EMFFs) is provided by the so-called ``Sachs'' form factors~\cite{Ernst:1960zza,Hand:1963zz} $G_E$ (electric) and $G_M$ (magnetic), defined as the following experimentally convenient independent linear combinations of $F_1$ and $F_2$,
\begin{eqnarray}
G_{E}& \equiv & F_{1}-\tau F_{2}\\
G_{M}& \equiv & F_1 + F_2,
\label{eq:gepgmp}
\end {eqnarray}
in which $\tau \equiv \frac{Q^2}{4M_p^2}$, with $M_p$ the mass of the proton. In terms of the Sachs form factors, the differential cross section for elastic $ep$ scattering in the Born approximation is given in the nucleon rest frame (which coincides with the lab frame in fixed-target experiments) by the Rosenbluth formula~\cite{Rosenbluth:1950yq}:
\begin{eqnarray}
  \frac{d\sigma}{d\Omega_e} = \left(\frac{d\sigma}{d\Omega_e}\right)_{Mott} \frac{\epsilon G_E^2 + \tau G_M^2}{\epsilon\left(1+\tau\right)}, \label{eq:csGEGM} \\
  \left(\frac{d\sigma}{d\Omega_e}\right)_{Mott} = \frac{\alpha^2 \cos^2\frac{\theta_e}{2}}{4 E_{e}^2 \sin^4\frac{\theta_e}{2}} \frac{E'_e}{E_e} \label{eq:Mott},
\end{eqnarray}
in which $\left(\tfrac{d\sigma}{d\Omega_e}\right)_{Mott}$ represents the theoretical Born cross section for electron scattering from a pointlike, spinless target of charge $e$, $E_e$ is the beam energy, $E'_e$ is the scattered electron energy, $\theta_e$ is the electron scattering angle, and $\epsilon \equiv \left({1+2(1+\tau)\tan^2 \frac{\theta_e}{2}}\right)^{-1}$ is the longitudinal polarization of the virtual photon. The expression~\eqref{eq:csGEGM} provides a simple technique for the extraction of $G_E^2$ and $G_M^2$ known as Rosenbluth or L/T (for longitudinal/transverse) separation, in which the differential cross section is measured at fixed $Q^2$ while varying the parameter $\epsilon$. 
A plot of the $\epsilon$ dependence of the ``reduced'' cross section, obtained by dividing the measured, radiatively corrected cross section by the Mott cross section and the kinematic factor in the denominator of Eq.~\eqref{eq:csGEGM}, yields a straight line with a slope (intercept) equal to $G_E^2$ ($\tau G_M^2$).

\begin{figure}
  \begin{center}
    \includegraphics[width=0.85\columnwidth]{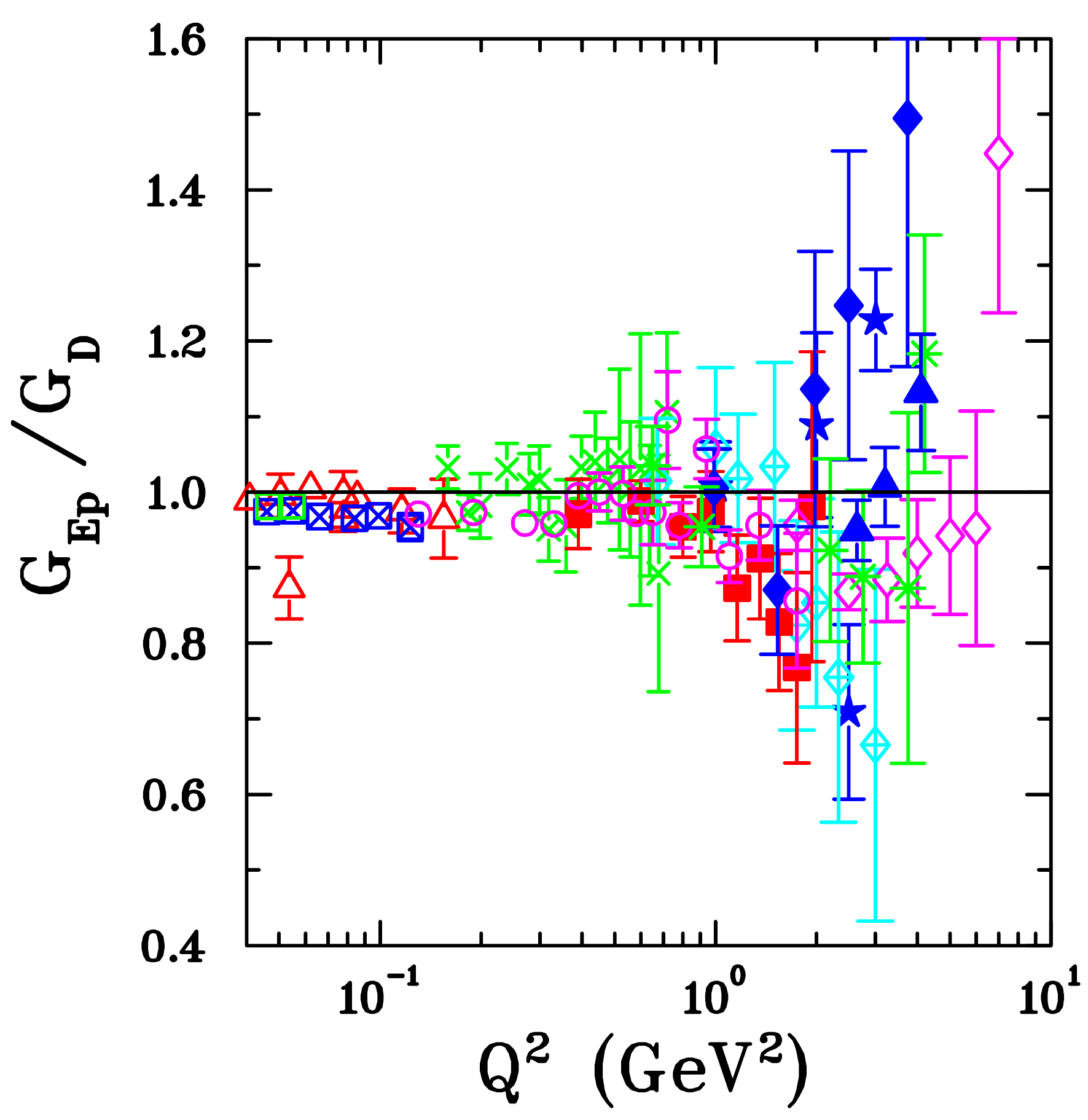}
    \caption{ $G_E^p/G_{D}$ extracted from cross section measurements versus $Q^2$. The data from before 1980 are:  open triangle \cite{Hand:1963zz}, 
      multiplication sign \cite{Janssens:1965kd}, open circle \cite{Price:1971zk}, filled diamond \cite{Litt:1969my}, filled square \cite{Berger:1971kr}, crossed 
      diamond \cite{Bartel:1973rf},  crossed square \cite{Borkowski:1974mb} and open square \cite{Simon:1980hu}.  The SLAC data from the 1990's are filled 
      star \cite{Walker:1993vj}  and open diamond \cite{Andivahis:1994rq}. The JLab data are asterisk \cite{Christy:2004rc} and  filled triangle \cite{Qattan:2004ht}. 
      Figure adapted from Fig. (3) of Ref.~\cite{Punjabi:2015bba}.}
    \label{fig:gepgd}
  \end{center}
\end{figure}

\begin{figure}
  \begin{center}
    \includegraphics[width=0.85\columnwidth]{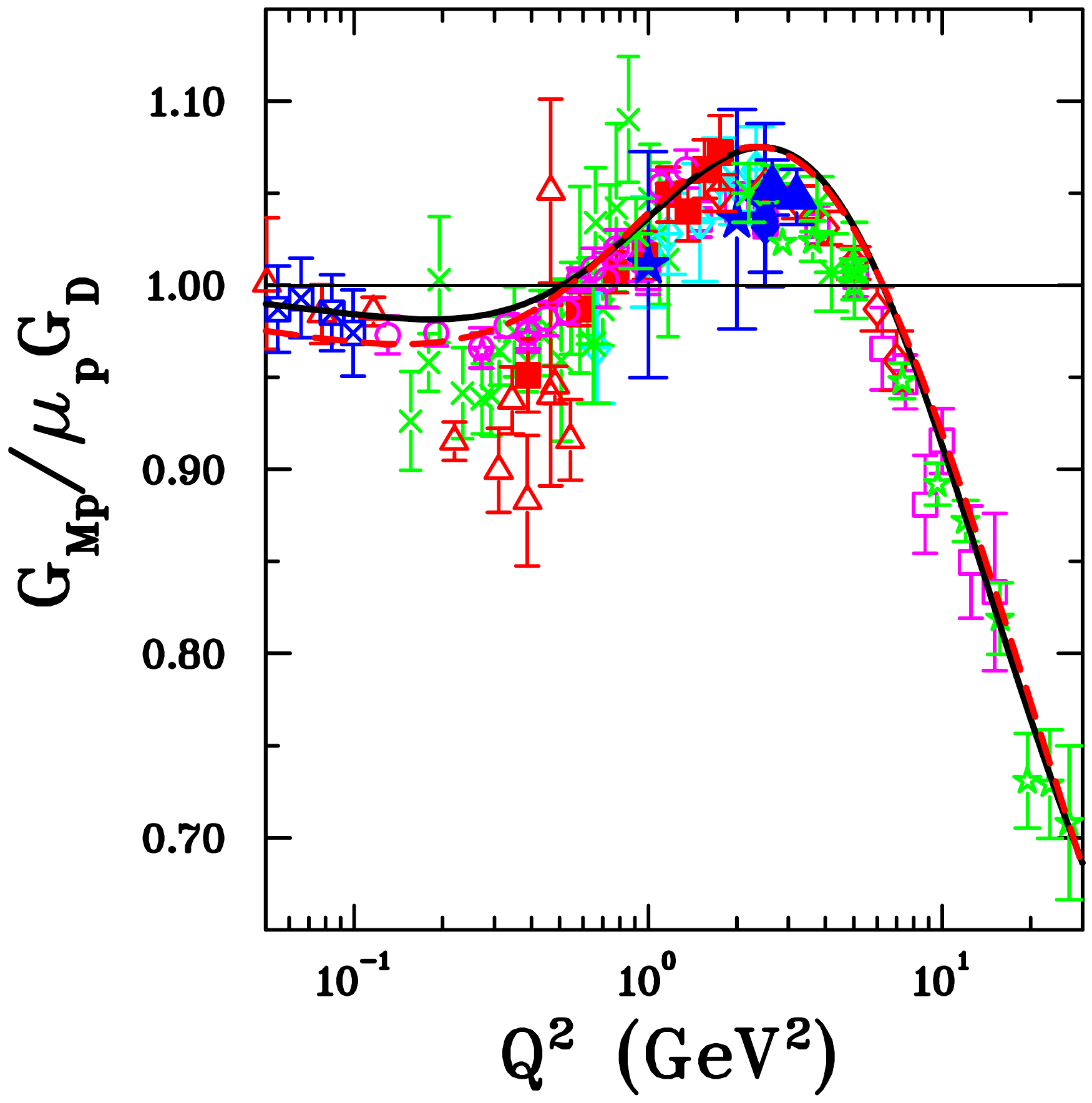}
    
    \caption{ $G_M^p/\mu_p G_{D}$ extracted from cross section measurements versus $Q^2$. The symbols are the same as in Fig.~\ref{fig:gepgd}. 
      Additional data points at the highest $Q^2$, open square \cite{Kirk:1972xm} and open  star \cite{Sill:1992qw}, were extracted from cross sections 
      assuming $\mu_p$$G_E^p$/$G_M^p~=~1$. The solid (dashed) line is a fit by Ref.~\cite{Kelly:2004hm} (Ref.~\cite{Brash:2001qq}). Figure adapted from Fig. (4) of Ref.~\cite{Punjabi:2015bba}.
    }
    \label{fig:gmpgd}
  \end{center}
\end{figure}
Until the late 1990s all (or most) form factor measurements suggested that both $G_E^p$ and $G_M^p$ decreased like $\frac{1}{Q^4}$ at large $Q^2$, and that the ratio $\mu_p G_{E}^p/G_{M}^p$ was approximately equal to one, regardless of $Q^2$. It also appeared that the dipole form $G_D \equiv \left(1+\frac{Q^2}{\Lambda^2}\right)^{-2}$, with $\Lambda^2 = 0.71$ GeV$^2$, provided a reasonable description of $G_E^p$, $G_M^p/\mu_p$ and $G_{M}^n/\mu_n$, as illustrated in Figs. \ref{fig:gepgd} and \ref{fig:gmpgd} (for $G_E^p$ and $G_M^p$). $G_{E}^n$ was expected to have an entirely different $Q^2$ dependence, given the zero net charge of the neutron, which 
imposes $G_{E}^n=0$ at $Q^2=0$.

The helicity structure of the single-photon-exchange amplitude also gives rise to significant double-polarization asymmetries, with different sensitivities to the form factors compared to the spin-averaged cross section. Non-zero asymmetries occur in the case where the electron beam is longitudinally polarized\footnote{The effects of transverse polarization of the electron beam are suppressed by factors of $m_e/E_e$, leading to asymmetries of order 10$^{-5}$ in experiments with ultra-relativistic electrons at GeV-scale energies. In the context of electromagnetic form factor measurements in the $Q^2$ regime of this work, these effects are negligible compared to the asymmetries for longitudinally polarized electrons and the precision with which they are measured.}, and either the target nucleon is also polarized or the polarization transferred to the recoiling nucleon is measured. The polarization transferred to the recoil proton in the scattering of longitudinally polarized electrons by 
unpolarized protons has only two non-zero components, longitudinal, $P_\ell$, and transverse, $P_t$, with respect to the momentum 
transfer and parallel to the scattering plane~\cite{AkhiezerRekalo1,*AkhiezerRekalo3,*AkhiezerRekalo2,Arnold:1980zj}:
\begin{eqnarray}
  \label{recoilformulas} P_t &=& -h P_e \sqrt{\frac{2\epsilon(1-\epsilon)}{\tau}}
  \frac{G_{E}G_{M}}{G_M^2+\frac{\epsilon}{\tau}G_{E}^2} \nonumber \\
  P_\ell &=& h P_e {\sqrt{1-\epsilon^2}}\frac{G_{M}^2}{G_M^2+\frac{\epsilon}{\tau}G_{E}^2} \\
  \frac{G_E}{G_M} &=& -\frac{P_t}{P_\ell}\sqrt{\frac{\tau(1+\epsilon)}{2\epsilon}}
  \nonumber. 
\end{eqnarray}
Here $h$ denotes the sign of the electron beam helicity, and $P_e$ is the electron beam polarization. The observables for scattering on a polarized proton target are related to those for polarization transfer by time-reversal symmetry~\cite{Dombey:1969wk,Donnelly:1985ry,Raskin:1988kc}. Specifically, the transverse asymmetry $A_t = P_t$, while the longitudinal asymmetry $A_\ell = -P_\ell$. The sign change between $A_\ell$ and $P_\ell$ is caused by the proton spin flip required to absorb transversely polarized virtual photons.

The interest in measuring these double-polarization observables is multi-faceted. First, the ratio $G_E/G_M$ is directly and linearly proportional to the ratio $P_t/P_\ell$ in the recoil polarization case or, equivalently, the ratio $A_t/A_\ell$ of the beam-target double-spin asymmetries in the polarized target case. Compared to the Rosenbluth method, polarization observables provide enhanced sensitivity to $G_E$ ($G_M$) at large (small) values of $Q^2$. Moreover, polarization observables provide an unambiguous determination of the relative sign of $G_E$ and $G_M$, whereas the Rosenbluth method is only sensitive to the \emph{squares} of the form factors. Finally, because of the ratio nature of 
the asymmetries, radiative corrections tend to be negligible, whereas they can and do affect the cross section measurements and Rosenbluth separations significantly, especially in kinematics where the relative contribution of either the $\epsilon G_E^2$ or the $\tau G_M^2$ term to the Born cross section~\eqref{eq:csGEGM} is small. The polarization transfer method in particular is highly attractive, as a simultaneous measurement of both recoil polarization components in a polarimeter facilitates a very precise measurement of $G_E/G_M$ in a single kinematic setting, with small systematic uncertainties resulting from cancellations of quantities such as the beam polarization, the polarimeter analyzing power, and the polarimeter instrumental asymmetry. 

\begin{figure}
  \begin{center}
    \includegraphics[width=0.85\columnwidth]{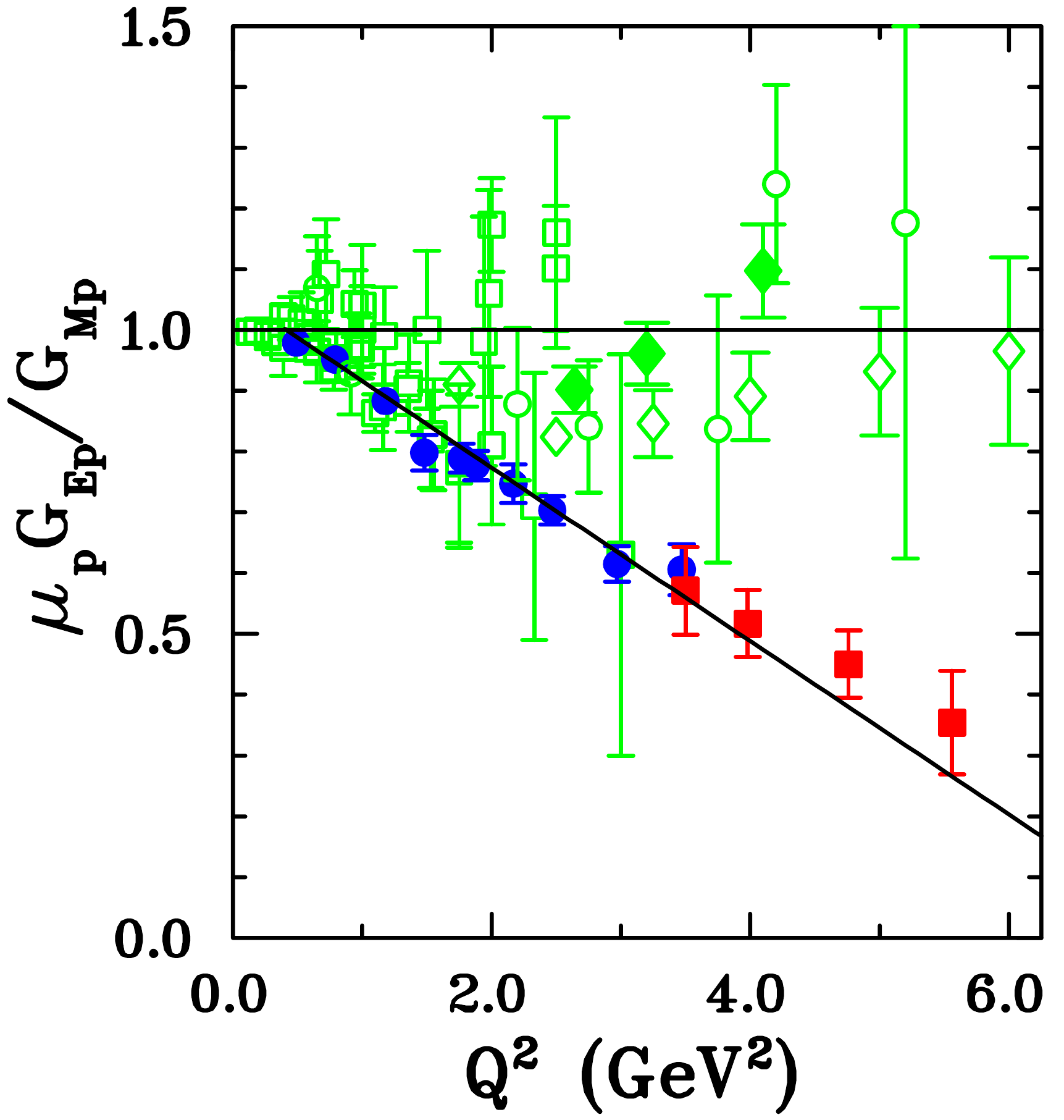}
    \caption{ The ratio $\mu_p G_E^p/G_M^p$ from the first two JLab experiments 
      filled circle \cite{Jones:1999rz,Punjabi:2005wq}, and filled square \cite{Gayou:2001qd,Puckett:2010ac}, compared to Rosenbluth separation results, 
      open diamond \cite{Andivahis:1994rq}, open circle \cite{Christy:2004rc}, filled diamond \cite{Qattan:2004ht}, and 
      open square~\cite{Hand:1963zz,Walker:1993vj,Hanson:1973vf,Bartel:1973rf,Berger:1971kr,Borkowski:1974mb,Simon:1980hu,Litt:1969my,Price:1971zk}. The curve shows the linear fit to the polarization data from Ref.~\cite{Gayou:2001qd}. Figure adapted from Fig. (9) of Ref.~\cite{Punjabi:2015bba}.} 
    \label{fig:gepgmp_pol_cs}
  \end{center}
\end{figure} 
In recent years the nucleon's elastic form factors have attracted steadily increasing attention, due in part to the
unexpected results of the first polarization transfer measurement of the ratio {\Ratio} at JLab.
This increasing attention is evident in the number of reviews of the subject published in the last 15 years~\cite{Gao:2003ag,HydeWright:2004gh,Perdrisat:2006hj,Arrington:2006zm,Cloet:2008re,scholar,Arrington:2011kb,Pacetti:2015iqa,Punjabi:2015bba}. The first measurement of $G_E^p/G_M^p$ by recoil polarization took place in 1994, at the MIT-Bates laboratory, at $Q^2$ values of 0.38 and 0.50 GeV$^2$, with 5\% statistical uncertainties~\cite{Milbrath:1997de}. 
The first two polarization transfer experiments at JLab, hereafter denoted GEp-I~\cite{Jones:1999rz,Punjabi:2005wq} and GEp-II~\cite{Gayou:2001qd,Puckett:2011xg}, consisted of measurements of the ratio $R \equiv \mu_p G_E^p/G_M^p$ for $0.5 \le Q^2$ (GeV$^2$) $ \le 5.6$. 
Together, the results of GEp-I and GEp-II, shown in Fig.~\ref{fig:gepgmp_pol_cs}, established conclusively that the concept of 
scaling of the proton form factor ratio had to be abandoned. There is a clear 
discrepancy between the values of $G_E^p/G_M^p$ extracted from double polarization 
experiments, and those obtained from cross section measurements.  
Among possible explanations for this discrepancy, the most thoroughly investigated is the hard two-photon exchange (TPEX) process, the amplitude for which does not ``factorize'' from the underlying nucleon structure information, cannot presently be calculated model-independently, and is neglected in the ``standard'' radiative corrections to experimental data. A recent overview of the theory, phenomenology and experimental knowledge of TPEX effects in elastic $ep$ scattering is given in Ref.~\cite{Afanasev:2017gsk}.

In the general case, elastic $eN$ scattering can be described in terms of three
complex amplitudes \cite{Guichon:2003qm,Afanasev:2005mp,Kivel:2012vs}, which can be written as $\tilde{G}_M$, $\tilde{G}_E$,
and $\tilde{F}_3$, the first two chosen as generalizations of the Sachs 
electric and magnetic form factors, $G_E$ and $G_M$, and the 
last one, $\tilde{F}_3$, being $\mathcal{O}(\alpha)$ relative to the Born terms and vanishing in the Born approximation. The ``generalized form factors'' $\tilde{G}_M$ and $\tilde{G}_E$ can be decomposed into sums of the real-valued Sachs form factors appearing in the Born amplitudes and depending only on $Q^2$, plus $\mathcal{O}(\alpha)$ complex-valued corrections that vanish in the Born approximation and depend on both $Q^2$ and $\epsilon$ as follows:
\begin{eqnarray}
\tilde{G}_M(Q^2,\epsilon)&\equiv&G_M(Q^2)+\delta \tilde{G}_M(Q^2,\epsilon) \\ 
\label{eq:regm}
\tilde{G}_E(Q^2,\epsilon)&\equiv&G_E(Q^2)+\delta \tilde{G_E}(Q^2, \epsilon).
\label{eq:rege} 
\end{eqnarray}
In terms of the generalized complex amplitudes, the reduced cross section $\sigma_R \equiv \frac{\epsilon(1+\tau)}{\tau} \sigma/\sigma_{Mott}$ and polarization observables are given at next-to-leading order in $\alpha$ by:
\begin{widetext}
  \begin{eqnarray}
    \sigma_{R} &=& G_M^2 + \frac{\epsilon}{\tau} G_E^2 + 2 G_M \Re\left(\delta \tilde{G}_M + \frac{\epsilon \nu }{M^2} \tilde F_3\right) + \frac{2\epsilon}{\tau} G_E \Re \left(\delta\tilde{G}_E + \frac{\nu}{M^2}\tilde F_3\right), \label{eq:siggen} \\
    P_t &=& -\frac{hP_e}{\sigma_R}\sqrt{\frac{2\epsilon(1-\epsilon)}{\tau}}\left[G_E G_M + G_M \Re \left(\delta \tilde{G}_E + \frac{\nu}{M^2} \tilde{F}_3\right) + G_E \Re\left(\delta \tilde{G}_M\right) \right], \label{eq:ptgen} \\
    P_\ell &=& \frac{hP_e}{\sigma_R}\sqrt{1-\epsilon^2}\left[G_M^2 + 2G_M \Re\left( \delta \tilde{G}_M +  \frac{\epsilon}{1+\epsilon}\frac{\nu}{M^2}\tilde{F}_3 \right)\right] \label{eq:plgen}, \\
    P_n &=&  \sqrt{\frac{2\epsilon(1+\epsilon)}{\tau}} \frac{1}{\sigma_R} \left[-G_M \Im\left( \delta \tilde{G}_E + \frac{\nu}{M^2} \tilde{F}_3\right) + G_E \Im \left(\delta \tilde{G}_M + \frac{2\epsilon}{1+\epsilon}\frac{\nu}{M^2}\tilde{F}_3\right)\right]\label{eq:pngen}, \\
    R &\equiv& -\mu_p \sqrt{\frac{\tau(1+\epsilon)}{2\epsilon}} \frac{P_t}{P_\ell} = \mu_p \frac{G_E}{G_M} \Re\left[1 - \frac{\delta \tilde{G}_M}{G_M} + \frac{\delta \tilde{G}_E}{G_E} + \frac{\nu\tilde{F}_3}{M^2}\left(\frac{(1+\epsilon)G_M - 2\epsilon G_E}{(1+\epsilon)G_EG_M}\right)\right], \label{eq:rgen}
  \end{eqnarray}
\end{widetext}
in which $\epsilon$ and $\tau$ are defined as above, the symbols $\Re$ and $\Im$ denote real and imaginary parts of the amplitudes, and 
\begin{eqnarray}
\frac{\nu}{M^2} \equiv \sqrt{\tau(1+\tau)\frac{1+\epsilon}{1-\epsilon}}.
\label{eq:nu2gamma}
\end{eqnarray}
The  reduced cross section
and the polarization transfer components $P_t$ and $P_{\ell}$ are defined only by the real parts
of the two-photon amplitudes. The  normal polarization transfer component, $P_n$, which is zero in the Born approximation, is defined by the imaginary parts of the two-photon exchange
amplitudes. 

There are several noteworthy features of Eqs.~\eqref{eq:siggen}-\eqref{eq:rgen}. The corrections to the reduced cross section beyond the Born approximation are additive with the Born terms, implying that even a small TPEX correction can seriously obscure the extraction of $G_E^2$ ($G_M^2$) at large (small) $Q^2$ when the relative contribution of either Born term to $\sigma_R^{Born}$ is small enough to be comparable to the TPEX correction. The ratio $R$ defined in Eq.~\eqref{eq:rgen}, on the other hand, is directly proportional to its Born value: $R = \mu G_E/G_M(1+\mathcal{O}(\alpha))$, and is subject only to \emph{relative} $\mathcal{O}(\alpha)$ TPEX corrections, in principle. In the limit $G_E \rightarrow 0$, however, the TPEX terms can become dominant even in the ratio $R$; the limit of Eq.~\eqref{eq:rgen} as $G_E \rightarrow 0$ is $R \rightarrow R_{Born} + \Re \left[\mu \frac{\delta \tilde{G}_E}{G_M} + \mu \frac{\nu}{M^2} \frac{ \tilde{F}_3}{G_M}\right]$, assuming $\delta \tilde{G}_M/G_M \ll 1$. 

Whereas the ratio $R$ measured in polarization transfer experiments only becomes significantly sensitive to TPEX corrections when $R_{Born}$ is comparable to $\alpha$, the reduced cross section becomes sensitive to TPEX corrections at relatively low $Q^2$ even for $R_{Born} \gg \alpha$. Given the superior sensitivity to $G_E$ at large $Q^2$ of the ratio $P_t/P_\ell$ and its relative robustness against radiative and TPEX corrections as compared to the Rosenbluth method, a general consensus has emerged that the polarization transfer data provide the most reliable determination of $G_E^p$ in the $Q^2$ range where cross section and polarization data disagree. 
Nevertheless, a large amount of experimental and theoretical effort is ongoing to understand the source of the discrepancy and develop a maximally model-independent prescription for TPEX corrections to elastic $ep$ scattering observables.



The subject of this article is the third dedicated series of polarization transfer measurements in elastic $\vec{e}p$ scattering at large $Q^2$, carried out in Jefferson Lab's (JLab's) Hall C from October, 2007
to June, 2008. Experiments E04-108 (GEp-III) and E04-019 (GEp-2$\gamma$) used the same apparatus and method to address two complementary physics goals. The goal of GEp-III was to extend the kinematic reach of the polarization transfer data for 
$G_E^p/G_M^p$ to the highest practically achievable $Q^2$, given the maximum electron
beam energy available at the time. The goal of GEp-2$\gamma$ was to
measure the $\epsilon$-dependence of $G_{E}^p/G_{M}^p$ at the fixed
$Q^2$ of 2.5~GeV$^2$ with small statistical and systematic
uncertainties, in order to test the polarization method and search for signatures of TPEX effects in two polarization observables. 

The results of GEp-III~\cite{Puckett:2010ac} and GEp-2$\gamma$~\cite{Meziane:2010xc} have already been published in short-form articles. The purpose of this article is to provide a detailed description of the apparatus and analysis methods common to both experiments and report the results of a full reanalysis of the data, carried out with the aim of reducing the systematic and, in the GEp-2$\gamma$ case, statistical uncertainties. Our reanalysis of the GEp-2$\gamma$ data includes the previously unpublished results of the full-acceptance analysis at $\epsilon = 0.632$ and $\epsilon = 0.783$, for which the acceptance-matching cuts applied to suppress certain systematic effects in the analysis of the originally published data~\cite{Meziane:2010xc} have been removed. The final results reported in this work supersede the originally published results. Section~\ref{sec:apparatus} describes the experiment apparatus and kinematics in detail. Section~\ref{sec:analysis} presents the details of the data analysis. Section~\ref{sec:results} presents the final results of both experiments and discusses the general features of the data. A brief overview of the theoretical interpretation of high-$Q^2$ nucleon FF data is given in Section~\ref{subsec:theory_sub_a}, while the implications of the GEp-2$\gamma$ data for the understanding of TPEX contributions in elastic $ep$ scattering and the discrepancy between cross section and polarization data for $G_E^p/G_M^p$ are discussed in Section~\ref{subsec:theory_sub_b}. Our conclusions are summarized in Section~\ref{sec:conclusions}.

\section{Experiment Description}
\label{sec:apparatus}

\begin{table*}
  \caption{\label{kintable} Central kinematics of the GEp-III
    and GEp-2$\gamma$ experiments. $Q^2$ denotes the central or nominal $Q^2$
    value, defined by the central momentum setting of the
    High Momentum Spectrometer (HMS) in which the proton was
    detected. $\epsilon$ is the value of the kinematic parameter defined in
    equation~\eqref{eq:csGEGM} computed from the incident beam energy (not corrected for energy loss in the target prior to scattering), and the central $Q^2$. $E_e$ is the incident beam energy, averaged over the duration of each running period. $E'_e$ is the scattered
    electron energy at the nominal $Q^2$. The central angle of BigCal is denoted $\theta_e$, and can differ slightly from the
    electron scattering angle at the central $Q^2$. $p_p$ is the HMS central momentum
    setting. $\theta_p$ is the HMS central angle. $\chi$ is the
    central spin precession angle in the HMS, $P_e$ is the average beam
    polarization, and $D_{cal}$ is the distance from the origin to
    the surface of BigCal.}
  \begin{ruledtabular}
    \begin{tabular}{lcccccccccc}
      Dates (mm/dd-mm/dd, yyyy) & $Q^2$ (GeV$^2$) & $\epsilon$ & $E_e$ (GeV) & $E'_e$
                                                                               (GeV) & $\theta_e$ ($^\circ$) & $p_p$ (GeV) & $\theta_p$
                                                                                                                             ($^\circ$) & $\chi$ ($^\circ$) & $P_e$ (\%) & $D_{cal}$ (m) \\
      \hline 
      11/27-12/08, 2007 & 2.50 & 0.154 & 1.873 & 0.541 & 105.2 & 2.0676 & 14.5 & 108.5 &
                                                                                         85.9 &  4.93 \\ 
      01/17-01/25, 2008 & 2.50 & 0.150 & 1.868 & 0.536 & 105.1 & 2.0676 & 14.5 & 108.5 &
                                                                                         85.5 & 4.94 \\ 
      12/09-12/16, 2007 & 2.50 & 0.633 & 2.847 & 1.515 & 44.9 & 2.0676 & 31.0 & 108.5 &
                                                                                        84.0 & 12.00 \\ 
      12/17-12/20, 2007 & 2.50 & 0.772 & 3.548 & 2.216 & 32.6 & 2.0676 & 35.4 & 108.5 &
                                                                                        85.8 & 11.16 \\
      01/05-01/11, 2008 & 2.50 & 0.789 & 3.680 & 2.348 & 30.8 & 2.0676 & 36.1 & 108.5 &
                                                                                        85.2 & 11.03 \\ \hline
      11/07-11/20, 2007 & 5.20 & 0.377 & 4.052 & 1.281 & 60.3 & 3.5887 & 17.9 & 177.2 & 
                                                                                        79.5 & 6.05 \\
      05/27-06/09, 2008 & 6.80 & 0.506 & 5.711 & 2.087 & 44.2 & 4.4644 & 19.1 & 217.9 & 
                                                                                        79.5 & 6.08 \\
      04/04-05/27, 2008 & 8.54 & 0.235 & 5.712 & 1.161 & 69.0 & 5.4070 & 11.6 & 262.2 &
                                                                                        80.9 & 4.30 
    \end{tabular}
  \end{ruledtabular}
\end{table*}
Longitudinally polarized electrons with energies up to 5.717 GeV produced by JLab's Continuous Electron Beam Accelerator Facility (CEBAF) were directed onto a liquid hydrogen target in experimental Hall C.  Elastically scattered protons were detected by the High
Momentum Spectrometer (HMS), equipped with a double Focal Plane
Polarimeter (FPP) to measure their polarization. Elastically scattered electrons were 
detected by a large-solid-angle electromagnetic calorimeter (BigCal) in coincidence
with the scattered protons. The main trigger for the event data acquisition (DAQ) was a coincidence between the single-arm triggers of the HMS and BigCal within a 50-ns window. Details of the coincidence trigger logic and the experiment data acquisition can be found in Ref.~\cite{Puckett:2015soa}. 
Table \ref{kintable} shows the central kinematics and running periods of the GEp-III and
GEp-2$\gamma$ experiments. The two running periods at $E_e \approx 1.87$ GeV were combined and analyzed together as a single kinematic setting. The same is true of the running periods at $E_e = 3.548$ GeV and $E_e = 3.680$ GeV. In both cases, the near-total overlap of the $Q^2$ and $\epsilon$ acceptances of two distinct measurements differing only slightly in beam energy and HMS central angle justifies combining the two settings into a single measurement\footnote{In this context, combining the data from two distinct measurements means combining all events from each of the two kinematically similar settings in a single unbinned maximum-likelihood extraction of $P_t$ and $P_\ell$, in which the small differences in central kinematics are accounted for event-by-event. This amounts to the assumption that $P_t$ and $P_\ell$ are the same for both settings. The data were also analyzed separately and found to be consistent with this assumption.}. The beam energy for each running period quoted in Table~\ref{kintable} represents the average incident beam energy during that period, and is not corrected for energy loss in the LH$_2$ target. The $\epsilon$ value quoted in Tab.~\ref{kintable} is computed from the average incident beam energy and central $Q^2$ value, and differs slightly from the acceptance-averaged value, hereafter referred to as $\left<\epsilon\right>$, and the ``central'' value $\epsilon_c$ quoted with the final GEp-2$\gamma$ results, which is computed from the central $Q^2$ value and the average\footnote{Where data from kinematically similar settings have been combined, the ``central'' $\epsilon$ value quoted with the final result represents a weighted average of the ``central'' values from each of the combined settings.} beam energy, corrected event-by-event for energy loss in the LH$_2$ target materials upstream of the reconstructed scattering vertex (see Tab.~\ref{tab:FinalResultsGEp2gamma} and \ref{tab:bcc}).

CEBAF consists of two antiparallel superconducting radio-frequency (SRF) linear accelerators (linacs), each capable (ca. 2007-2008) of approximately 600
MeV of acceleration, connected by nine recirculating magnetic arcs, with five at the north end and four at the south end. With this “racetrack” design, the
electron beam can be accelerated in up to five passes through both linacs, for a
maximum energy of approximately 6 GeV before extraction and delivery to the three experimental halls. Polarized electrons are excited from a ``superlattice'' GaAs
photocathode using circularly polarized laser light. Details of the CEBAF accelerator design and operational parameters are described in Refs.~\cite{Leemann:2001dg,Chao:2011za}, while more details specific to the running period of the GEp-III and GEp-2$\gamma$ experiments can be found in Ref.~\cite{Puckett:2015soa}. The typical beam current on target during the experiment was 60-100 $\mu$A, while the typical beam polarization was 80-86\%. The beam helicity was flipped pseudorandomly~\cite{Alcorn:2004sb} at a frequency of 30 Hz throughout the experiment.

During normal operations, the Hall C arc magnets, which steer the beam extracted from the CEBAF accelerator to Hall C, are operated in an achromatic tune.
For a measurement of the beam energy, the arc magnets are operated
in a dispersive tune. The central bend angle of the arc is 34.3$^{\circ}$.  The field integral of the arc magnets has been measured as a function of the power supply current. The beam position and arc magnet current setting information are used in the feedback system which stabilizes the beam energy and position. This system has been calibrated using dedicated arc beam energy measurements from Halls A and C, and is used for continuous monitoring of the beam momentum.

\begin{table}
  \caption{ \label{tab:emeas} Arc measurements of the beam energy ($E_{arc}$) taken during the GEp-III and GEp-2$\gamma$ experiments. No dedicated Hall C arc measurement was performed during the period from December 17-20, 2007, during which the nominal beam energy was 3.548 GeV. The data at a central $Q^2$ of 6.8 GeV$^2$ were collected at the same nominal beam energy as the $Q^2 = 8.5$ GeV$^2$ data during April-June, 2008.} 
  \begin{tabular}{lccc }
    \hline \hline   \\
    Date & $Q^2$ (GeV$^2$) & Number of passes &$E_{arc}$ (MeV) \\ \hline
    11/19/2007 &5.2& 5 &4052.34 $\pm$ 1.38\\
    11/28/2007 &2.5& 3 &1873.02 $\pm$ 1.09\\
    12/11/2007 &2.5& 4 &2847.16 $\pm$ 1.19\\
    1/6/2008 &2.5 &4 &3680.23 $\pm$ 1.31\\
    1/23/2008 &2.5& 2 &1868.13 $\pm$ 1.09\\
    4/6/2008 &8.5 &5 &5717.32 $\pm$ 1.64 \\\hline \hline 
  \end{tabular}
\end{table}
Table~\ref{tab:emeas} shows the Hall C arc measurements of the beam energy performed during the GEp-III and GEp-2$\gamma$ experiments. The arc energy of $E_e = 5.717$ GeV measured at the beginning of the $Q^2 = 8.5$ GeV$^2$ running in April 2008 differs slightly from the average beam energy for this run period and the subsequent 6.8 GeV$^2$ running, shown in Table~\ref{kintable}. During the 8.5 GeV$^2$ running, a number of slight changes in accelerator tune to optimize the performance of CEBAF in the context of simultaneous delivery of longitudinally polarized beam to Halls A and C at different passes resulted in several slight changes in beam energy at the 1-2 MeV level. While no additional arc energy measurements were performed, the small, occasional changes in beam energy were detected by the online beam energy monitoring system, and also confirmed by shifts in the elastic peak position in the variables used for elastic event selection in the offline analysis (see section~\ref{sec:elastic_selection}). These small changes were included in the final beam energy database for the offline analysis. Except for the first few days at 8.5 GeV$^2$, during which the beam energy was 5.717 GeV, the actual incident beam energy varied between 5.710 and 5.714 GeV during most of the 8.5 GeV$^2$ running, averaging 5.712 GeV. The incident beam energy was stable at 5.711 GeV during the 6.8 GeV$^2$ running. As discussed in section~\ref{sec:elastic_selection} and Ref.~\cite{Puckett:2017egz}, the contribution of the systematic uncertainty in the beam energy to the total systematic uncertainties in the polarization transfer observables is a small fraction of the total. 

The target system used for this experiment consists of several different solid targets
and a three-loop cryogenic target system for liquid hydrogen (LH$_2$). The solid targets include thin foils of Carbon and/or Aluminum used for spectrometer optics calibrations and to measure the contribution of the walls of the cryotarget cell to the experiment
background. The spectrometer optics calibrations and systematic studies are described in detail in Ref.~\cite{Puckett:2017egz}, while details of the solid targets are described in Ref.~\cite{Puckett:2015soa}. For the first kinematic point taken from Nov. 7-20, 2007, 
a 15-cm LH$_2$ cryotarget cell was used. For all of the other production kinematics of both experiments,
a 20-cm cryotarget cell was used. The center of the 20-cm cell was offset 3.84~cm downstream of the
origin along the beamline to allow electrons scattered by up to 120 degrees
to exit through the thin scattering chamber exit window and be detected by the calorimeter. 
The liquid hydrogen targets were operated at a constant temperature of 19 K and nominal density of $\rho \approx 0.072$ g/cm$^3$ throughout the experiment. The size of the beam spot on target was enlarged to a transverse size of typically $2 \times 2$~mm$^2$ by the Hall C fast raster magnet system, to minimize localized heating and boiling of the liquid hydrogen and resulting fluctuations in target density and luminosity. More details of the cryogenic target system can be found in Ref.~\cite{Puckett:2015soa}. 

\subsection{Hall C HMS}

\label{subsec:HMS}
The High Momentum Spectrometer (HMS) is part of the standard experimental equipment in JLab's Hall C. It is a superconducting magnetic spectrometer with three quadrupoles and one dipole arranged in a QQQD layout. The HMS has a 25-degree central vertical bend angle and point-to-point focusing in both the dispersive and non-dispersive planes when operated in its ``standard'' tune. The HMS dipole field is regulated by an NMR probe and is stable at the 10$^{-5}$ level, while the quadrupole magnet power supplies are regulated by current and are stable at the 10$^{-4}$ level. The HMS solid angle acceptance is approximately 6.74 msr when used with the larger of its two retractable, acceptance-defining octagonal collimators, as it was in this experiment. The HMS momentum acceptance is approximately $\pm 9\%$ relative to the central momentum setting. The maximum central momentum setting is 7.4 GeV/c. The HMS detector package and superconducting magnets are supported on a common carriage that rotates on concentric rails about the central pivot of Hall C. The detector package is located inside a concrete shield hut supported on a separate carriage from the detector and magnet supports. With the exception of small air gaps between the scattering chamber exit window and the HMS entrance window and between the HMS dipole exit window and the first HMS drift chamber, the entire flight path of charged particles through the HMS is under vacuum, minimizing energy loss and multiple scattering prior to the measurement of charged particle trajectories. 

As shown in Fig.~\ref{fig:fppschematic}, the HMS detector package was modified by removing the gas Cherenkov counter and the two rearmost planes of scintillator hodoscopes from the standard HMS detector package to accommodate the Focal Plane Polarimeter (FPP), leaving only the two upstream planes of scintillators (``S1X'' and ``S1Y'') to form a fast trigger. The HMS calorimeter was not removed, and its signals were recorded to the data stream, but it was not used either in the trigger or in the offline analysis, except for crude pion rejection in the analysis of the HMS optics calibration data, for which the HMS was set with negative polarity for electron detection. The standard HMS drift chambers, described in detail in Ref.~\cite{Baker:1995ky}, were used to measure the trajectories of elastically scattered protons. The measured proton tracks were then used to reconstruct the event kinematics at the target and to define the incident trajectory for the secondary polarization-analyzing scattering in the CH$_2$ analyzers of the FPP. Because the two rear planes of scintillators had been removed, the ``S1X'' and ``S1Y'' planes could not, by themselves, provide an adequately selective trigger for most kinematic settings of the experiment. To overcome this challenge, two additional 1 cm-thick plastic scintillator paddles were installed between the exit window of the HMS vacuum and the first HMS drift chamber, with sufficient area to cover the envelope of elastically scattered protons for all kinematic settings. These two paddles were collectively referred to as ``S0''. The S0 plane reduced the trigger rate to a manageable level by restricting the acceptance to the region populated by elastically scattered protons and suppressing triggers due to inelastic processes that occur at a much higher rate for large $Q^2$ values.  During most of the experiment, the HMS trigger required at least one paddle to fire in each of the ``S1X'', ``S1Y'' and ``S0'' planes. During part of the measurement at $E_e = 2.847$ GeV and the entire duration of the measurements at $E_e = 3.548$ GeV and $E_e = 3.680$ GeV, for which the HMS was located at relatively large scattering angles, the trigger was based on ``S1X'' and ``S1Y'' only, as the rates were low enough to use this less-selective trigger in coincidence with the electron calorimeter. The price to pay for installing the S0 trigger plane upstream of the drift chambers is that the angular resolution of the HMS was significantly degraded due to the additional multiple scattering in S0~\cite{Puckett:2015soa}. More details of the custom HMS trigger logic used for these experiments are given in Ref.~\cite{Puckett:2015soa}.


\subsection{Focal Plane Polarimeter}
\label{sec:FPP}

\begin{figure}
  \begin{center}
    \includegraphics[width=.98\columnwidth]{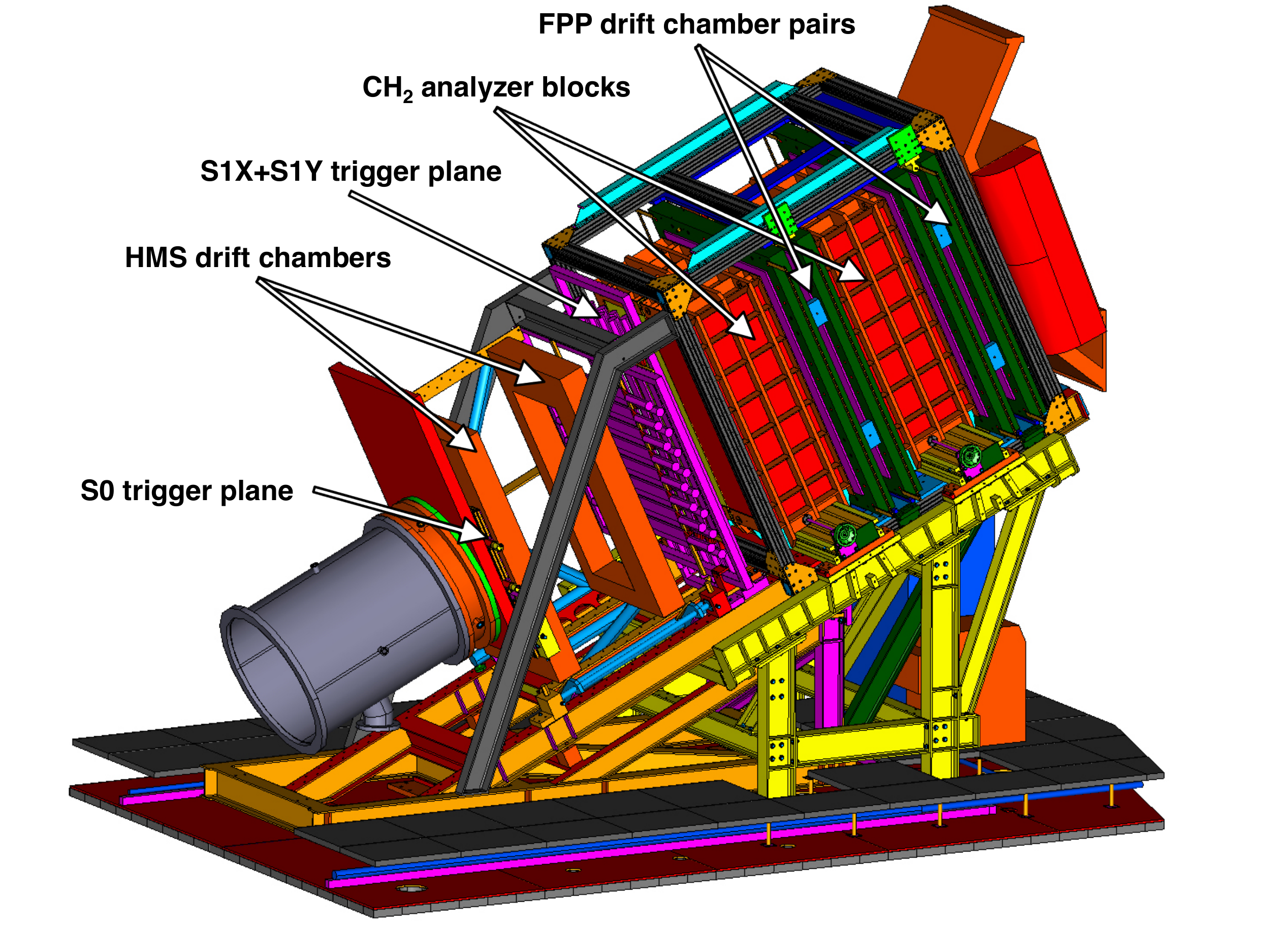}
  \end{center}
  \caption{ Design drawing of the FPP installed in the HMS detector
    package, with the HMS drift chambers and the trigger planes. \label{fig:fppschematic}}
\end{figure}

A new focal plane polarimeter (FPP) was designed, built and installed in the HMS to measure the polarization of the recoiling protons.
It consists of two CH$_{2}$ analyzer blocks arranged in series to increase the efficiency, each followed by a pair 
of drift chambers. A design drawing of the HMS detector package with
the FPP, the HMS drift chambers and the trigger scintillator planes is shown
in Figure \ref{fig:fppschematic}.

\subsubsection{FPP Analyzer}
The FPP analyzer is made of polyethylene (CH$_2$). It consists of two retractable
doors, each made of two blocks, allowing for the collection of ``straight-through'' trajectories for calibration and 
alignment studies. Each pair is 145 cm (tall)$\times$ 111 cm (wide)$\times$ 55 cm 
(thick) and made of several layers of CH$_{2}$ held together by an outer aluminum frame. To reduce the 
occurrence of leakage through the seam when the doors are inserted, an
overlapping step was designed into the edge of both doors. Given their substantial
weight, the CH$_2$ blocks were supported on a different frame than the detector and 
attached directly to the floor of the shield hut, ensuring that the 
other detectors did not move while inserting or retracting the doors.

The choice of CH$_2$ as the analyzer material was driven by a compromise among the analyzing power and optimal thickness of the material on the one hand, and the cost and space constraints within the HMS hut on the other. Measurements of the analyzing power of the reaction $\vec{p}+$CH$_2 \rightarrow X$ at 
Dubna \cite{Azhgirei:2005cs} showed that the overall figure of merit of the polarimeter does not increase when the analyzer thickness is increased beyond the nuclear collision length $\lambda_T$ of CH$_2$. With this result in mind, the HMS FPP was designed as a double polarimeter with two analyzers, each approximately one $\lambda_T$ thick and followed by pairs of drift chambers to measure the angular distribution of scattered protons. The analyzers and the drift chambers were designed to be large enough to have $2\pi$ azimuthal angular acceptance for transverse momenta $p_T \equiv p\sin \vartheta$ up to 0.7 GeV/c, beyond which the polarimeter figure of merit essentially saturates.  

\subsubsection{FPP drift chambers}
The tracking system of the FPP consists of two drift chamber pairs, one after each analyzer
block. All four chambers are identical in design and construction. The active area of each chamber is 164 cm (tall) $\times$ 132 cm (wide). Each chamber contains three detection planes sandwiched between cathode layers. Each detection
layer consists of alternating sense wires and field wires with a
spacing of 2 cm between adjacent sense wires (1 cm between a sense
wire and its neighboring field wires). The wire spacing in the cathode
layers, located 0.8 cm above and below the detection layers, is 3 mm.
The characteristics of the different wires are given in Table~\ref{fppwires}. 
\begin{table}
  \caption{\label{fppwires} Characteristics of the wires used in the 
    FPP drift chambers. The sense wires are gold-plated tungsten, while the cathode and field wires are made of a beryllium-bronze alloy.}
  \begin{ruledtabular}
    \begin{tabular}{ccc}
      Type &   Diameter ($\mu$m)  & Tension (g)  \\ \hline
      Sense &   30        &   70     \\ 
      Field   &  100       &   150    \\ 
      Cathode &  80        &   120    
    \end{tabular}
  \end{ruledtabular}
\end{table}
The sense wire planes have three different
orientations, denoted ``U'', ``V'', and ``X''. The stacking order along the $z$ axis of
the planes in each chamber is VXU. The ``V'' wires are strung along the $+45^\circ$ line relative to the $x$ axis and thus measure the coordinate along the $-45^\circ$-line; i.e., $v \equiv \frac{x-y}{\sqrt{2}}$. The ``X'' wires are strung perpendicular to the $x$ axis
and thus measure the $x$ coordinate. The ``U'' wires are strung along the $-45^\circ$ line relative to the $x$ axis and thus measure the coordinate $u \equiv \frac{x+y}{\sqrt{2}}$. 
The U and V layers have 104 sense wires each, while the X layers have 83 sense wires. Each layer within each chamber has a sense wire passing through the point $(x,y) = (0,0)$, the geometric center of the chamber active area\footnote{The symmetry created by this common intersection point and the relative lack of redundancy of coordinate measurements, with only six coordinate measurements along each track, creates an essentially unresolvable left-right ambiguity for a small fraction of tracks passing through the region near the center of the chambers at close to normal incidence, for which two mirror-image solutions of the left-right ambiguity exist with identical combinations of drift distances that are basically indistinguishable in terms of $\chi^2$.}. 

Each drift chamber is enclosed by 30 $\mu$m-thick aluminized mylar gas windows and a rigid aluminum frame. Each pair of chambers 
is attached to a common set of rigid spacer blocks (two on each side of the chamber frame) by a set of two aligning bolts per block penetrating each chamber. Each of the two spacer blocks along both the top and bottom sides of the chamber frame is also attached to a third threaded steel rod that goes through both chambers in the pair. The chamber pair is then mounted to the FPP support frame via C-shaped channels machined into the top spacer blocks that mate with a cylindrical Thomson rail attached to the top of the support frame, and via protrusions of the bottom spacer blocks with guide wheels that slide into a ``U'' channel on the bottom of the FPP support frame. After installation, each chamber pair was bolted to a hard mechanical stop built into the support frame. The design ensures that the relative positioning of the two chambers within a pair is fixed and reproducible. 

The FPP drift chambers used the same 50\%/50\% argon/ethane gas mixture as the HMS drift chambers. The basic drift cell in the FPP drift chambers has the same aspect ratio as the HMS drift cell, but the dimensions are twice as large. The cathode and field wires were maintained at a constant high voltage of -2400 V, while the sense wires were at ground potential.  This operational configuration gives the FPP drift chambers similar, but not identical, electric field and drift velocity characteristics to the HMS drift chambers. The main difference is that the HMS drift chambers were operated with a different electric field configuration in which three different high voltage settings were applied to the field and cathode wires according to their distance from the nearest sense wire, leading to nearly cylindrical equipotential surfaces surrounding each sense wire. This in turn means that the drift time measured by the HMS chambers is a function of the distance of closest approach of the track to the wire, rather than the in-plane track-wire distance. Since the tracks of interest in the HMS drift chambers are very nearly perpendicular to the wire planes, the difference between these two distances is small in any case. 

The FPP wire signals are processed by front-end amplifier/discriminator (A/D) cards attached directly to the chambers. Each A/D card processes the signals from eight sense wires. The amplified, discriminated FPP signals are digitized by TDCs located close to the chambers within the HMS shield hut. 
A significant advantage of the Hall C FPP DAQ system compared to previous experiments using the Hall A FPP~\cite{Punjabi:2005wq,Puckett:2011xg} is that each sense wire was read out individually by a dedicated multi-hit TDC channel, whereas the straw chamber signals in the Hall A FPP were multiplexed in groups of eight wires by the front-end electronics to reduce the number of readout channels required, effectively preventing the resolution of multi-track events in which two or more tracks create simultaneous signals on straws located within the same group of eight. As discussed in Sec.~\ref{subsubsec:Ay}, the ability to isolate true single-track events significantly increased the effective analyzing power of the Hall C FPP relative to the Hall A FPP for equivalent analyzer material and thickness. From the start of the experiment in October 2007 to February 2008, VME-based F1 TDC modules~\cite{F1TDC} housed in a pair of VME crates in the HMS shield hut were used to read out the FPP signals. For the high-$Q^2$ data collection from April to early June of 2008, the FPP signals were read out using LeCroy 1877-model Fastbus TDCs. The FPP data acquisition was changed from VME to Fastbus TDCs due to relatively frequent malfunctions of the VME DAQ system encountered during the GEp-2$\gamma$ production running, especially for the data taken at the relatively forward HMS central angle of 14.5 degrees, for which the detector hut was fairly close to the beam dump and the hit rates in the FPP chambers were relatively high. Since no such problems were observed with the Fastbus TDCs used concurrently to read out the HMS drift chambers, a second Fastbus crate equipped with LeCroy 1877 TDC modules was installed in the HMS shield hut during the planned two-month accelerator shutdown\footnote{The purpose of this accelerator down was to install refurbished cryomodules in CEBAF to reach the maximum beam energy of 5.7 GeV needed for the high-$Q^2$ running of GEp-III.} in February and March of 2008 in preparation for the high-$Q^2$ running at an HMS angle of 11.6 degrees. As expected based on the experience with the HMS drift chamber readout, the Fastbus TDC readout for the FPP drift chambers functioned fairly smoothly throughout the 2008 high-$Q^2$ running.

\subsection{Electron Calorimeter}

Elastically scattered electrons were detected by an electromagnetic calorimeter, named BigCal, built specifically for this experiment. The calorimeter was made of 1,744 lead-glass blocks (TF1-0 type) stacked 
with a frontal area of $122 \times 218$~cm$^2$. The array was constructed from blocks of two different sizes. The bottom part of the calorimeter consisted of a $32 \times 32$ array of blocks with dimensions of $3.8 \times 3.8 \times 45$~cm$^3$ originating from the IHEP in Protvino, Russia, while the top part of the calorimeter consisted of a $30 \times 24$ array of blocks with dimensions of $4 \times 4 \times 40$~cm$^3$ from the Yerevan Physics Institute in Yerevan, Armenia, used previously in a Compton scattering measurement in Hall A~\cite{Shahinyan:2007qg}. The 45-cm (40-cm) depth of the Protvino (Yerevan) blocks corresponds to 16.4 (14.6) radiation lengths, sufficient to absorb the total energy of elastically scattered electrons. The Cherenkov light created in the glass by relativistic particles 
from the electromagnetic cascade was registered by photomultiplier tubes (PMTs) of type FEU-84, coupled optically to the end of each block with a $5$~mm-thick transparent silicon "cookie" to compensate for a possible misalignment between the two elements. The blocks were optically isolated from each other via an aluminized mylar wrapping. For each kinematic setting, the calorimeter was positioned at an angle corresponding to the central $Q^2$ value and beam energy. The distance from the origin to the surface of BigCal was chosen to be as large as possible, consistent with matching between the solid angle acceptance of BigCal for elastically scattered electrons and the fixed solid angle of the HMS for elastically scattered protons. For the kinematics at $E_e = 3.548$ GeV and 3.680 GeV (see Tab.~\ref{kintable}), BigCal was placed closer to the target than the acceptance-matching distance due to limitations imposed by the signal cable length and the location of the BigCal readout electronics, as well as the available space in Hall C. At $Q^2 = 8.5$ GeV$^2$, the electron solid angle for acceptance matching was 143 msr, or about twenty times the solid angle acceptance of the HMS.
\begin{figure}
\centering
\includegraphics[width=.85\columnwidth]{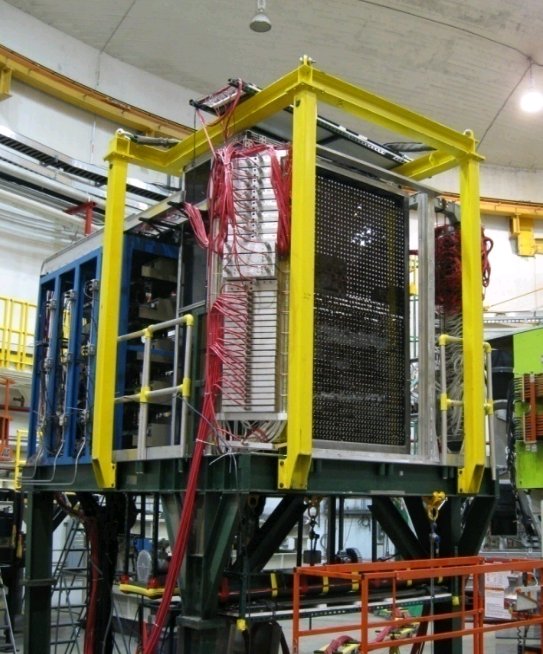}
\caption{ BigCal calorimeter with its front aluminum shielding plates removed, exposing the stack of 1744 lead glass blocks.
\label{fig:BigCal}}
\end{figure}
Fig.~\ref{fig:BigCal} shows BigCal with the front shielding plates removed, revealing the array of lead-glass blocks.

The analog signals from the PMTs were sent to specialized NIM modules for amplification and summing, with eight input channels each. The outputs included copies of the individual input signals amplified by a factor of 4.2, and several copies of the analog sum of the eight input signals. The amplified analog signals from the individual PMTs were sent to LeCroy model 1881M charge-integrating Fastbus ADCs for readout. One copy of each ``first level'' sum of eight blocks was sent to a fixed-threshold discriminator, the output of which was then sent to a TDC for timing readout. Additional copies of each sum of eight were combined with other sums-of-eight into ``second-level'' sums of up to 64 blocks using identical analog summing modules. These ``level 2'' sums, of which there are a total of 38, were also sent to fixed-threshold discriminators, and a global ``OR'' of all the second-level discriminator outputs was used to define the trigger for BigCal. The groupings of blocks for the ``level 2'' sums were organized with partial overlap to avoid regions of trigger inefficiency, as detailed in Ref.~\cite{Puckett:2015soa}. Because there was no overlap in the trigger logic between the left and right halves of the calorimeter, the trigger threshold was limited to slightly less than half of the average elastically scattered electron energy. A higher threshold would have resulted in significant efficiency losses at the boundary between the left and right halves of the calorimeter. 

Four one-inch thick aluminum plates 
(for a total of about one radiation length) were installed in front
of the glass to absorb low-energy photons and mitigate radiation damage to the glass.
This additional material degrades the energy resolution, but does not significantly affect the position resolution. 
All four aluminum plates were used for all kinematics except the lowest $\epsilon $
point of the GEp-2$\gamma$ experiment, for which only one plate was used. 
For this setting, the calorimeter was placed at the backward angle of $\theta_e \approx 105^\circ$, for which the elastically scattered electron energy was only $E'_e \approx 0.54$ GeV, the radiation dose rate in the lead-glass was low enough that the additional shielding was not needed, and the better energy resolution afforded by removing three of the four plates was needed to maintain high trigger efficiency at the operating threshold.

The glass transparency gradually deteriorated throughout the experiment due to accumulated radiation damage. The effective gain/signal strength in the BigCal blocks was monitored \emph{in situ} throughout the experiment using the known energy of elastically scattered electrons, reconstructed precisely from the measured proton kinematics. The PMT high voltages were periodically increased to compensate for the gradual decrease in light yield and maintain a roughly constant absolute signal size, in order to avoid drifts in the effective trigger threshold and other deleterious effects. However, as discussed in Ref.~\cite{Puckett:2017egz}, the reduced photoelectron yield caused the energy resolution to deteriorate. With the four-inch-thick aluminum absorber in place, the energy resolution worsened from about $10.9\%/\sqrt{E}$ following the initial calibration to roughly $22\%/\sqrt{E}$ at the end of the experiment. During the early 2008 accelerator shutdown, the glass was 
partially annealed using a UV lamp system but it did not fully recover to its initial transparency and energy resolution prior to the start of the high-$Q^2$ running in April 2008, at which point the transparency resumed its gradual deterioration. The achieved energy resolution, while relatively poor for this type of detector and dramatically worsened by radiation damage, was nonetheless adequate for triggering with the threshold set at half the elastically scattered electron energy or less. In contrast to the energy resolution, the position resolution of BigCal, estimated to be roughly 6 mm using the $Q^2 = 6.8$ GeV$^2$ data collected at the end of the experiment~\cite{Puckett:2017egz,Puckett:2015soa}, did not change noticeably during the experiment. The achieved coordinate resolution of BigCal was significantly better than needed given the experimentally realized angular, momentum and vertex resolution of the HMS, and proved essential for the suppression of the inelastic background, especially at high $Q^2$, as discussed in section~\ref{sec:elastic_selection}. More details of the calibration and event reconstruction procedures for BigCal can be found in Refs.~\cite{Puckett:2017egz,Puckett:2015soa}.

\section{Data analysis}
\label{sec:analysis}

The analysis of the data proceeds in three phases: 
\begin{enumerate}
\item Decoding of the raw data and the reconstruction of events 
\item The selection of elastic $ep$ events and the estimation of the residual contamination of the final sample by inelastic backgrounds and accidental coincidences
\item The extraction of the polarization transfer observables from the measured angular distributions of protons scattered in the FPP. 
\end{enumerate}
The raw data decoding and the event reconstruction procedure, including detector calibrations and reconstruction algorithms, are described in the technical supplement to this article~\cite{Puckett:2017egz} as well as the Ph.D. thesis~\cite{Puckett:2015soa}. The elastic event selection and background estimation procedure are discussed in Sec.~\ref{sec:elastic_selection}. The extraction of polarization observables is presented in Sec.~\ref{sec:PolarizationAnalysis}. The detailed evaluation of systematic uncertainties is presented in Refs.~\cite{Puckett:2017egz,Puckett:2015soa}.

\subsection{Elastic event selection}
\label{sec:elastic_selection}
Elastic events were selected using the two-body kinematic correlations between the electron and the proton. Accidental coincidences were suppressed by applying a loose, $\pm 10$ ns cut to the time-of-flight-corrected difference $\Delta t$ between the timing signals associated with the electron shower in BigCal and the proton trigger in the fast scintillator hodoscopes of the HMS. The resolution of the coincidence time difference $\Delta t$ is dominated by the timing resolution of BigCal, which varied from $1.5-2$ ns depending on the electron energy. 
The contamination of the data by accidental coincidences within the $\pm 10$ ns cut region was less than 10\% before applying the exclusivity cuts described below, and negligible after applying the cuts. The transferred polarization components for the accidental coincidence events were found to be similar to those of the real coincidence events for the inelastic background~\cite{Luo:2011uy}, such that the accidental contamination of the inelastic background sample at the level of $10\%$ or less did not noticeably affect the corrections to the elastic $ep$ signal polarizations, which were essentially negligible except at $Q^2 = 8.5$ GeV$^2$. 
\begin{figure*}
  \begin{center}  
    \includegraphics[width=\textwidth]{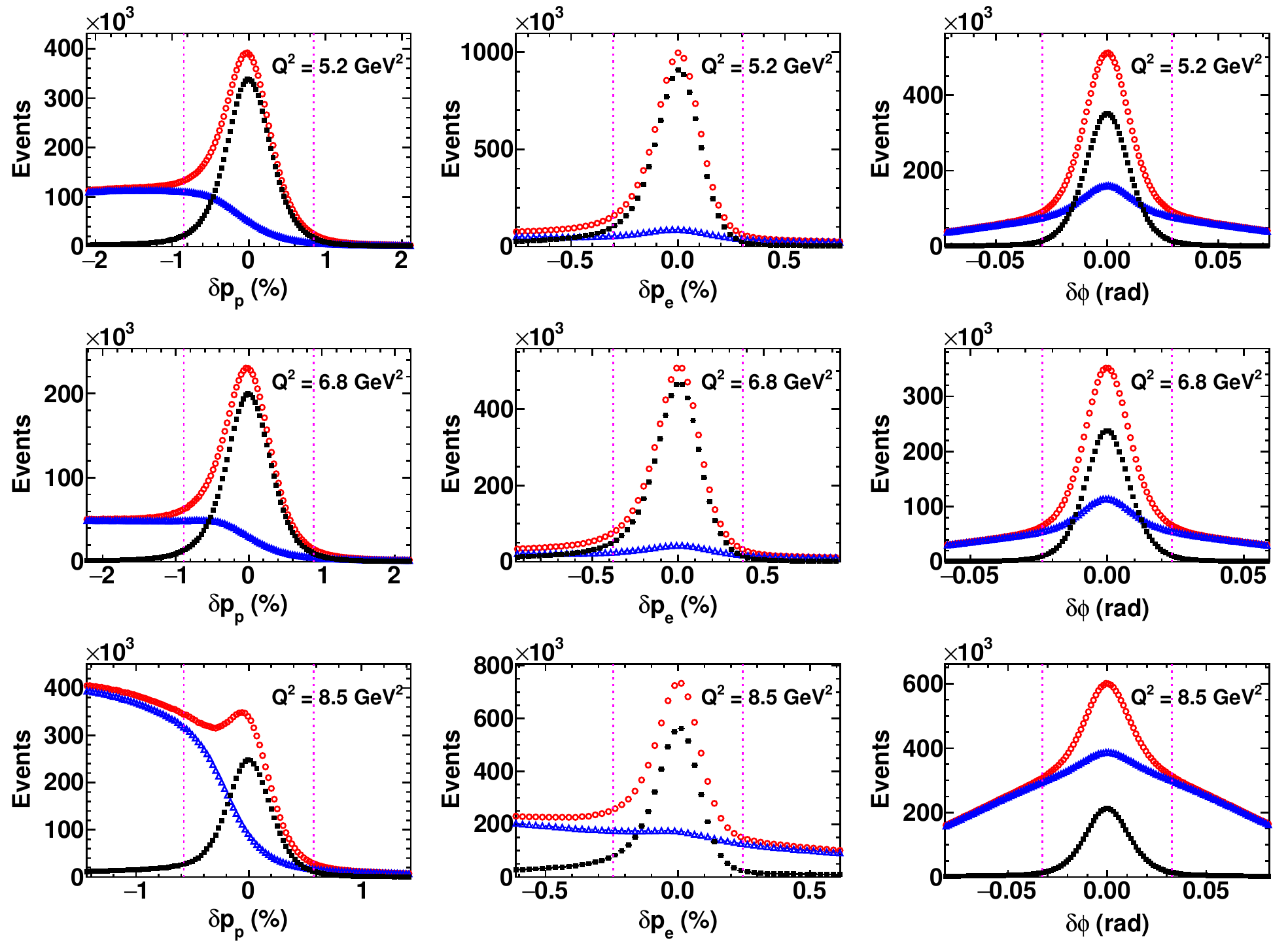}
  \end{center}      
  \caption{\label{fig:exclusivity_cuts_gep3}  Simplified illustration of elastic event selection for the GEp-III kinematics: $Q^2 = 5.2$ GeV$^2$ (top row), $Q^2 = 6.8$ GeV$^2$ (middle row) and $Q^2 = 8.5$ GeV$^2$ (bottom row). Exclusivity cut variables are $\delta p_p \equiv 100 \times \frac{p_p - p_p(\theta_p)}{p_0}$ (left column), $\delta p_e \equiv 100 \times \frac{p_p - p_p(\theta_e)}{p_0}$ (middle column), and $\delta \phi \equiv \phi_e - \phi_p - \pi$ (right column). The distribution of each variable is shown for all events (red empty circles), events selected by applying $\pm3\sigma$ cuts of fixed width to \emph{both} of the other two variables (black filled squares), and events rejected by these cuts (blue empty triangles). Vertical dotted lines indicate the $\pm3\sigma$ cut applied to each variable. Similar plots for the GEp-2$\gamma$ kinematics can be found in Ref.~\cite{Puckett:2017egz}. Note that the horizontal axis range in each plot is a fixed multiple of the elastic peak width, which varies with $Q^2$ and $E_e$.}
\end{figure*}

The beam energy is known with an absolute accuracy $\Delta E/E \lesssim 5 \times 10^{-4}$ from the standard Hall C ``arc'' measurement technique. The ``per-bunch'' beam energy spread under normal accelerator operating conditions is typically less than $3 \times 10^{-5}$ and is continuously monitored using synchrotron light interferometry~\cite{Chevtsov:2005ed}, while the CEBAF fast energy feedback system maintains the ``long term'' stability of the central beam energy at the 10$^{-4}$ level~\cite{Krafft:2000nm}. The spread and systematic uncertainty in the electron beam energy is significantly smaller than the HMS momentum resolution of $\sigma_p/p \approx 10^{-3}$, and its contribution to the systematic uncertainty in the determination of the reaction kinematics is small. 

The scattering angles and energies/momenta of both outgoing particles are measured in each event. Because the energy resolution of BigCal was too poor to provide meaningful separation between elastic and inelastic events for any cut with a high efficiency for elastic events, no cuts were applied to the measured energy of the electron, beyond the hardware threshold imposed by the BigCal trigger and the software threshold imposed by the clustering algorithm. This leaves the proton momentum and the polar and azimuthal scattering angles of the electron and proton as useful kinematic quantities for the identification of elastic events. 

Figure~\ref{fig:exclusivity_cuts_gep3} shows a simplified version of the procedure for isolating elastic $ep$ events in the GEp-III data using the two-body kinematic correlations between the electron detected in BigCal and the proton detected in the HMS. Similar plots for the GEp-2$\gamma$ kinematics can be found in Ref.~\cite{Puckett:2017egz}. The proton momentum $p_p$ and scattering angle $\theta_p$ in elastic scattering are related by:  
\begin{eqnarray}
  p_p(\theta_p) &=& \frac{2M_pE_e(M_p+E_e)\cos (\theta_p)}{M_p^2 + 2M_p E_e + E_e^2 \sin^2 (\theta_p)} \label{eq:pp_ptheta}.
\end{eqnarray}
The difference $\delta p_p \equiv 100 \times \frac{p_p-p_p(\theta_p)}{p_0}$, where $p_0$ is the central momentum of the HMS, provides a measure of ``inelasticity'' for the detected proton independent of any measurement of the electron kinematics. The $\delta p_p$ spectra exhibit significant inelastic backgrounds before applying cuts based on the measured electron scattering angles, especially at $Q^2 = 8.5$ GeV$^2$. 

The scattered electron's trajectory is defined by the straight line from the reconstructed interaction vertex to the measured electron impact coordinates at the surface of BigCal. 
The correlation between the electron polar scattering angle $\theta_e$ and the proton momentum $p_p$ was expressed in terms of the difference $\delta p_e \equiv 100 \times \frac{p_p - p_p(\theta_e)}{p_0}$, where $p_p(\theta_e)$ is calculated from elastic kinematics as follows:
\begin{eqnarray}
  E'_e(\theta_e) &=& \frac{E_e}{1+\frac{E_e}{M_p}(1-\cos \theta_e)}, \nonumber \\
  Q^2(\theta_e) &=& 2E_e E'_e(\theta_e) (1-\cos \theta_e), \nonumber \\
  p_p (\theta_e) &=& \sqrt{Q^2(\theta_e)\left(1+\tau(\theta_e)\right)},
\end{eqnarray}
with $\tau(\theta_e) \equiv \frac{Q^2(\theta_e)}{4M_p^2}$. Finally, coplanarity of the outgoing electron and proton is enforced by applying a cut to $\delta \phi \equiv \phi_e - \phi_p - \pi$. The azimuthal angles of the detected particles are defined in a global coordinate system in which the distribution of $\phi_e$ ($\phi_p$) is centered at $+\pi/2$ ($-\pi/2$), such that co-planarity implies $\phi_e = \phi_p + \pi$ for all elastic $ep$ events within the detector acceptances.

The simplified elastic event selection procedure shown in Fig.~\ref{fig:exclusivity_cuts_gep3} corresponds to fixed-width, $\pm3\sigma$ cuts centered at zero for all variables. It should be noted, however, that for the final analysis, cuts of variable width (mean) were applied to $\delta p_p$ ($\delta \phi$) to account for observed variations of the width (position) of the elastic peak within the HMS acceptance (for details, see~\cite{Puckett:2017egz}). While the differences in statistics and analysis results between the full procedure and the simple procedure of Fig.~\ref{fig:exclusivity_cuts_gep3} are small for sufficiently wide cuts, the full procedure optimizes the effective signal-to-background ratio and efficiency of the elastic event selection procedure, and suppresses cut-induced systematic bias in the reconstructed proton kinematics. 

\begin{figure}
  \begin{center}
    \includegraphics[width=0.85\columnwidth]{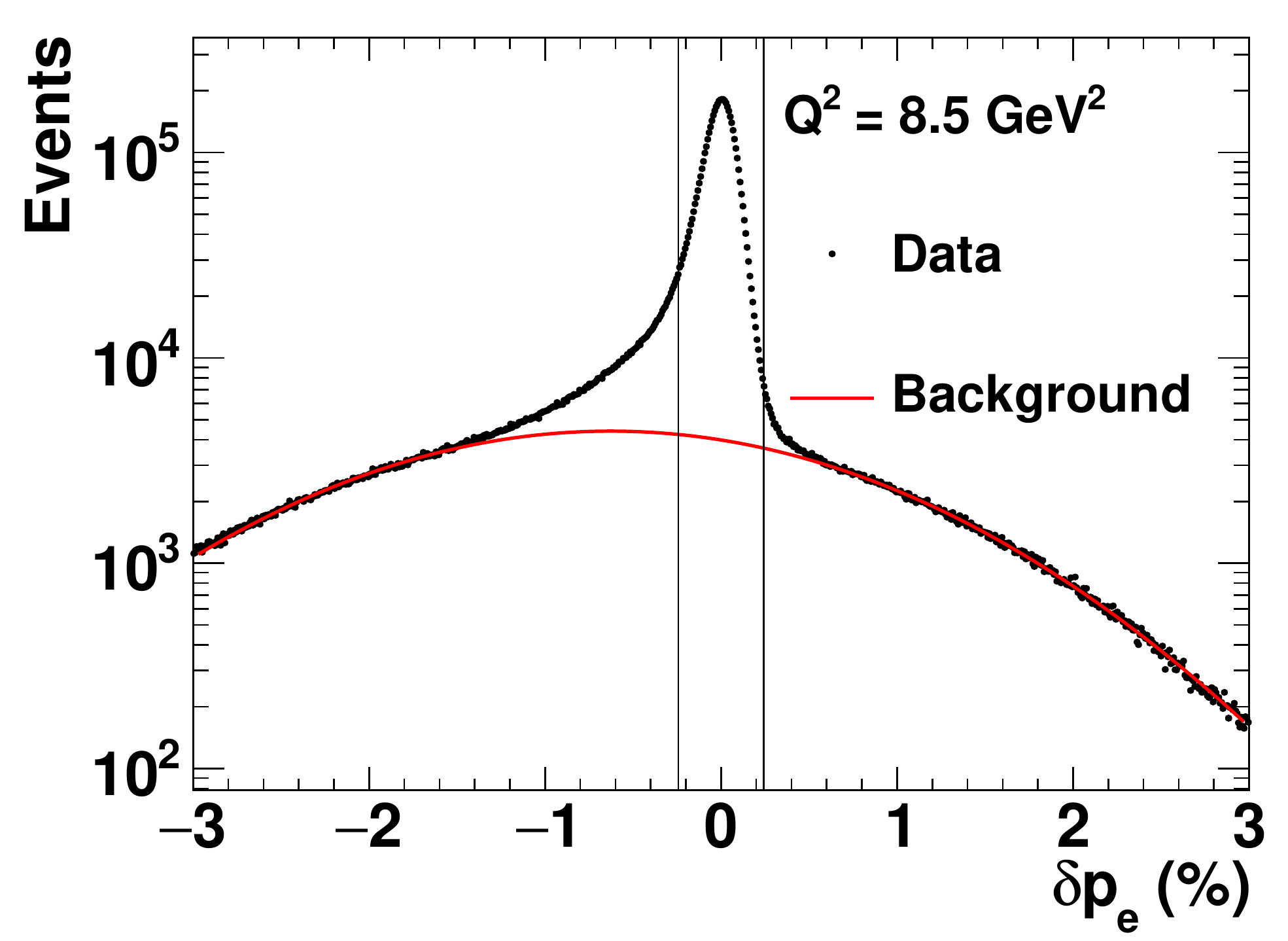}
  \end{center}
  \caption{\label{fig:backgroundestimationexample}  Example of the Gaussian sideband fit of the $\delta p_e$ distribution used to estimate the residual background contamination of the final elastic event selection cuts at $Q^2 = 8.5$ GeV$^2$. Data (black filled circles) are shown after applying $\pm3\sigma$ cuts to both $\delta p_p$ and $\delta \phi$. In this example, the estimated fractional background contamination, integrated within the $\pm3\sigma$ cut region (black vertical lines), is $f \equiv \frac{B}{S+B} = (4.89 \pm 0.01) \%$, where $S$ and $B$ refer to the signal and the background, respectively, and the quoted uncertainty is statistical only. See text for details.}
\end{figure}

In contrast to $\delta p_p$ and $\delta \phi$, the resolution of $\delta p_e$ is approximately constant within the acceptance, and mostly dominated by the HMS momentum resolution. 
In general, the observed correlations of $\delta p_e$ with the reconstructed proton kinematics are small compared to experimental resolution. Moreover, the extracted polarization transfer observables are generally less sensitive to the systematic error in the reconstructed proton momentum than to the errors in the reconstructed proton angles, which dominate the experimental resolution of $\delta p_p$ and $\delta \phi$. The results are thus less susceptible to systematic bias induced by the $\delta p_e$ cut than that induced by the $\delta p_p$ and $\delta \phi$ cuts, given the experimentally realized angular and momentum resolution of the HMS. Therefore, a fixed-width, $\pm3\sigma$ cut centered at zero was applied to $\delta p_e$ for all kinematics, which has the added benefit of simplifying the estimation of the residual background contamination of the final elastic event sample, as shown in Fig.~\ref{fig:backgroundestimationexample} and discussed below.  

For electron scattering from hydrogen, elastically scattered protons have the highest kinematically allowed momenta for positively charged particles at a given $\theta_p$. Events at $\delta p_p < 0$ are dominated by inelastic reactions on hydrogen, including $\pi^0$ photoproduction ($\gamma p \rightarrow \pi^0 p$) near the Bremsstrahlung end point ($E_\gamma \rightarrow E_e$), with one or both $\pi^0$ decay photons detected by BigCal, and, to a lesser extent, $\pi^0$ electroproduction ($ep \rightarrow e' \pi^0 p$) near threshold, with the scattered electron detected in BigCal. At the multi-GeV energies characteristic of these experiments, the kinematic separation between the $ep$ and $\pi^0p$ reactions in terms of $\delta p_p$ is comparable to the experimental resolution, such that there is significant overlap between the $\pi^0p$ and $ep$ reactions in the vicinity of the elastic peak. The 20-cm liquid hydrogen target is itself a $\sim2.2\%$ radiator, creating a significant ``external'' Bremsstrahlung flux along the target length in addition to the real and virtual photon flux present in the electron beam independent of the target thickness\footnote{For example, at $Q^2 = 8.5$ GeV$^2$, the observed fractional contamination by inelastic backgrounds of the final sample of events selected as elastic increases by a factor of 1.6 from the upstream end of the target to its downstream end.}. 

Events at positive $\delta p_p$ (the so-called ``super-elastic'' region) originate from quasi-elastic and inelastic scattering in the aluminum entry and exit windows of the liquid hydrogen target cell, and from non-Gaussian tails of the HMS angular and/or momentum resolution. Because the aluminum window thickness is only $\sim$5\% of the total target thickness by mass (12\% by radiation length), and the exclusivity cut variables are smeared by Fermi motion of the nucleons in aluminum, the contribution of scattering from the target end windows to the total event yield is essentially negligible ($\lesssim 10^{-3}$) after the cuts. 


The residual peaks at zero in the $\delta p_e$ and $\delta \phi$ spectra of rejected events result from radiative effects and non-Gaussian tails of the experimental resolution. In particular, the remnant peaks in the $\delta \phi$ distributions of rejected events contain significant contributions from the elastic radiative tail, because events affected by radiation from the incident electron beam (coherent or incoherent with the hard scattering amplitude) are strongly suppressed by both the $\delta p_e$ and $\delta p_p$ cuts without affecting the co-planarity of the outgoing particles. 

Figure~\ref{fig:backgroundestimationexample} illustrates the procedure for estimating the residual background contamination in the final sample of elastic events. By far the worst case for background contamination after applying exclusivity cuts is $Q^2 = 8.5$ GeV$^2$, for which the contamination approaches 5\% for $\pm3\sigma$ cuts. The $\delta p_e$ distribution of the background in the vicinity of the elastic peak after applying cuts to $\delta p_p$ and $\delta \phi$ is well approximated by a Gaussian distribution, as was confirmed by examining the events rejected by the $\delta p_p$ and/or $\delta \phi$ cuts, as well as by Monte Carlo simulations of the main background processes. The shape of the elastic $ep$ radiative tail in the $\delta p_e$ distribution was also well-reproduced by Monte Carlo simulations with radiative corrections to the unpolarized cross section following the formalism described in Ref.~\cite{Ent:2001hm}. 
In Fig.~\ref{fig:backgroundestimationexample}, the $\delta p_e$ distribution of the background was fitted with a Gaussian by excluding the region $-1.6\% \le \delta p_e \le 0.5\%$ in which the elastic peak and radiative tail contributions are significant. The residual background contamination was then estimated by extrapolating the Gaussian fit of the background into the elastic peak region.

\begin{figure}
  \begin{center}
    \includegraphics[width=0.85\columnwidth]{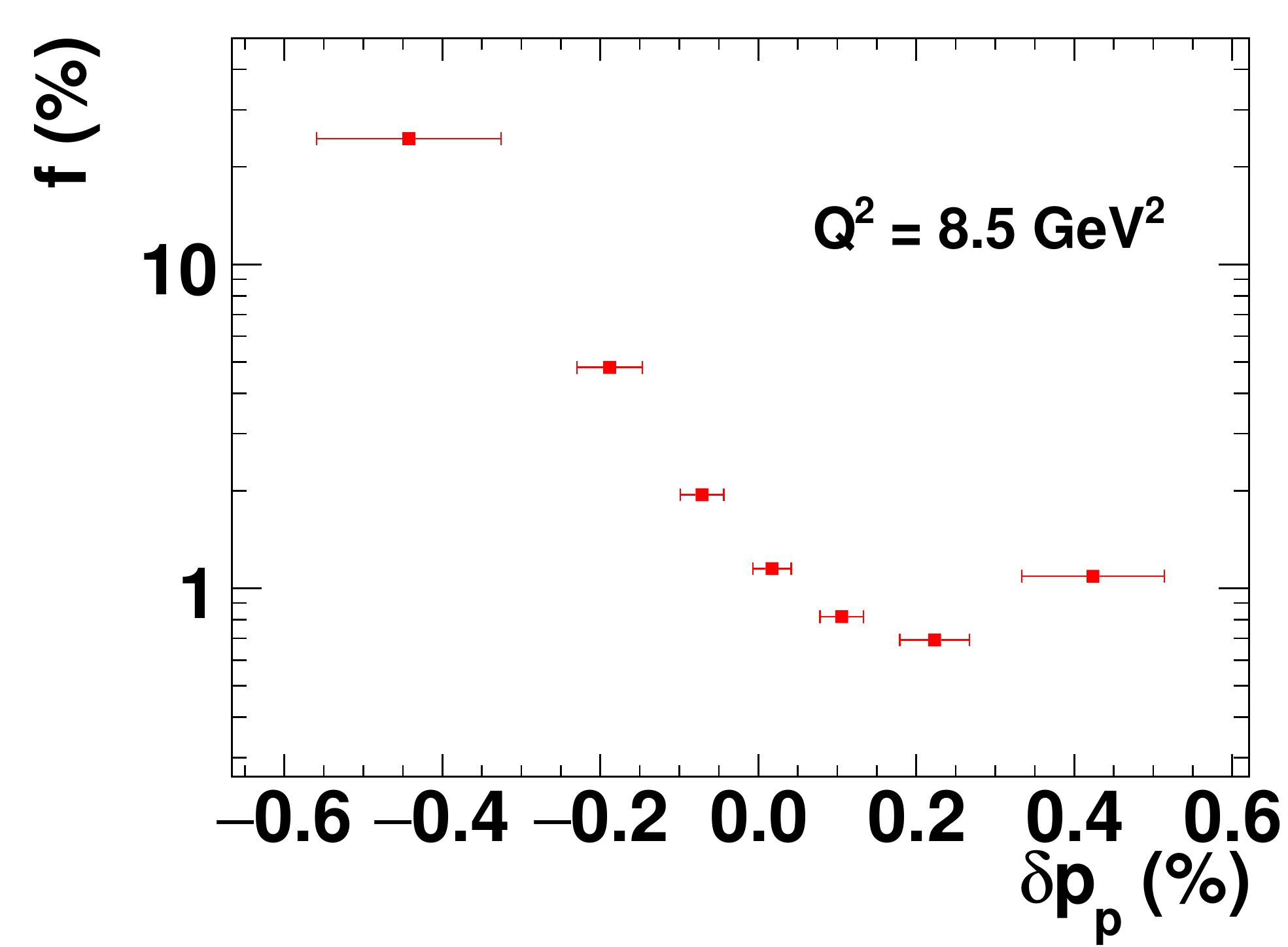}
  \end{center}
  \caption{\label{fig:bg_vs_dpp85} Fractional background contamination $f$ as a function of $\delta p_p$ at $Q^2 = 8.5$ GeV$^2$, with $\pm3\sigma$ cuts applied to $\delta p_e$ and $\delta \phi$. The horizontal ``error bars'' represent the \emph{rms} deviation from the mean of the $\delta p_p$ values of all events in each bin. The uncertainties $\Delta f$ are statistical only and are smaller than the data points.}
\end{figure}

\begin{table}
  \caption{\label{tab:frac_bg}  Estimated fractional background contamination $f \equiv \frac{B}{S+B}$ (where $B$ and $S$ refer to the background and the signal, respectively) within the final, $\pm3\sigma$ cut region of the $\delta p_e$ distribution, for all the kinematics of the GEp-III and GEp-2$\gamma$ experiments. The estimates shown are obtained after applying $\pm3\sigma$ cuts to $\delta p_p$ and $\delta \phi$. The quoted uncertainties are statistical only. The quoted beam energy $E_e$ is the value from Table~\ref{kintable}, which is averaged over the duration of the running period, and \emph{not} corrected for energy loss in the LH$_2$ target. }
  \begin{center}
    \begin{ruledtabular}
      \begin{tabular}{ccc}
        $Q^2$ (GeV$^2$) & $E_e$ (GeV) & $(f \pm \Delta f_{stat})$ (\%) \\ \hline
        2.5 & 1.873 & $0.435 \pm 0.002$ \\
        2.5 & 1.868 & $0.512 \pm 0.001$ \\
        2.5 & 2.847 & $0.161 \pm 0.002$ \\
        2.5 & 3.548 & $0.198 \pm 0.002$ \\
        2.5 & 3.680 & $0.208 \pm 0.001$ \\
        5.2 & 4.052 & $1.018 \pm 0.004$ \\
        6.8 & 5.711 & $0.748 \pm 0.004$ \\
        8.5 & 5.712 & $4.89 \pm 0.01$ 
      \end{tabular}
    \end{ruledtabular}
  \end{center}
\end{table}
Table~\ref{tab:frac_bg} shows the estimated, acceptance-averaged fractional background contamination of the final, $\pm3\sigma$ cuts used for all six kinematics. 
The inelastic contamination estimates shown in Table~\ref{tab:frac_bg} are determined directly from the data, but are not used directly in the final analysis, because the background contamination and the transferred polarization components of the background both vary strongly as a function of $\delta p_p$ within the final cut region, as the dominant background process evolves from $\pi^0p$ photo/electro-production to quasielastic Al$(e,e'p)$. The $\pi^0 p$ contribution rises rapidly for negative $\delta p_p$ values as the kinematic threshold is crossed, whereas the $\delta p_p$ distribution of the (very small) target endcap contamination is relatively uniform within the cut region. The recoil proton polarization for the inelastic $\pi^0p$ reaction on hydrogen generally differs strongly from that of the elastic $ep$ process, while the proton polarization in quasi-elastic Al$(e,e'p)$ is generally similar to elastic $ep$, since it is basically the same process embedded in a nucleus (see Figs.~\ref{fig:Pinel_gep2g} and~\ref{fig:Pinel_gep3}). Fig.~\ref{fig:bg_vs_dpp85} shows the $\delta p_p$ dependence of the fractional background contamination $f$ for $Q^2 = 8.5$ GeV$^2$, the setting with (by far) the greatest residual background contamination. Details of the background subtraction procedure are given in section~\ref{sec:PolarizationAnalysis} and the systematic uncertainties associated with the background subtraction are presented in Ref.~\cite{Puckett:2017egz}.

The stability of the transferred polarization components with respect to the width of the elastic event selection cuts and the amount of background included in the final event sample was checked by varying the width of the $\delta p_p$, $\delta \phi$, and $\delta p_e$ cuts independently between $\pm2.5\sigma$ and $\pm3.5\sigma$ and observing the variations in the background-corrected results. The observed variations of $P_t$, $P_\ell$, and the ratio $P_t/P_\ell$ were compatible with purely statistical fluctuations for all kinematics. Therefore, no additional systematic uncertainty contributions were assigned. The cut sensitivity study also confirmed that the application of cuts that were adequately loose and carefully centered with respect to the elastic peak eliminated any cut-induced systematic bias of the reconstructed proton kinematics. This was a non-trivial concern for this analysis given the exaggerated effect of multiple-scattering in ``S$_0$'' on the event-by-event errors in the reconstructed proton angles and the very high sensitivity of the spin transport calculation to systematic errors in these angles, particularly the non-dispersive-plane angle $\phi_{tar}$ (see Ref.~\cite{Puckett:2017egz} for a detailed discussion).

\subsection{Extraction of Polarization Transfer Observables}
\label{sec:PolarizationAnalysis}

\subsubsection{FPP Angular Distribution}
An expression for the general angular distribution in the polarimeter is given by:
\begin{eqnarray}
  N^\pm(p,\vartheta,\varphi) &=& N_0^\pm
  \frac{\varepsilon(p,\vartheta)E(\vartheta,\varphi)}{2\pi} \times  \nonumber \\
  & & \left[ 1 \pm A_y (P_{y,tr}^{FPP}\cos \varphi - P_{x,tr}^{FPP} \sin \varphi) \right. \nonumber \\
  & & \left.+ A_y (P_{y,ind}^{FPP} \cos \varphi - P_{x,ind}^{FPP} \sin \varphi )\right] \label{angulardistribution},
\end{eqnarray}
where $N_0^{\pm}$ is the number of incident protons corresponding to a ${\pm}1$ beam helicity state, $\varepsilon(p,\vartheta)$ is the fraction of protons of momentum $p$ scattered at a polar angle $\vartheta$ and producing one single track, $E(\vartheta,\varphi)$ represents the angular dependence of the combined effective polarimeter acceptance/detection efficiency, which factorizes from the differential nuclear scattering cross section, $A_y = A_{y}(p,\vartheta)$ represents the analyzing power of $\vec{p}+CH_{2} \rightarrow \text{one charged particle} + X$ scattering, and $P_{x,tr/ind}^{FPP}$ and $P_{y,tr/ind}^{FPP}$ are the transverse components of the proton polarization at the focal plane, with $P_{tr}$ ($P_{ind}$) denoting transferred (induced) polarization. As explained below, only the $\varphi$ dependence of the detector acceptance/efficiency is relevant for polarimetry.

Note that for all kinematics of the GEp-III and GEp-2$\gamma$ experiments, $N_0^+ = N_0^- = N_{total}/2$ to within statistical uncertainties. This is a consequence of the beam-helicity independence of the elastic $\vec{e}p$ scattering cross section for an unpolarized target (in the one-photon-exchange approximation), and the rapid (30 Hz) helicity reversal, which cancels the effects of slow drifts in experimental conditions such as luminosity and detection efficiency.

As described in~\cite{Puckett:2015soa}, the azimuthal scattering angle $\varphi$ was defined in a coordinate system that is comoving with the incident proton, in which the HMS track defines the $z$ axis, the $y$ axis is chosen to be perpendicular to the HMS track, but parallel to the $yz$ plane of the fixed TRANSPORT coordinate system (see Ref.~\cite{Puckett:2017egz}), and the $x$ axis is defined by $\hat{x} = \hat{y} \times \hat{z}$. In this coordinate system, $\varphi$ is the azimuthal angle of the scattered proton trajectory measured clockwise from the $x$ axis toward the $y$ axis. Note that this convention for the definition of $\varphi$ differs from the convention used in the analysis of the GEp-I and GEp-II experiments~\cite{Punjabi:2005wq,Puckett:2011xg}. In the GEp-I and GEp-II analyses, $\varphi$ was defined such that $\varphi = 0$ for scattering along the $+y$ axis, and $\varphi$ was measured counterclockwise from the $y$ axis toward the $x$ axis (see Eq. (4) of Ref.~\cite{Puckett:2011xg}). With $\varphi$ defined as in the GEp-III/GEp-2$\gamma$ analysis, the $\sin(\varphi)$ asymmetry is dominant, whereas the $\cos(\varphi)$ asymmetry is dominant using the GEp-I/GEp-II convention.

In the one-photon-exchange approximation in elastic $ep$ scattering, the induced polarization terms are identically zero due to time reversal invariance. When two photons are exchanged, a non-zero induced polarization of elastically scattered protons can occur at subleading order in $\alpha$ due to the interference between the one-photon and two-photon-exchange amplitudes. Because it is subleading order in $\alpha$, it is not expected to exceed $\simeq$1-2\% in magnitude~\cite{Afanasev:2005mp}, and must be normal to the $ep$ scattering plane due to parity invariance of the electromagnetic interaction. The helicity-independent azimuthal asymmetry resulting from a small induced polarization at this level is smaller yet as the analyzing power does not exceed roughly 20\% at any $(p,\vartheta)$ in these experiments. 

The ``false'' or instrumental asymmetry resulting from the effective acceptance/efficiency function $E(\varphi)$ can be expressed in terms of its Fourier expansion:
\begin{eqnarray}
  E(\varphi) &=& C\left[1+\sum_{m=1}^{\infty} (c_m \cos(m\varphi) + s_m \sin(m\varphi))\right] \nonumber \\
  &\equiv& C\left[1 + \mu_0(\varphi)\right],
\end{eqnarray}
with an overall multiplicative constant $C$ that is ultimately absorbed into the overall normalization of the distributions when integrating over the dependence on kinematic variables other than $\varphi$. A clean extraction of the transferred polarization components is obtained from the difference and/or the difference/sum ratio between the angular distributions for positive and negative beam helicities, integrated over all momenta within the HMS acceptance and a limited $\vartheta$ range chosen to exclude small-angle Coulomb scattering and large-angle scatterings for which $A_y \approx 0$. The helicity difference and sum distributions are given by:
\begin{eqnarray}
f^{+}-f^{-} &\equiv& \frac{\pi}{\Delta \varphi}\left[\dfrac{N^{+}(\varphi)}{N_{0}^{+}}-\dfrac{N^{-}(\varphi)}{N_{0}^{-}}\right] 
\nonumber \\
            &=& \bar{A}_y \left[P_{y,tr}^{FPP}\cos \varphi-P_{x,tr}^{FPP}\sin \varphi \right] \times \nonumber \\ 
  & & \left[1+\mu_0(\varphi)\right] \label{diffdistri} \nonumber \\
  &\approx & \bar{A}_y \left[P_{y,tr}^{FPP}\cos \varphi - P_{x,tr}^{FPP} \sin \varphi \right] \\
f^{+}+f^{-}  &\equiv& \frac{\pi}{\Delta \varphi}\left[\frac{N^{+}(\varphi)}{N_{0}^{+}}+\frac{N^{-}(\varphi)}{N_{0}^{-}}\right]
                      \nonumber \\
&=& \left[1+\mu_0(\varphi)\right] \times \nonumber \\ 
  & & \left[1+ \bar{A}_y(P_{y,ind}^{FPP} \cos \varphi - P_{x,ind}^{FPP} \sin \varphi)\right] \nonumber \\
  &\approx & 1 + \mu_0(\varphi) \label{sumdistri}
\end{eqnarray}
where $\Delta \varphi$ is the bin width in $\varphi$ and $\bar{A}_y$ is the average analyzing power within the range of $\vartheta$ considered\footnote{Note also that in the context of Eqs.~\eqref{diffdistri}-~\eqref{sumdistri}, $N_0^\pm$ is the total number of incident protons corresponding to beam helicity $\pm 1$ producing a detected scattering event within the accepted $\vartheta$ range.}. 

The difference-sum ratio is given by
\begin{eqnarray}
  \frac{f_+ - f_-}{f_+ + f_-} &=& \frac{\bar{A}_y \left(P_{y,tr}^{FPP} \cos \varphi - P_{x,tr}^{FPP} \sin \varphi \right)}{1 + \bar{A}_y\left(P_{y,ind}^{FPP} \cos \varphi - P_{x,ind}^{FPP} \sin \varphi \right)} \nonumber \\ 
  &\approx & \bar{A}_y \left(P_{y,tr}^{FPP} \cos \varphi - P_{x,tr}^{FPP} \sin \varphi \right) \label{diffsumratio} \\
  \frac{2f_\pm}{f_+ + f_-} &=& 1 \pm \frac{\bar{A}_y \left(P_{y,tr}^{FPP} \cos \varphi - P_{x,tr}^{FPP} \sin \varphi \right)}{1 +\bar{A}_y\left(P_{y,ind}^{FPP} \cos \varphi - P_{x,ind}^{FPP} \sin \varphi \right)} \nonumber \\
  &\approx & 1 \pm \bar{A}_y \left(P_{y,tr}^{FPP} \cos \varphi - P_{x,tr}^{FPP} \sin \varphi \right) \label{fplusminussumratio}
\end{eqnarray}
where in Eqs.~\eqref{diffsumratio}-\eqref{fplusminussumratio}, the induced polarization terms in the denominator are neglected. Equations~\eqref{diffdistri}-\eqref{fplusminussumratio} show that the false asymmetries and/or the induced polarization terms are cancelled by the beam helicity reversal in the different asymmetry observables. The helicity-difference distribution cancels the induced polarization terms but is sensitive at second order to the false asymmetry $\mu_0$, while the difference-sum ratio cancels the false asymmetry terms, but is sensitive at second order to any induced polarization terms. The helicity-sum distribution cancels the transferred polarization terms, but includes contributions from false asymmetries and any induced polarization terms, if they exist. The transferred polarizations, the induced polarizations, and the false asymmetry terms can all be rigorously separated, in principle, via Fourier analysis of the distributions \eqref{diffdistri}-\eqref{fplusminussumratio}, assuming infinite statistical precision. In practice, however, it is very statistically and systematically challenging to separate the induced polarization terms from the false asymmetry terms when both are ``small'', as is the case in this experiment, especially for the induced polarization terms. For the transferred polarization components, on the other hand, it can be shown~\cite{Besset:1979sh} that the false asymmetry effects are cancelled exactly to all orders by the beam helicity reversal in the linearized maximum-likelihood estimators for $P_t$ and $P_\ell$ defined in section~\ref{sec:likelihood} below, given sufficient statistical precision that the sums over all events entering the maximum-likelihood estimators are a good approximation to the corresponding weighted integrals over the azimuthal angular distribution discussed in~\cite{Besset:1979sh}.

\subsubsection{FPP event selection criteria}
\label{subsubsec:fppselect}
Useful scattering events for polarimetry were selected according to several criteria, detailed in Ref.~\cite{Puckett:2017egz}. First, only single-track events were included in the analysis of each polarimeter, as the analyzing power for events with two or more reconstructed tracks in either polarimeter was found to be much lower than that of the single-track events, such that even a separate analysis of the multi-track events did not meaningfully improve the polarimeter figure-of-merit in a weighted average with the single-track events. Secondly, cuts were applied to the parameters $s_{close}$, defined as the distance of closest approach between incident and scattered tracks, and $z_{close}$, defined as the $z$-coordinate of the point of closest approach between incident and scattered tracks. A loose, $\sim 10\sigma$ upper limit for $s_{close}$ was chosen to optimize the statistical precision of the analysis, by excluding events at large $s_{close}$ values with low analyzing power. The $z_{close}$ ranges considered for FPP1 and FPP2 events correspond to the physical extent of the CH$_2$ analyzers ($L_{CH_2} = 55$ cm) plus a small additional tolerance ($\Delta z = \pm 2.5$ cm) to allow for the resolution of $z_{close}$ while excluding the ``unphysical'' region close to (and including) the drift chambers themselves. 

A ``cone test'' was applied to each candidate scattering event, to minimize instrumental asymmetries in the $\varphi$ distribution arising from the geometrical acceptance of the FPP, and to guarantee full $2\pi$ azimuthal acceptance over the full range of $(\vartheta, z_{close})$ values included in the analysis. Simply defined, the cone test requires that the projection of the cone of opening angle $\vartheta$ from the reconstructed interaction vertex $z_{close}$ to the rearmost wire plane of the FPP drift chamber pair that detected the track lie entirely within the active area of the chamber for all possible azimuthal scattering angles $\varphi$. This in turn guarantees that the effective range of $\vartheta$ integration is the same for all $\varphi$ values, such that the average analyzing power is $\varphi$-independent. As a result, the analyzing power, which depends strongly on $\vartheta$, cancels reliably in the ratio of polarization components $P_y^{FPP}/P_x^{FPP}$ at the focal plane and $P_t/P_\ell$ at the target, regardless of the range of $\vartheta$ included in the analysis. Due to the large active area of the FPP drift chambers, the efficiency of the cone test is close to 100\% for scattering angles up to about 30 degrees. The details of the cone test calculation are given in~\cite{Puckett:2015soa}.

The useful range of $\vartheta$ varies with $Q^2$, because the width of the multiple-Coulomb-scattering peak at small $\vartheta$ and the angular distributions of both the scattering probability and the analyzing power are observed to scale approximately as $1/p_p$. The useful range of $\vartheta$ was selected for each $Q^2$ by applying a cut to the ``transverse momentum'' $p_T \equiv p_p \sin \vartheta$, where $\vartheta$ is the proton's polar scattering angle in the FPP, and $p_p$ is the incident proton momentum. The value of $p_p$ used in the definition of $p_T$ is corrected for the mean energy loss along the path length in CH$_2$ traversed by the incident proton prior to the scattering. 
For all three $\epsilon$ values at $Q^2 = 2.5$ GeV$^2$, the range of $p_T$ included in the analysis was 0.06 GeV $\le p_T \le$ 1.2 GeV for both polarimeters. A slightly wider range 0.05 GeV $\le p_T \le$ 1.5 GeV was used for the GEp-III kinematics, for which the uncertainties are statistics-limited. For all kinematics, the low-$p_T$ cutoff is large compared to the intrinsic angular resolution of the FPP drift chambers, which is about 1.9 (2.1) mrad in the $x$ ($y$) direction. In the worst case, at 8.5 GeV$^2$, the 0.05 GeV minimum $p_T$ corresponds to a minimum $\vartheta$ of about 9 mrad or $4.5\sigma$. More details of the FPP event selection criteria, $p_T$ distributions, track multiplicities per event, and closest approach parameters can be found in Ref.~\cite{Puckett:2017egz}.

\subsubsection{Focal plane azimuthal asymmetries}
\begin{figure}
\begin{center}
  \includegraphics[width=0.85\columnwidth]{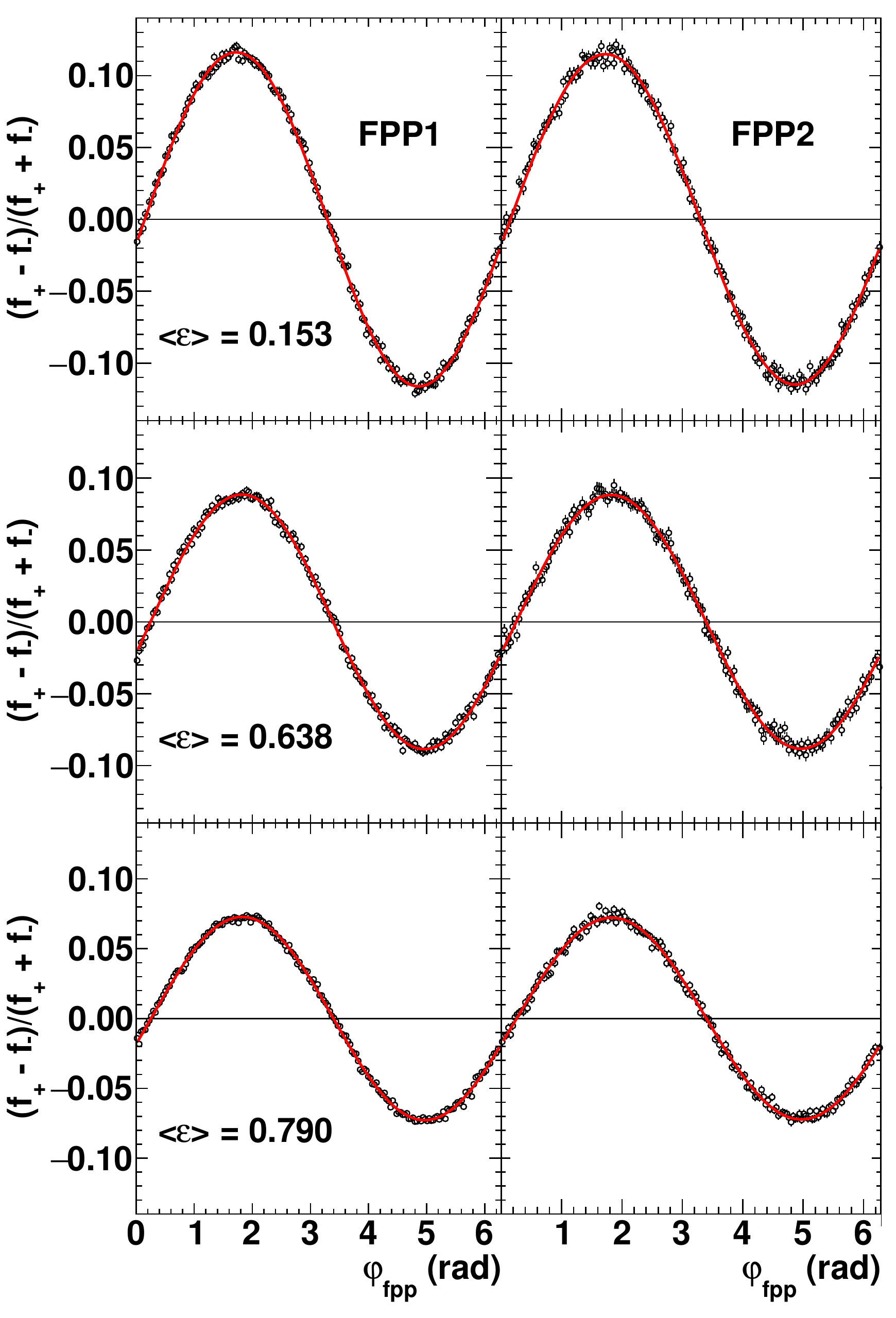}
\end{center}
\caption{\label{fig:fppasym_2gamma}  Focal-plane helicity difference/sum ratio asymmetry $(f_+ - f_-)/(f_+ + f_-)$, defined as in Eq.~\eqref{diffsumratio}, for the GEp-2$\gamma$ ($Q^2 = 2.5$ GeV$^2$) kinematics, for single-track events selected according to the criteria discussed in Sec.~\ref{subsubsec:fppselect}. $\left<\epsilon\right>$ is the acceptance-averaged value of $\epsilon$. The left (right) column shows the asymmetries for events scattering in the first (second) analyzer. Asymmetries are shown for $\left<\epsilon\right> = 0.153$ (top), $\left<\epsilon\right> = 0.638$ (middle) and $\left<\epsilon\right> = 0.790$ (bottom). Red curves are fits using $(f_+ - f_-)/(f_+ + f_-) = c \cos(\varphi) - s \sin(\varphi)$. Asymmetry fit results are shown in Table~\ref{tab:asymresults}.}
\end{figure}

\begin{figure}
  \begin{center}
    \includegraphics[width=0.85\columnwidth]{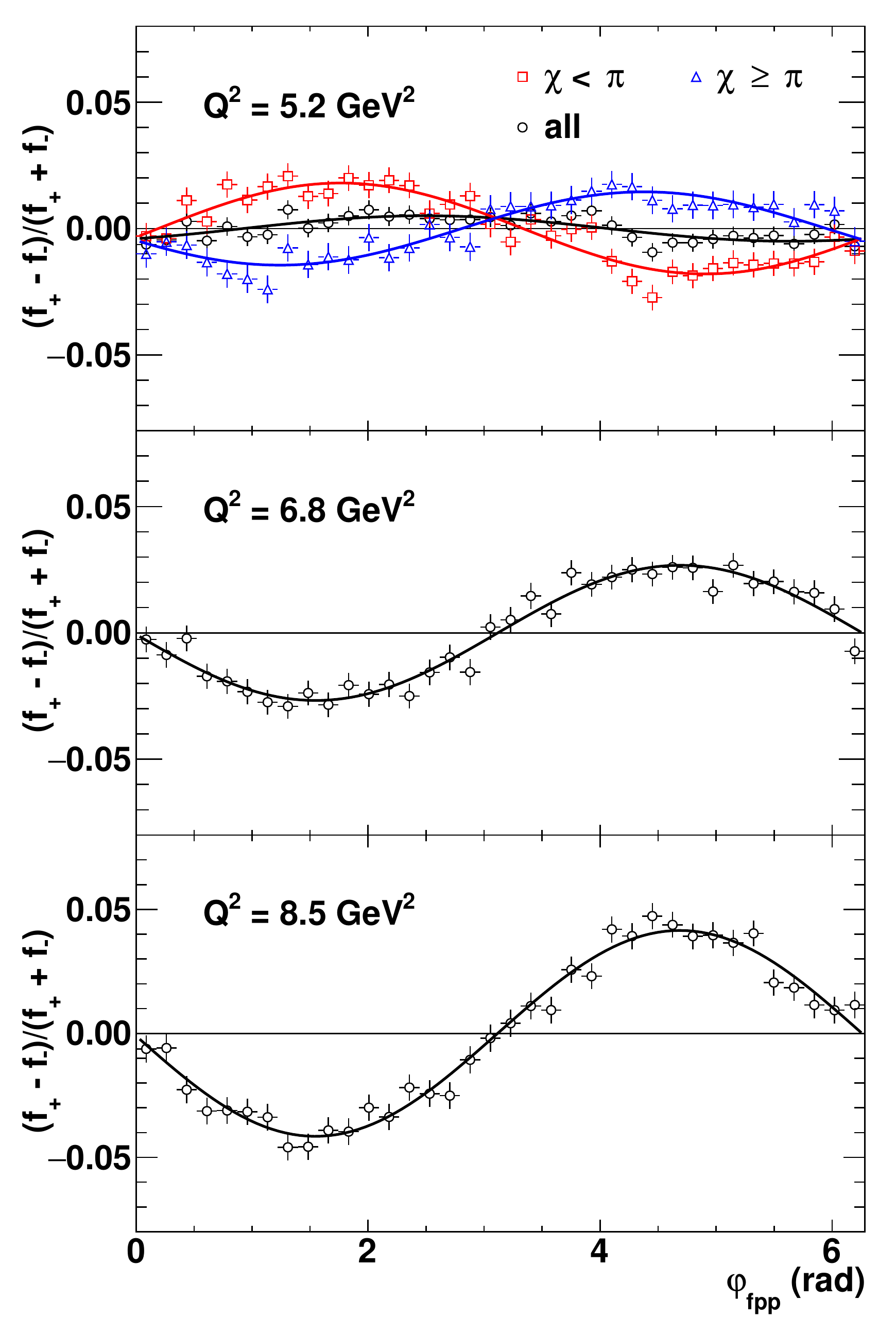}
  \end{center}
  \caption{\label{fig:fppasym_gep3}  Focal plane helicity difference/sum ratio asymmetry $(f_+ - f_-)/(f_+ + f_-)$, defined as in Eq.~\eqref{diffsumratio}, for the GEp-III kinematics, for FPP1 and FPP2 data combined,  for single-track events selected according to the criteria discussed in Sec.~\ref{subsubsec:fppselect}. Asymmetry fit results are shown in Table~\ref{tab:asymresults}. The asymmetry at $Q^2 = 5.2$ GeV$^2$ is also shown separately for events with precession angles $\chi < \pi$ and $\chi \ge \pi$, illustrating the expected sign change of the $\sin(\varphi)$ term.}
\end{figure}

\begin{figure}
  \begin{center}
    \includegraphics[width=0.85\columnwidth]{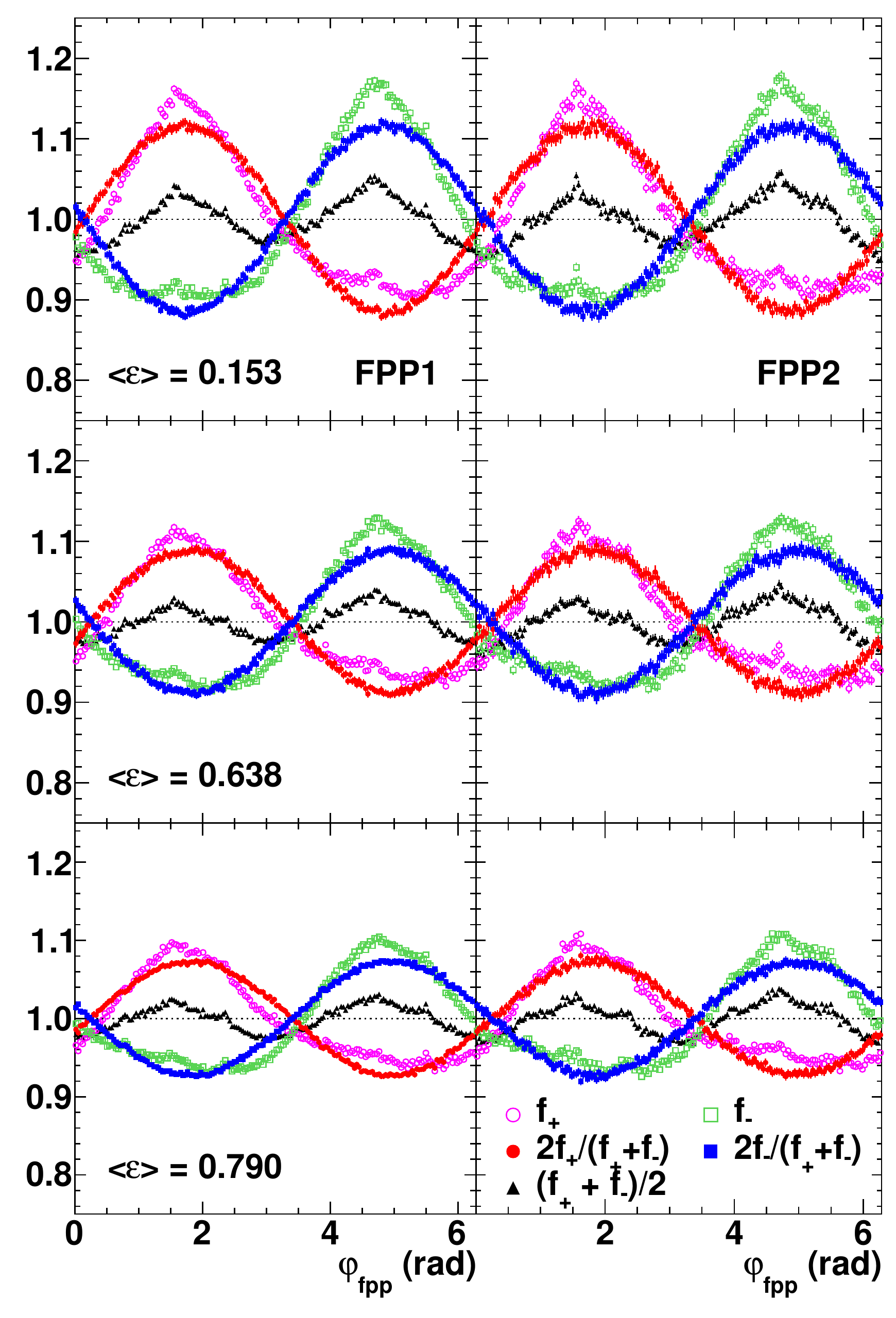}
  \end{center}
   \caption{\label{fig:phisumplusminusdiff_gep2gamma}  Azimuthal angular distributions for FPP1 (left column) and FPP2 (right column) for the GEp-2$\gamma$ kinematics, for events selected according to the criteria discussed in Sec.~\ref{subsubsec:fppselect}. The helicity-sum distribution (black filled triangles) cancels the asymmetry due to the proton's transferred polarization. The raw $\varphi$ distributions for the $+$ (pink empty circles) and $-$ (green empty squares) helicity states include contributions from the transferred polarization and the false asymmetry. The corrected $\varphi$ distributions $2f_+/(f_+ + f_-)$ (red filled circles) and $2f_-/(f_+ + f_-)$ (blue filled squares) exhibit pure sinusoidal behavior, and include only contributions from the transferred polarization terms, assuming the induced polarization terms are small.}
\end{figure}
Figure~\ref{fig:fppasym_2gamma} shows the ratio of the helicity-difference and helicity-sum azimuthal distributions $A \equiv (f_+(\varphi) - f_-(\varphi))/(f_+(\varphi)+f_-(\varphi))$, defined in Eq.~\eqref{diffsumratio}, for each of the GEp-2$\gamma$ kinematics, for each polarimeter separately, fitted with a function $A = c \cos \varphi - s \sin \varphi$. The fit results are shown in Table~\ref{tab:asymresults}. The asymmetries are consistent with a pure sinusoidal $\varphi$ dependence, and Fourier analysis including a constant term and higher harmonics up to 8$\varphi$ showed no statistically significant evidence for the presence of terms other than $\cos \varphi$ and $\sin \varphi$, as expected from Eq.~\eqref{diffsumratio}. This suggests that the beam helicity reversal does an excellent job of suppressing the instrumental asymmetries, which are significant at certain values of $\vartheta$ and $z_{close}$. The FPP1 and FPP2 asymmetries are mostly consistent with each other, and are always consistent in terms of the ratio $c/s = P_y^{FPP}/P_x^{FPP}$, or equivalently, in terms of the phase of the asymmetry, since the analyzing power cancels in this ratio. 
For the GEp-2$\gamma$ kinematics, the use of identical event selection criteria for all three $\epsilon$ values eliminates, in principle, point-to-point systematic variations of the effective average analyzing power arising from the cuts on the scattering parameters $\vartheta, s_{close}$, and $z_{close}$. 

Figure~\ref{fig:fppasym_gep3} shows the difference/sum ratio asymmetry $(f_+ - f_-)/(f_+ + f_-)$ for the GEp-III kinematics, for both polarimeters combined. For the GEp-III kinematics, the combined asymmetries are also compatible with a purely sinusoidal $\varphi$ dependence, albeit with much lower statistical precision. The asymmetry amplitude at $Q^2 = 8.5$ GeV$^2$ is larger than for the other two kinematics despite the lower analyzing power, because of the precession of the proton spin in the HMS. The central precession angle at $Q^2 = 8.5$ GeV$^2$ is close to 270 degrees, and the asymmetry magnitude is maximal at $\sin \chi = \pm 1$. In contrast, the central precession angle for $Q^2 = 5.2$ GeV$^2$ is close to 180 degrees, such that the acceptance-averaged asymmetry is close to zero. However, as shown in Fig.~\ref{fig:axfp_chi} and discussed below, the $\chi$ acceptance of the HMS for each $Q^2$ point is wide enough to provide sufficient sensitivity to $P_\ell$, and the precision of the form factor ratio extraction is not dramatically affected by the unfavorable precession angle, since $P_\ell$ is quite large (58\%-98\%) in all the kinematics of these experiments.
\begin{table*}
  \caption{\label{tab:asymresults} Focal plane helicity difference/sum ratio asymmetry fit results for GEp-2$\gamma$ ($Q^2 = 2.5$ GeV$^2$, top) and GEp-III kinematics (bottom). The fit function is $\tfrac{f_+ - f_-}{f_+ + f_-} = c \cos(\varphi) - s\sin(\varphi)$. To first order, $c = \bar{A}_yP_y^{FPP}$ and $s = \bar{A}_yP_x^{FPP}$. FPP1 and FPP2 asymmetries are shown separately for GEp-2$\gamma$, while the combined asymmetries are shown for GEp-III. $\left<\epsilon\right>$ is the acceptance-averaged value of $\epsilon$.}
  \begin{ruledtabular}
    \begin{tabular}{lccc}
      Nominal $Q^2$ & 2.5 & 2.5 & 2.5  \\
      $\left< \epsilon \right>$ & 0.153 & 0.638 & 0.790  \\ \hline
      FPP1 $c \pm \Delta c_{stat}$ & $-0.01728 \pm 0.00033$ & $-0.02179 \pm 0.00031$ & $-0.01831 \pm 0.00022$  \\
      FPP1 $s \pm \Delta s_{stat}$ & $-0.11489 \pm 0.00033$ & $-0.08586 \pm 0.00030$ & $-0.07046 \pm 0.00022$  \\
      FPP1 $\chi^2/ndf$ & $162/178$ & $188/178$ & $137/178$ \\
      FPP2 $c \pm \Delta c_{stat}$ & $-0.01760 \pm 0.00046$ & $-0.02183 \pm 0.00047$ & $-0.01827 \pm 0.00032$  \\
      FPP2 $s \pm \Delta s_{stat}$ & $-0.11360 \pm 0.00045$ & $-0.08553 \pm 0.00046$ & $-0.06990 \pm 0.00032$  \\
      FPP2 $\chi^2/ndf$ & $173/178$ & $145/178$ & $167/178$ 
    \end{tabular}
    \begin{tabular}{lcccc}
      Nominal $Q^2$ (GeV$^2$) & $\left<\epsilon\right>$ & Combined $c \pm \Delta c_{stat}$ & Combined $s \pm \Delta s_{stat}$ & Combined $\chi^2/ndf$ \\ \hline
      5.2 (all) & 0.382 & $-0.0040 \pm 0.0009$ & $-0.0030 \pm 0.0009$ & $23.9/34$ \\
      5.2 ($\chi < \pi$) & 0.382 &$-0.0034 \pm 0.0012$ & $-0.0177 \pm 0.0012$ & $29.8/34$ \\
      5.2 ($\chi \ge \pi$) & 0.382 &$-0.0047 \pm 0.0013$ & $0.0137 \pm 0.0013$ & $22.2/34$ \\
      6.8 & 0.519 & $-0.0006 \pm 0.0012$ & $0.0267 \pm 0.0012$ & $26.1/34$ \\
      8.5 & 0.243 & $-0.0010 \pm 0.0013$ & $0.0415 \pm 0.0012$ & $29.5/34$
    \end{tabular}
  \end{ruledtabular}
\end{table*}
Table~\ref{tab:asymresults} summarizes the focal-plane helicity-difference asymmetry fit results. For each of the $Q^2 = 2.5$ GeV$^2$ kinematics, the FPP1 and FPP2 asymmetries are fitted separately, while the results shown for the GEp-III kinematics are for FPP1 and FPP2 combined. For $Q^2 = 5.2$ GeV$^2$, the asymmetry results are also fitted separately for precession angles $\chi < \pi$ and $\chi \ge \pi$, illustrating the expected sign change of $s$, the $-\sin(\varphi)$ coefficient of the asymmetry. If $Q^2$ were chosen such that the HMS acceptance were centered \emph{exactly} at $\chi = \pi$, and if the effects of quadrupole precession were absent, we would expect the values of $s$ for $\chi < \pi$ and $\chi \ge \pi$ to be equal and opposite. However, the central value of $\chi$ for $Q^2 = 5.2$ GeV$^2$ is 177.2$^\circ$, such that the HMS acceptance extends to slightly greater $\left|\sin(\chi)\right|$ for $\chi < \pi$ than for $\chi \ge \pi$ (see also Fig.~\ref{fig:axfp_chi}). Moreover, as discussed in Ref.~\cite{Puckett:2017egz}, the mixing of $P_t$ and $P_\ell$ due to quadrupole precession shifts the ``expected'' location of the zero crossing of the $-\sin(\varphi)$ coefficient of the asymmetry to about 180.4 degrees instead of the nominal 180 degrees. Both of these effects lead to the expectation of a slightly larger $\sin(\varphi)$ asymmetry for $\chi < \pi$ than for $\chi \ge \pi$, as observed. 

Figure~\ref{fig:phisumplusminusdiff_gep2gamma} shows the raw $\varphi$ distributions $f_+$, $f_-$, $f_+ + f_-$ and $2f_\pm/(f_+ + f_-)$ for the GEp-2$\gamma$ kinematics. Similar results with lower statistical precision are obtained for the GEp-III kinematics. The normalized distributions $2f_\pm/(f_+ + f_-)$ are consistent with the pure sinusoidal behavior predicted by Eq.~\ref{fplusminussumratio} for all kinematics and for both polarimeters separately. The helicity sum distribution $f_+ + f_-$, which cancels the asymmetry due to the transferred polarization, exhibits a characteristic instrumental asymmetry with several notable features common to all kinematics. The dominant feature of the false asymmetry is a $\cos(2\varphi)$ term that is roughly independent of kinematics, negative, and about 2-3\% in magnitude when averaged over the useful $\vartheta$ acceptance at $Q^2 = 2.5$ GeV$^2$. This asymmetry appears at small $\vartheta$ as a consequence of the $x/y$ resolution asymmetry of the FPP drift chambers and at large $\vartheta$ due to acceptance/edge effects, and is generally small at intermediate $\vartheta$ values near the maximum of the analyzing power distribution (see Sec.~\ref{subsubsec:Ay}). Although the ``cone test'' (see Section~\ref{subsubsec:fppselect}) is designed to eliminate acceptance-related false asymmetries, it cannot do so completely because it is applied based on the \emph{reconstructed} parameters of the incident and scattered tracks, which are affected in a $\varphi$-dependent way by the FPP $x/y$ resolution asymmetry. 

The other prominent feature of the false asymmetry is the presence of small peaks at 45-degree intervals corresponding to the FPP drift chamber wire orientations. The peaks are absent at $\varphi = 0$ deg., 180 deg.,  and 360 deg., angles corresponding to scattering along the dispersive ($x$) direction. These artificial peaks are caused by incorrect solutions of the left-right ambiguity due to the irreducible ambiguity in the drift chambers' design, resulting from the symmetry of the wire layout and the lack of redundancy of coordinate measurements. These incorrect solutions occur primarily for small-angle tracks traversing the chambers at close to normal incidence near the center of the drift chambers, where the $x$, $u$, and $v$ wires share a common intersection point in the $xy$ plane. When an incorrect left-right assignment occurs for events in the Coulomb peak of the $\vartheta$ distribution, the reconstructed track position at one or both sets of drift chambers is incorrectly placed on the opposite side of all three wires that fired in that drift chamber. If the left-right assignment of the hits in one chamber (but not the other) in a pair is incorrect, the reconstructed point of closest approach ``collapses'' to the location of the chamber for which the left-right combination was \emph{correctly} assigned, and the value of $\varphi$ ``collapses'' to one of the three different wire orientations depending on the topology of the event and the measured drift distances of the incorrectly assigned hits. The overwhelming majority of these mistracked events are rejected by the $z_{close}$ cut, which excludes the unphysical region corresponding to the drift chambers themselves. However, for $z_{close}$ values within the analyzer region but close to the chambers, some of these mistracked events leak into the ``good'' event sample due to detector resolution, producing the pattern of small, residual artificial peaks observed in $(f_+ + f_-)(\varphi)$. These ``mistracked'' events have low/zero analyzing power and tend to dilute the asymmetry in the $z_{close}$ region closest to the drift chambers. In principle, they can be further suppressed by excluding the part of the analyzer region closest to the drift chambers. In practice, this is unnecessary, because the instrumental asymmetry they generate is cancelled by the beam helicity reversal, and the resulting dilution of the effective average analyzing power cancels in the ratio of polarization components, such that they cause no systematic effect whatsoever on the extraction of $R$. The effect of the mistracked events on the average analyzing power, which is important for the extraction of the $\epsilon$ dependence of $P_\ell/P_\ell^{Born}$, is measured and accounted for, and is the same for all three $\epsilon$ values at 2.5 GeV$^2$. The sensitivity of the measured $P_\ell/P_\ell^{Born}$ ratio to the range of $z_{close}$ and $p_T$ included in the analysis was examined and found to be small compared to the statistical and systematic uncertainties in this observable.

\subsubsection{FPP efficiency}

\begin{table*}
  \caption{\label{tab:FPPefficiency} Experimentally realized effective global FPP efficiencies. ``Total elastic events'' is the number of events passing the elastic event selection cuts, including the requirement that a definite beam helicity state was recorded for the event. The FPP1 (FPP2) efficiency is the fraction of the total number of elastic events passing all the event selection criteria from Section~\ref{subsubsec:fppselect} for FPP1 (FPP2). Note that the efficiencies quoted here do not include single-track events in FPP2 reconstructed as having scattered in the first analyzer, that failed the event selection criteria for FPP1. These events were included in the GEp-III analysis, but excluded from the GEp-2$\gamma$ analysis. Note also that the ``efficiencies'' are not corrected for data runs that were rejected due to data quality issues in either FPP1, FPP2 or both. See text for details.}
  \begin{ruledtabular}
    \begin{tabular}{llcccc}
      $Q^2$ (GeV$^2$) & $\left<\epsilon\right>$ & Total elastic events ($\times 10^6$) & FPP1 efficiency (\%) & FPP2 efficiency (\%) & Combined efficiency (\%) \\ \hline 
      2.5 & 0.153 & 99.2 & 20.5 & 11.5 & 32.0 \\ 
      2.5 & 0.638 & 96.8 & 23.8 & 10.6 & 34.4 \\ 
      2.5 & 0.790 & 161.2 & 26.1 & 12.8 & 38.9 \\
      5.2 & 0.382 & 9.15 & 16.8 & 8.6 & 25.4 \\
      6.8 & 0.519 & 4.96 & 17.1 & 8.0 & 25.1 \\
      8.5 & 0.243 & 5.01 & 15.0 & 7.0 & 22.0 
    \end{tabular}
  \end{ruledtabular}
\end{table*}
Table~\ref{tab:FPPefficiency} summarizes the total elastic $ep$ statistics collected and the effective ``efficiency'' of the FPP, defined as the fraction of incident protons producing a useful secondary scattering for polarimetry. The raw FPP wire efficiencies and angular distributions were examined on a run-by-run basis, and runs with data quality issues in either FPP1 or FPP2 (or both) were rejected for the polarimeter in question. The total number of elastic events shown in Table~\ref{tab:FPPefficiency}, which serves as the denominator for the efficiency determination, is \emph{not} corrected for runs rejected from the analysis because of FPP data quality issues. In other words, FPP-specific data losses due to transient malfunctioning of the data acquisition system for either set (or both sets) of FPP drift chambers during runs of otherwise good data quality are included in the effective efficiencies shown\footnote{Since the FPP1 and FPP2 drift chambers were read out by different VME crates during most of the experiment prior to the switch to Fastbus DAQ during the high-$Q^2$ running from April-June of 2008, a somewhat common occurrence was a data acquisition run with only one of the two sets of drift chambers providing usable data.}. These data losses are responsible for reducing the experimentally realized efficiency for FPP1 by several percent for the $Q^2 = 2.5$ GeV$^2$ data at $\left<\epsilon\right> = 0.153$, compared to what it would have been in the ideal case, and by smaller amounts for other kinematic settings. Overall, the combined efficiency per incident proton for producing a scattering event passing all the event selection criteria ranges from 22\% at $Q^2 = 8.5$ GeV$^2$ to nearly 39\% at $(Q^2, \left<\epsilon\right>)=(2.5$ GeV$^2, 0.79)$. 
The use of two polarimeters in series, each with an analyzer thickness of one nuclear interaction length $\lambda_T$, leads to an efficiency gain of approximately 50\% relative to the use of a single polarimeter with one $\lambda_T$ analyzer thickness, regardless of $p_p$. 

\subsubsection{HMS Spin Transport}
The asymmetries measured by the FPP are proportional to the transverse components of the proton polarization at the HMS focal plane (see equation~\eqref{diffsumratio}), which are related to the reaction-plane transferred polarization components $P_t$ and $P_\ell$ by a rotation due to the precession of the proton polarization in the HMS magnetic field. The HMS is a focusing spectrometer characterized by its 25-degree central vertical bend angle and relatively small angular acceptance in both the dispersive and non-dispersive directions. The precession of the polarization of charged particles with anomalous magnetic moments moving relativistically in a magnetic field is described by the Thomas-BMT equation~\cite{BMTequation}. The spin transport for protons (anomalous magnetic moment $\kappa_p \approx 1.79$) through the HMS is dominated by a rotation in the dispersive plane by an angle $\chi \equiv \gamma \kappa_p \theta_{bend}$ relative to the proton trajectory, where $\gamma \equiv E_p/M_p$ is the usual relativistic $\gamma$ factor and $\theta_{bend}$ is the trajectory bend angle in the dispersive plane. In this so-called ``ideal dipole'' approximation, the dispersive plane component of the proton polarization precesses by an angle $\chi$, and the non-dispersive plane component does not rotate, such that $P_x^{FPP} \approx -\sin \chi P_e P_\ell$ and $P_y^{FPP} = P_e P_t$. 

\begin{figure}
  \begin{center}
    \includegraphics[width=0.85\columnwidth]{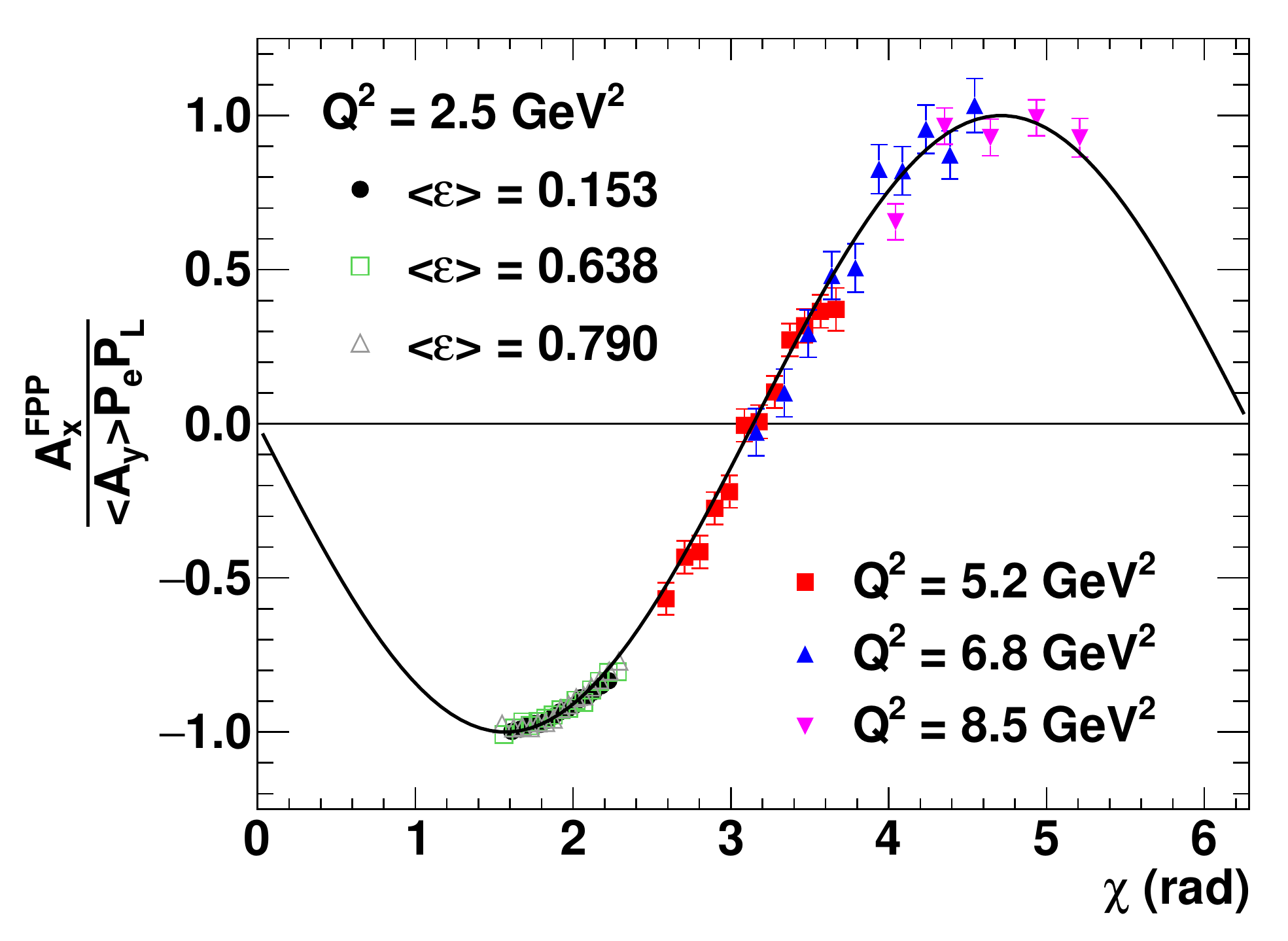}
  \end{center}
  \caption{\label{fig:axfp_chi}  Normalized focal-plane asymmetry $\frac{A_x^{FPP}}{\left< A_y \right> P_e P_\ell}$, integrated over the accepted $\vartheta$ range, vs. precession angle $\chi \equiv \gamma \kappa_p \theta_{bend}$, for all six kinematics. The black solid curve is $-\sin \chi$, which is the expected value of the normalized asymmetry in the ideal dipole approximation to the spin transport in the HMS.}
\end{figure}
Figure~\ref{fig:axfp_chi} illustrates the dominant dipole precession of the proton spin using the ratio $\frac{A_x^{FPP}}{\left<A_y\right>P_e P_\ell}$, where $A_x^{FPP} = A_y P_x^{FPP}$ is the $-\sin \varphi$ coefficient of the asymmetry $(f_+ - f_-)/(f_+ + f_-)$ (see Eq.~\eqref{diffsumratio}), $\left<A_y\right>$ is the average analyzing power within the accepted range of scattering angles, and $P_e$ is the average beam polarization. In the ideal dipole approximation, $A_x^{FPP} = -\left<A_y\right> P_e P_\ell \sin \chi$. The ideal dipole approximation accounts for most of the observed $\chi$ dependence of the asymmetry. The $Q^2 = 5.2$ GeV$^2$ data show how it is possible to achieve good precision on the ratio $P_t/P_\ell$ even when the acceptance-averaged asymmetry is close to zero; due to the large value of $P_\ell$ and the relatively large $\chi$ acceptance of the HMS, the relative statistical uncertainty $\Delta P_\ell/P_\ell$ is almost a factor of four smaller than $\Delta P_t/P_t$.

Deviations from the ideal dipole approximation arise due to the quadrupole magnets and the finite angular acceptance of the HMS. The forward spin transport matrix of the HMS depends on all the parameters of the proton trajectory at the target and must be calculated for each event. Owing to the relatively simple magnetic field layout and small angular and momentum acceptance of the HMS, it was not necessary to perform a computationally expensive numerical integration of the BMT equation for each proton trajectory. Instead, a general fifth-order expansion of the forward spin transport matrix in terms of all the trajectory parameters at the target was fitted to a sample of random test trajectories populating the full acceptance of the HMS that were propagated through a detailed COSY~\cite{Makino:2006sx} model of the HMS including fringe fields. Unlike the parameters describing charged particle transport through the HMS, which are independent of the central momentum setting for the standard tune, the spin transport coefficients had to be computed separately for each central momentum setting, since the spin precession frequency relative to the proton trajectory is proportional to $\gamma$. The use of the same central momentum setting for all three kinematics at $Q^2 = 2.5$ GeV$^2$ ensures that the magnetic field and the spin transport matrix are the same for all three kinematics. This in turn minimizes the point-to-point systematic uncertainties in the polarization transfer observables, which is essential in the accurate determination of their $\epsilon$ dependence. 

The fitted expansion coefficients of the COSY spin transport model express the matrix elements of the absolute total rotation of the proton spin in the fixed TRANSPORT coordinate system. The total rotation of the proton spin relevant to the extraction of the polarization transfer observables also includes a rotation from the reaction plane coordinate system in which $P_t$ and $P_\ell$ are defined to the TRANSPORT coordinate system that is fixed with respect to the HMS optical axis at the target, and a rotation from the fixed TRANSPORT coordinate system at the focal plane to the coordinate system comoving with the proton trajectory, in which the polar and azimuthal scattering angles $\vartheta$ and $\varphi$ are defined. Details of the calculation of the total rotation matrix for each event are given in Ref.~\cite{Puckett:2015soa}. 
\subsubsection{Maximum-Likelihood Extraction of $P_t$, $P_\ell$, and $R$}
\label{sec:likelihood}
The transferred polarization components $P_t$ and $P_\ell$ are extracted from the measured FPP angular distributions using an unbinned maximum-likelihood estimator. Neglecting induced polarization terms, the likelihood function is defined up to an overall normalization constant independent of $P_t$ and $P_\ell$ as
\begin{eqnarray}
  \mathcal{L}(P_t, P_\ell) &=& \prod_{i=1}^{N_{event}} \frac{E(\varphi_i)}{2\pi} \times \nonumber \\ 
                           & & \bigg\{ 1 + h_i P_e A_y^{(i)} \Big[ \left(S_{yt}^{(i)} P_t + S_{y\ell}^{(i)} P_\ell \right) \cos \varphi_i \nonumber \\
                           & &  - \left(S_{xt}^{(i)} P_t + S_{x\ell}^{(i)} P_\ell \right)\sin \varphi_i \Big] \bigg\}, \label{likelihoodfunc}
\end{eqnarray}
where $E(\varphi_i) \propto 1 + \sum_n \left[c_n \cos (n\varphi_i) + s_n \sin(n\varphi_i)\right]$ is the false/instrumental asymmetry for the $i$-th event\footnote{In principle, the false asymmetry Fourier coefficients can depend on $\vartheta$, $p$ and any other parameters of the event such as $s_{close}$, $z_{close}$}, $h_i = \pm 1$ is the beam helicity state for the $i$-th event, $P_e$ is the beam polarization, $A_y^{(i)} \equiv A_y(p_i, \vartheta_i)$ is the analyzing power of $\vec{p} + $CH$_2$ scattering, which depends on the proton momentum $p_i$ and scattering angle $\vartheta_i$, and the $S_{jk}^{(i)}$'s are the forward spin transport matrix elements relating polarization component $P_k$ in reaction-plane coordinates to component $P_j$ in the comoving coordinate system of the secondary analyzing reaction measured by the FPP. 

The product over all events in Eq.~\eqref{likelihoodfunc} was converted to a sum by taking the logarithm, and then the problem of maximizing $\ln \mathcal{L}$ as a function of the parameters $P_t$ and $P_\ell$ was linearized by truncating the expansion of $\ln(1+x)$ at quadratic order in $x$; i.e., $\ln(1+x) = x - \frac{x^2}{2} + \mathcal{O}x^3$. In this context, ``$x$'' represents the sum of all the $\varphi$-dependent terms in Eq.~\eqref{likelihoodfunc}. The largest acceptance-averaged helicity-dependent ``raw'' asymmetry observed in either experiment was about 0.12 (see Fig.~\ref{fig:fppasym_2gamma}), while the largest raw asymmetry observed at any $\vartheta$ was about 0.16. The acceptance-averaged helicity-independent false/instrumental asymmetries are at the few-percent level\footnote{The magnitude of the $\cos(2\varphi)$ false asymmetry arising from the $x/y$ resolution asymmetry and the acceptance of the FPP drift chambers rises to the $\sim 10\%$ level at the extremes of the accepted $\vartheta$ range.}. It is therefore estimated that the maximum truncation error in $\Delta (\ln \mathcal{L})/\ln \mathcal{L}$ due to the linearization procedure is $\left|\frac{x - x^2/2}{\ln(1+x)}-1\right| \lesssim 0.82\%$ at any $\vartheta$, and smaller when averaged over the full $\vartheta$ acceptance. 

The linearized maximum-likelihood estimators for $P_t$ and $P_\ell$ are given by the solution of the following linear system of equations:
\begin{eqnarray}
  \sum_{i} \left[\begin{array}{cc} \left(\lambda_t^{(i)}\right)^2 & \lambda_t^{(i)}\lambda_\ell^{(i)} \\ \lambda_t^{(i)}\lambda_\ell^{(i)} & \left(\lambda_\ell^{(i)}\right)^2\end{array}\right]\left[\begin{array}{c} \hat{P}_t \\ \hat{P}_\ell \end{array}\right] = \nonumber \\
  \sum_i \left[\begin{array}{c} \lambda_t^{(i)}\\ \lambda_\ell^{(i)}\end{array}\right], \label{MLestimators}
\end{eqnarray}
where $\lambda_t^{(i)}$ and $\lambda_\ell^{(i)}$ given by
\begin{eqnarray}
  \lambda_t^{(i)} &\equiv & h_i P_e A_y^{(i)}\left(S_{yt}^{(i)} \cos \varphi_i - S_{xt}^{(i)} \sin \varphi_i \right) \nonumber \\
  \lambda_\ell^{(i)} &\equiv & h_i P_e A_y^{(i)} \left(S_{y\ell}^{(i)}\cos \varphi_i - S_{x\ell}^{(i)} \sin \varphi_i \right) \label{MLmatrixelements}
\end{eqnarray}
are the coefficients of $P_t$ and $P_\ell$ in the equation for the likelihood function \eqref{likelihoodfunc}.

Note that the false asymmetry $E(\varphi)$ does not enter the definition of the estimators. Up to the effects of spin precession, the estimators defined by Eq.~\eqref{MLestimators} are equivalent to the ``weighted sum'' estimators of Ref.~\cite{Besset:1979sh}. In Ref.~\cite{Besset:1979sh} it was shown that these estimators are unbiased and efficient, and in particular that the instrumental asymmetries are cancelled to all orders by the beam helicity reversal, which provides an effective detection efficiency that is 180-degree symmetric; i.e., $E(\varphi) = E(\varphi + \pi)$. The equation for the maximum likelihood estimators can be rewritten in matrix form as $A\mathbf{P} = \mathbf{b}$, the solution of which is $\mathbf{P} = A^{-1}\mathbf{b}$. The symmetric $2\times 2$ matrix $A^{-1}$, with $A$ defined by the $2\times 2$ matrix on the LHS of Eq.~\eqref{MLestimators}, is the covariance matrix of the parameters $\mathbf{P}$. The standard statistical variances in $P_t$ and $P_\ell$ are given by the diagonal elements of $A^{-1}$, while the covariance of $P_t$ and $P_\ell$ is given by the off-diagonal element:
\begin{eqnarray}
  \Delta P_t &=& \sqrt{\left(A^{-1}\right)_{tt}} \\
  \Delta P_\ell &=& \sqrt{\left(A^{-1}\right)_{\ell \ell}} \\
  \text{cov}(P_t,P_\ell) &=& \left(A^{-1}\right)_{t\ell} = \left(A^{-1}\right)_{\ell t}
\end{eqnarray}
The ratio $R \equiv -\mu_p \sqrt{\frac{\tau(1+\epsilon)}{2\epsilon}} \frac{P_t}{P_\ell} \equiv -K \frac{P_t}{P_\ell}$, which equals $\mu_p \frac{G_E^p}{G_M^p}$ in the one-photon-exchange approximation, is computed from the results of the maximum-likelihood analysis for $P_t$ and $P_\ell$. The uncertainty in $R$ is computed using the standard prescription for error propagation using the covariance matrix $A^{-1}$ discussed above:
\begin{eqnarray}
  \left(\frac{\Delta R}{R}\right)^2 &=& \left(\frac{\Delta P_t}{P_t}\right)^2 + \left(\frac{\Delta P_\ell}{P_\ell}\right)^2 - \frac{2\text{cov}(P_t, P_\ell)}{P_t P_\ell} \label{eq:dRstat}
\end{eqnarray}
Although it is not immediately obvious from Eq.~\eqref{MLestimators}, both the beam polarization and the analyzing power cancel in the ratio $P_t/P_\ell$. All of the matrix elements on the LHS of~\eqref{MLestimators} are proportional to $\left(P_eA_y\right)^2$, while the components of the vector on the RHS of~\eqref{MLestimators} are proportional to $P_e A_y$. The estimators $\hat{P}_t$ and $\hat{P}_\ell$ are thus proportional to $\left(P_e A_y \right)^{-1}$, and the statistical variances in $P_t$ and $P_\ell$ are proportional to $\left(P_e A_y\right)^{-2}$. Strictly speaking, the cancellation of $A_y$ in the ratio $P_t/P_\ell$ requires that the effective range of integration in $\vartheta$ (or equivalently $p_T$) be independent of $\varphi$, which is guaranteed in principle by the application of the cone test. According to the $\chi^2$ of a constant fit, the extracted ratio $R$ showed no statistically significant $p_T$ dependence for any of the six kinematic settings, as detailed in Ref.~\cite{Puckett:2017egz}.


\begin{figure}
  \begin{center}
    \includegraphics[width=0.85\columnwidth]{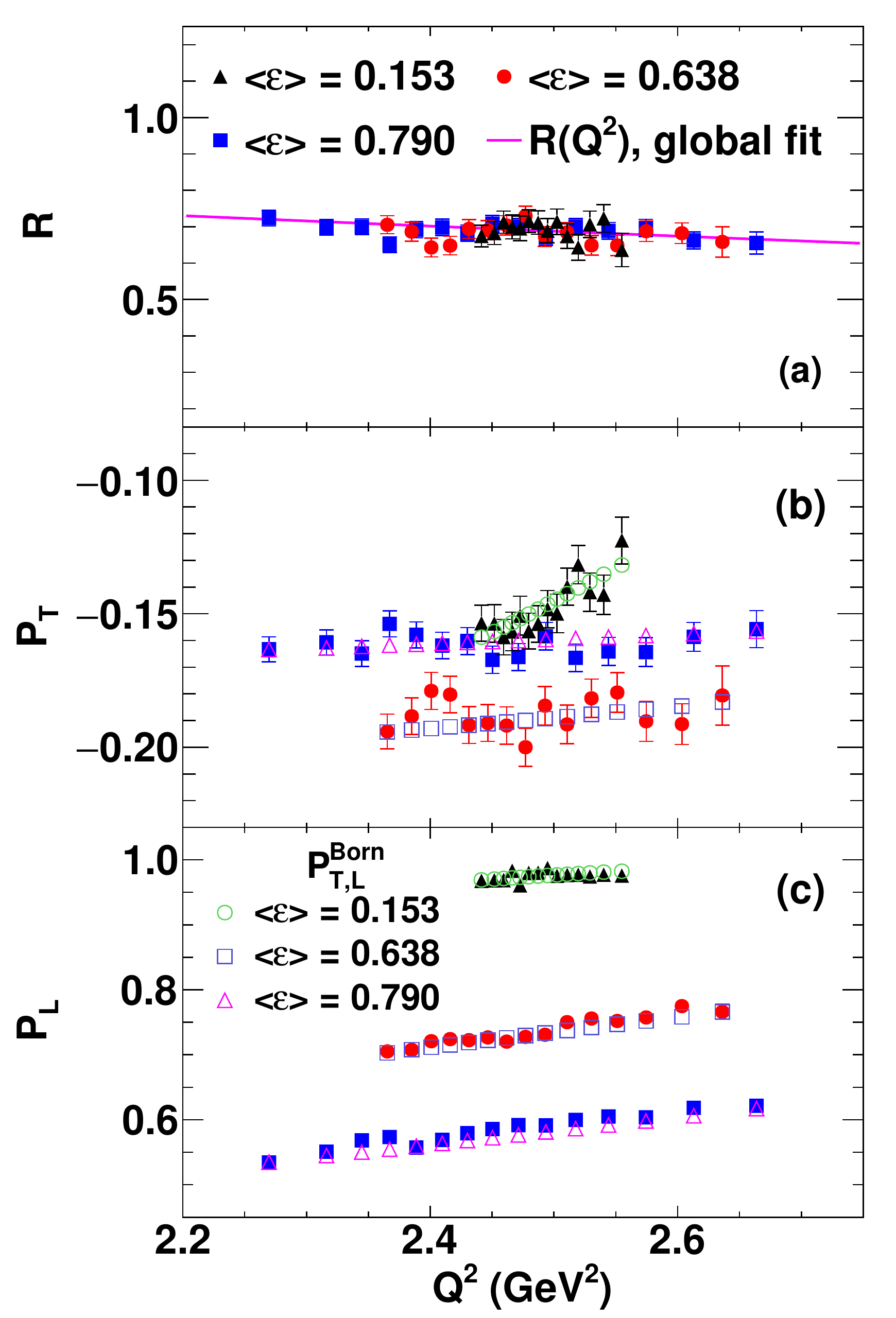}
  \end{center}
  \caption{\label{fig:R_PT_PL_Q2_gep2g}  $Q^2$ dependence of $R$ (panel (a)), $P_t$ (panel (b)) and $P_\ell$ (panel (c)) for the GEp-2$\gamma$ kinematics. Uncertainties shown are statistical only. $R$ is compared to $R(Q^2)$ from ``Global Fit II'' described in appendix~\ref{sec:globalfit}, which includes the GEp-2$\gamma$ results reported in this work. Transferred polarization components are compared to their Born approximation values, which are also computed in each bin from ``Global Fit II''. Note that for $\left<\epsilon\right> = 0.153$, $P_t$ and $P_\ell$ are equal to their Born approximation values by definition, because this setting is used to determine the analyzing power $A_y$. See text for details.}
\end{figure}
Figure~\ref{fig:R_PT_PL_Q2_gep2g} shows the $Q^2$ dependence of $P_t$, $P_\ell$ and $R$ within the HMS acceptance for the GEp-2$\gamma$ kinematics. As discussed below, $P_t$ and $P_\ell$ are equal to their Born approximation values for $\left<\epsilon\right> = 0.153$ by definition, since this point is used for the analyzing power calibration at 2.5 GeV$^2$. At a fixed central $Q^2$ of 2.5 GeV$^2$, the fixed angular acceptance of the HMS corresponds to a $Q^2$ acceptance that is roughly three times greater at $\left<\epsilon\right> = 0.790$ than at $\left<\epsilon\right> = 0.153$. The observed $Q^2$ dependence of $R$ within the acceptance of each kinematic setting is statistically compatible with both the expected $R(Q^2)$ and a constant $R$ value. The observed $Q^2$ dependences of $P_t$ and $P_\ell$ are also similar to those of $P_t^{Born}$ and $P_\ell^{Born}$, providing important added confirmation that both the HMS spin transport and the momentum dependence of $A_y$ (see Sec.~\ref{subsubsec:Ay}) are accounted for correctly. The $P_\ell$ values at $\left<\epsilon\right> = 0.790$ show a clear excess over $P_\ell^{Born}$. The curve for $R(Q^2)$ shown in Fig.~\ref{fig:R_PT_PL_Q2_gep2g} is obtained from the global fit to proton form factor data described in appendix~\ref{sec:globalfit}.


The kinematic factors $\tau$, $\epsilon$ and $K$ are computed from the beam energy $E_e$ and the proton momentum $p_p$ for each event. The value of $K$ is averaged over all events in computing the acceptance-averaged $R$ value from the acceptance-averaged unbinned maximum-likelihood estimators for $P_t$ and $P_\ell$. $Q^2$ and $\epsilon$ are one-to-one correlated within the acceptance at each setting due to the fixed beam energy. $P_t$ and $P_\ell$ depend on both $Q^2$ and $\epsilon$, and can vary significantly within the acceptance of a single measurement. $R$ depends only on $Q^2$ (in the Born approximation), and its expected variation within the acceptance is generally smaller than that of $P_t$ or $P_\ell$. The correlated $(\epsilon, Q^2)$ acceptances of all kinematic settings are small enough that, to within experimental precision, $P_t$, $P_\ell$, and $R$ vary linearly with $Q^2$ within the acceptance, and the acceptance-averaged values of $P_t$, $P_\ell$ and $R$ equal their values at the acceptance-averaged kinematics. 

The choice to use the measured quantities $E_e$ and $p_p$ to compute $Q^2$ and $\epsilon$ is not unique; the reaction kinematics in $ep \rightarrow ep$ are fixed by choosing any two of $E_e, E'_e, \theta_e, \theta_p, p_p$. The choice of any two of these five variables gives equivalent results; radiative effects on the average kinematics of the final elastic $ep$ sample are suppressed to a negligible level by the tight exclusivity cuts. The use of the beam energy $E_e$ and the proton momentum $p_p$ to compute the event kinematics is optimal because the beam energy is known with a high degree of certainty, $Q^2$ depends only on the proton momentum in elastic $ep$ scattering, and the systematic uncertainty in $p_p$ is easily quantifiable and independent of $\epsilon$ for a fixed HMS central momentum setting/nominal $Q^2$ value. 


\subsubsection{Analyzing Power Calibration}
\label{subsubsec:Ay}
The analyzing power $A_y$ is not \emph{a priori} known. However, the elastic $ep$ process is ``self-calibrating'' with respect to the analyzing power, as it can be extracted directly from the measured asymmetries, provided the beam polarization is known. The M$\o$ller measurement of the electron beam polarization is subject to a global uncertainty of approximately 1\% and point-to-point uncertainties of $\left(\Delta P_e/P_e\right)_{ptp} = 0.5\%$. The ratio $R = -K \frac{P_t}{P_\ell}$ does not depend on the beam polarization or the analyzing power because both of these quantities cancel in the ratio $P_t/P_\ell$. Moreover, in the one-photon-exchange approximation, the values of $P_t$ and $P_\ell$ depend only on the ratio $r \equiv G_E/G_M \equiv R/\mu_p$, and not on $G_E$ or $G_M$ separately. In terms of $r$, Eq.~\eqref{recoilformulas} becomes (for $P_e = 1$):
\begin{eqnarray}
  P_t^{Born} &=& -\sqrt{\frac{2\epsilon(1-\epsilon)}{\tau}} \frac{r}{1+\frac{\epsilon}{\tau}r^2} \nonumber \\
  P_\ell^{Born} &=& \frac{\sqrt{1-\epsilon^2}}{1+\frac{\epsilon}{\tau}r^2} \label{eq:PtBornPlBorn}
\end{eqnarray} 
The average analyzing power in a particular bin of $\vartheta$ and/or $p_p$ is determined by computing the maximum-likelihood estimators $\hat{P}_t$ and $\hat{P}_\ell$ assuming $A_y = 1$, with $P_e$ taken from the M$\o$ller measurements, and forming the ratios to $P_t^{Born}$ and $P_\ell^{Born}$:
\begin{eqnarray}
  \hat{P}_t^{(A_y = 1)} &=& \bar{A}_y P_t \\
  \hat{P}_\ell^{(Ay=1)} &=& \bar{A}_y P_\ell \\
  \bar{A}_y &=& \frac{\hat{P}_t^{(A_y=1)}}{P_t^{Born}} = \frac{\hat{P}_\ell^{(A_y=1)}}{P_\ell^{Born}} \label{eq:Ayformula}
\end{eqnarray}
The value of $A_y$ in any kinematic bin is computed from a weighted average of $A_y(\hat{P}_t) \equiv \hat{P}_t/P_t^{Born}$ and $A_y(\hat{P}_\ell) \equiv \hat{P}_\ell/P_\ell^{Born}$, that is usually dominated by $A_y(\hat{P}_\ell)$. Although $P_t$ and $P_\ell$ are determined with comparable \emph{absolute} precision, $P_\ell$ is determined with a much better \emph{relative} precision than $P_t$, because the magnitude of $P_\ell$ is several times greater than that of $P_t$ for all kinematics.

\begin{figure}
\begin{center}
  \includegraphics[width=0.85\columnwidth]{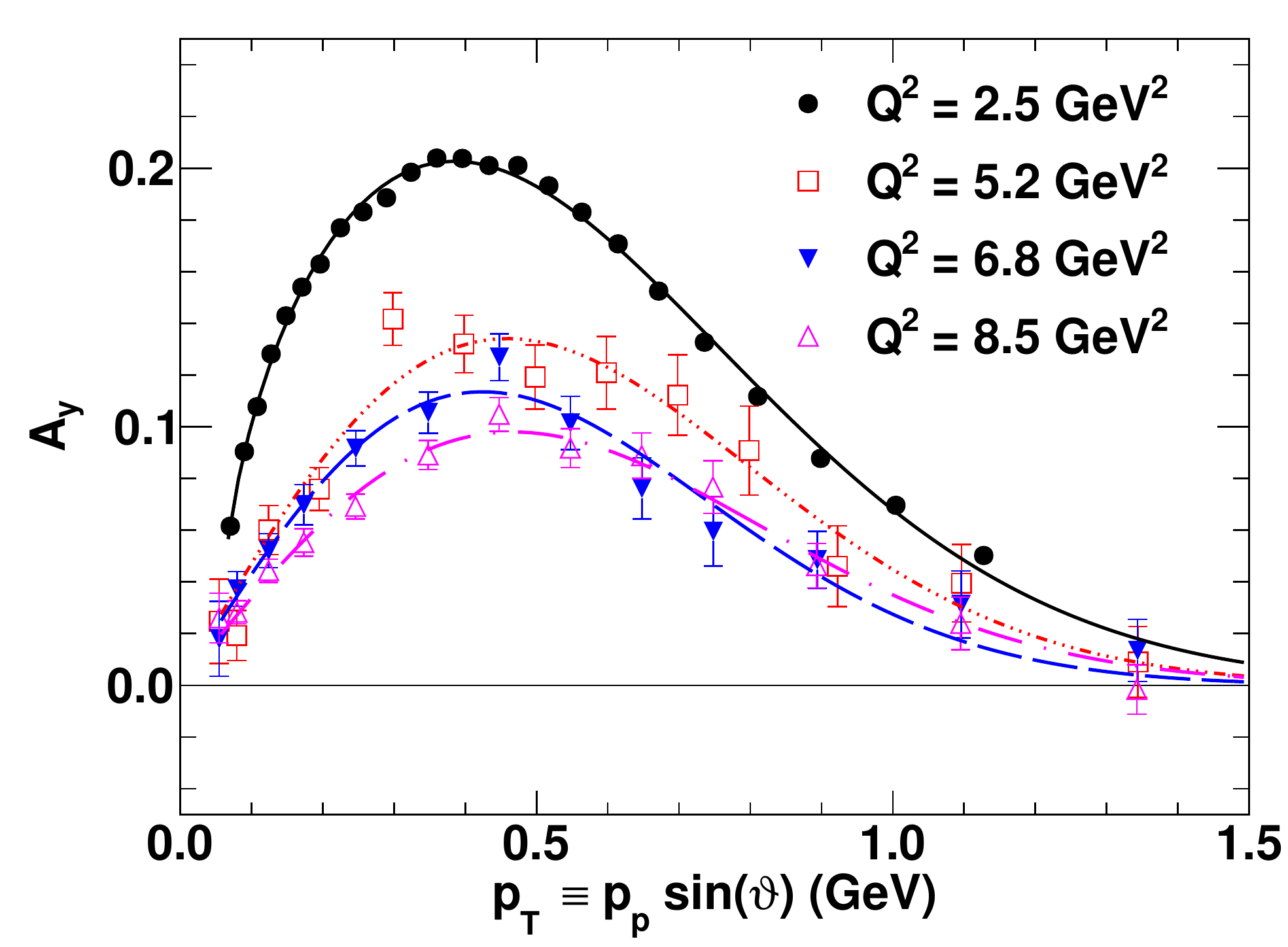}
\end{center}
\caption{\label{fig:AypT}  Analyzing power $A_y$ vs. $p_T \equiv p_p \sin \vartheta$ for the four different $Q^2$ values from GEp-III/2$\gamma$. Data are from both polarimeters combined. Curves are the fits to the data, used to estimate the position and value of the maximum in $A_y(p_T)$. See text for details.}
\end{figure}

Figure~\ref{fig:AypT} shows the angular dependence of $A_y$, expressed in terms of the ``transverse momentum'' $p_T$, for all four $Q^2$ values. The shape of $A_y(p_T)$ is qualitatively similar for all four HMS central momentum settings. The maximum $A_y^{max}(p_T = p_T^{max})$ was estimated by fitting each $A_y(p_T)$ curve with the following simple parametrization:
\begin{eqnarray}
  A_y(p_T) &=& \left\{ \begin{array}{rl} N (p_T - p_T^0)^\alpha e^{-b(p_T-p_T^0)^\beta}, & p_T \ge p_T^0 \\ 0, & p_T < p_T^0 \end{array} \right\},\nonumber \\
  & & \label{eq:Ayfitfunc}
\end{eqnarray}
which is positive-definite, vanishes at asymptotically large $p_T$ and $p_T < p_T^0$, and is sufficiently flexible to describe the data with reasonable accuracy. An adequate description of the GEp-III data is achieved by fixing the exponents $(\alpha, \beta) = (1,2)$ and the zero-offset $p_T^0 = 0$ and varying only the normalization constant $N$ and the ``slope'' $b$ of the exponential. The data at 2.5 GeV$^2$ are precise enough that all five parameters had to be varied to achieve a good description, and in contrast to the GEp-III data, strongly favor a ``zero offset'' $p_T^0 \approx 0.05$ GeV.
\begin{table*}
  \caption{\label{tab:Ayfitresults} $A_y(p_T)$ fit results. Fit parametrization is as in Eq.~\eqref{eq:Ayfitfunc}. Uncertainties in fit parameters are statistical only. Parameters with no uncertainties are fixed at the quoted values. $A_y^{max}$ is the maximum value of $A_y$ occuring at $p_T = p_T^{max}$. Uncertainties in derived quantities $A_y^{max}$ and $p_T^{max}$ are computed from the full covariance matrix of the fit result and the gradient of the fit function with respect to the parameters evaluated at the maximum. $\bar{A}_y$ values are for $0.06\le p_T$ (GeV)$\le 1.2$.}
  \begin{ruledtabular}
    \begin{tabular}{rcccc}
      $Q^2$ (GeV$^2$) & 2.5 ($\left<\epsilon\right>=0.153$) & 5.2 & 6.8 & 8.5 \\ \hline
      $N$ & $0.44 \pm 0.02$ & $0.48 \pm 0.03$ & $0.44 \pm 0.02$ & $0.35 \pm 0.01$ \\ 
      $\alpha$ & $0.48 \pm 0.02$ & 1 & 1 & 1 \\
      $\beta$ & $1.89 \pm 0.05$ & 2 & 2 & 2 \\
      $b$ & $2.05 \pm 0.03$ & $2.4 \pm 0.2$ & $2.8 \pm 0.2$ & $2.3 \pm 0.1$ \\
      $p_T^0$ (GeV) & $0.053 \pm 0.002$ & 0 & 0 & 0 \\ 
      $A_y^{max}$ & $0.2027 \pm 0.0006$ & $0.134 \pm 0.005$ & $0.114 \pm 0.004$ & $0.098 \pm 0.003$\\
      $\bar{A}_y$ & $0.1471 \pm 0.0003$ & $0.085 \pm 0.003$ & $0.073 \pm 0.003$ & $0.061 \pm 0.002$ \\
      $p_T^{max}$ (GeV) & $0.38 \pm 0.03$ & $0.46 \pm 0.09$ & $0.42 \pm 0.08$ & $0.47 \pm 0.08$  
    \end{tabular}
  \end{ruledtabular}
\end{table*}
Table~\ref{tab:Ayfitresults} shows the best-fit parameters and their uncertainties, and the resulting values for $A_y^{max}$ and $p_T^{max}$, defined, respectively, as the maximum value of the analyzing power and the $p_T$ value at which it occurs. The $p_T^{max}$ values exhibit some variation with $Q^2$ but are statistically compatible with a constant value $p_T^{max} \approx 0.4$ GeV (see Tab.~\ref{tab:Ayfitresults}).

\begin{figure}
\begin{center}
  \includegraphics[width=0.85\columnwidth]{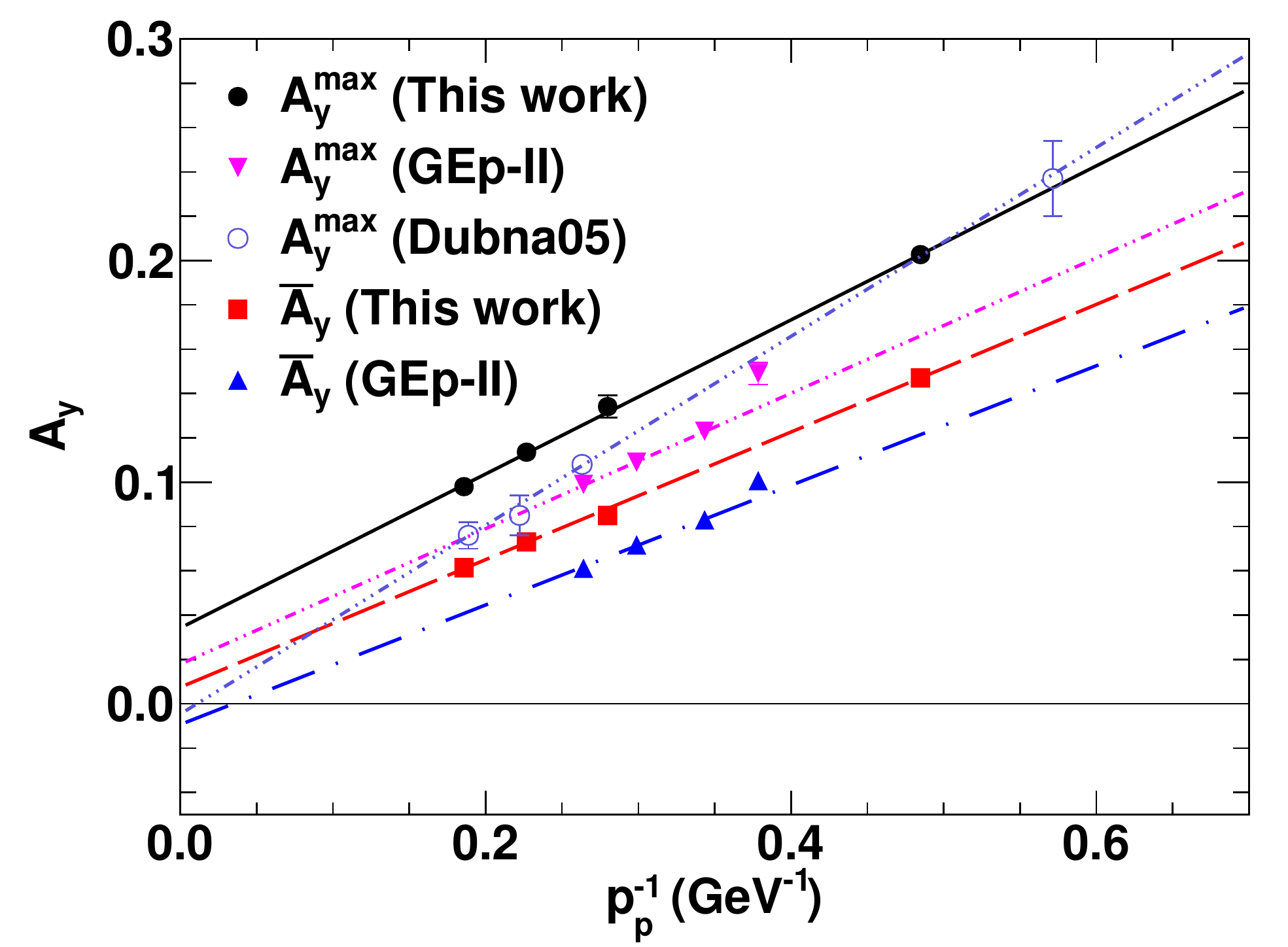}
\end{center}
\caption{\label{fig:Aypinv}  Maximum ($A_y^{max}$) and average ($\bar{A}_y$) analyzing power as a function of $\left<p_p\right>^{-1}$, compared to existing data from Refs.~\cite{Puckett:2011xg} (GEp-II) and \cite{Azhgirei:2005cs} (Azhgirey 2005). The average $A_y$ values for GEp-III/GEp-2$\gamma$ and GEp-II are computed for $0.06 \le p_T (\text{GeV}) \le 1.2$. Curves are linear fits to the data. See text for details.}
\end{figure}
Figure~\ref{fig:Aypinv} shows the proton momentum dependence of $A_y^{max}$ and $\bar{A}_y$, the average analyzing power within the accepted $p_T$ range, compared to selected existing measurements in $\vec{p} + $CH$_2$ scattering in the few-GeV momentum range, including GEp-II~\cite{Puckett:2011xg} in JLab's Hall A and dedicated measurements performed at the JINR in Dubna, Russia~\cite{Azhgirei:2005cs}. It is worth remarking that the GEp-II data were obtained with two different analyzer thicknesses; the lowest-$Q^2$ (largest $p_p^{-1}$) measurement used a CH$_2$ thickness of 58 cm, which is similar to the thickness used in Hall C for each of the two FPPs, while the three measurements at higher $Q^2$ used a thickness of 100 cm, leading to an apparent reduction in $A_y$ that was at least partially offset by an increase in the efficiency. The linear fits to the GEp-II data shown in Fig.~\ref{fig:Aypinv} only include the three highest-$Q^2$ points, which used the same analyzer thickness. It is also worth noting that the Dubna measurements~\cite{Azhgirei:2005cs} do not correspond to constant analyzer thickness; the Dubna measurement at $p_p = 1.75$ GeV used a CH$_2$ thickness of 37.5 g/cm$^2$, significantly less than the thickness used in either the Hall C or Hall A polarimeters. The Dubna measurements at higher proton momenta correspond to a range of analyzer thicknesses generally lying between the $\sim$50 g/cm$^2$ thickness used for each of the two analyzers in the Hall C double-FPP and the $\sim$90 g/cm$^2$ thickness of the GEp-II polarimeter. While the Dubna data appear to have a significantly greater slope than the JLab data, the difference is not statistically significant, given the large uncertainty of the Dubna measurement at $p_p = 1.75$ GeV, and the fact that this measurement corresponds to a CH$_2$ thickness of approximately half the thickness used for the other measurements at higher $p_p$.

For the GEp-III/2$\gamma$ experiments, $A_y^{max}$ and $\bar{A}_y$ depend linearly on $p_p^{-1}$. Notably, the extrapolated values of $A_y^{max}$ and $\bar{A}_y$ at asymptotically large proton momentum ($1/p_p \rightarrow 0$) are non-zero and positive for the conditions of GEp-III/2$\gamma$, although in the case of $\bar{A}_y$, the asymptotic value is only $\sim 3\sigma$ different from zero. The experimentally realized effective analyzing power for the Hall C double-FPP is substantially greater than that of the GEp-II or Dubna polarimeters at similar $p_p$. The difference is attributable to the Hall C drift chambers' ability, given their overall performance characteristics and the trigger and DAQ conditions specific to Hall C, to separate true single-track events from multiple-track events, revealing the significantly higher analyzing power for true single-track events compared to the totally inclusive sample. In the straw chambers of the Hall A FPP, for example, groups of eight adjacent wires in a plane were multiplexed into a single readout channel by the front-end electronics~\cite{Punjabi:2005wq}, preventing the resolution of multiple-track events in which two or more tracks pass through the same group of eight straws within a plane simultaneously.  

The effective analyzing power of a given sample of $\vec{p}+$CH$_2$ scattering events is clearly sensitive to experimental details such as the analyzer thickness, the momentum distribution of incident protons, the tracking resolution/efficiency, the background rate/occupancy of the detectors, the trigger and data acquisition conditions, and the cuts applied to select events. For this reason, it is generally not possible to predict $A_y$ using previous measurements such as~\cite{Azhgirei:2005cs,Punjabi:2005wq,Puckett:2011xg} with sufficient accuracy for an \emph{absolute} determination of $P_\ell$ commensurate with the statistical precision of the GEp-2$\gamma$ data. 

Nonetheless, the \emph{relative} variation of $P_\ell/P_\ell^{Born}$ with $\epsilon$ can be precisely extracted from the GEp-2$\gamma$ data by exploiting the fact that the experimental conditions which influence the effective average analyzing power are the same across all three kinematics measured at $Q^2 = 2.5$ GeV$^2$. In particular, the application of identical cuts on the FPP scattering parameters ensures that the effective average analyzing power is the same for all three $\epsilon$ values, up to differences in the momentum distribution of incident protons. As shown in Fig.~\ref{fig:Aypinv}, the average analyzing power for a given $p_T$ range is inversely proportional to the proton momentum $p_p$, while the $p_T$ distribution of the analyzing power is approximately independent of $p_p$. Given these experimental realities, the momentum dependence of $A_y$ can be accounted for on an event-by-event basis in the maximum-likelihood analysis by assuming that the overall momentum dependence factorizes from the $\vartheta$ and/or $p_T$ dependence:
\begin{eqnarray}
  A_y(p_p, p_T) &=& A_y^0(p_T) \frac{\bar{p_p}}{p_p}, \label{eq:Ayfppfactorized}
\end{eqnarray} 
where $A_y^0(p_T)$ and $\bar{p}_p$ are, respectively, the acceptance-averaged values of $A_y(p_T)$ and $p_p$. For the extraction of the ratio $R$, the analyzing power calibration is only relevant insofar as it optimizes the statistical figure-of-merit of the maximum-likelihood analysis by properly weighting events according to $A_y$. The extraction of the $\epsilon$ dependence of $P_\ell/P_\ell^{Born}$ in the GEp-2$\gamma$ analysis relies on the assumption that $A_y$ is the same for all three measurements, up to a global $p_p^{-1}$ scaling that factorizes from the $p_T$ dependence according to Eq.~\eqref{eq:Ayfppfactorized}. The lowest-$\epsilon$ data ($\left<\epsilon\right> = 0.153$) were used to determine the common $A_y^0(p_T)$ for the GEp-2$\gamma$ analysis for several reasons. First, the value of $P_\ell^{Born}$ approaches one as $\epsilon \rightarrow 0$ as a simple consequence of angular momentum conservation, and is highly insensitive to $r$ at $\left<\epsilon\right> = 0.153$, such that the relative statistical uncertainty $\Delta P_\ell^{Born}/P_\ell^{Born}$ due to the uncertainty in $r$ is more than three times smaller at the lowest $\epsilon$ than at either of the two higher $\epsilon$ values, and negligibly small compared to the statistical uncertainty in $P_\ell$ itself (see Tab.~\ref{tab:FinalResultsGEp2gamma}). Moreover, despite the fact that the measurement at $\left<\epsilon\right> = 0.153$ has the worst relative statistical precision for the \emph{ratio} $P_t/P_\ell$, it has the best relative precision for $P_\ell$ due to the large magnitude of $P_\ell$. 

\subsubsection{Background Subtraction}
The maximum-likelihood estimators are modified by the residual inelastic contamination of the elastic $ep$ sample as follows:
\begin{eqnarray}
  \sum_{i} \left[\begin{array}{cc} \left(\lambda_t^{(i)}\right)^2 & \lambda_t^{(i)}\lambda_\ell^{(i)} \\ \lambda_t^{(i)}\lambda_\ell^{(i)} & \left(\lambda_\ell^{(i)}\right)^2\end{array}\right]\left[\begin{array}{c} \hat{P}_t \\ \hat{P}_\ell \end{array}\right] = \nonumber \\
  \sum_i \left[\begin{array}{c} \lambda_t^{(i)}(1-\lambda_{bg}^{(i)}) \\ \lambda_\ell^{(i)}(1-\lambda_{bg}^{(i)} )\end{array}\right]. \label{MLestimators_bg}
\end{eqnarray}
The coefficients $\lambda_t^{(i)}$, $\lambda_\ell^{(i)}$ defined by Eq.~\eqref{MLmatrixelements} become:
\begin{eqnarray}
  \lambda_t^{(i)} &\rightarrow& (1-f_{bg}^{(i)}) \lambda_t^{(i)} \nonumber \\
  \lambda_\ell^{(i)} &\rightarrow & (1-f_{bg}^{(i)})\lambda_\ell^{(i)} \label{MLMEwithbackground},
\end{eqnarray}
with $f_{bg}^{(i)}$ denoting the fractional background contamination evaluated at the reconstructed kinematics of the $i$-th event, estimated according to the procedure discussed in section~\ref{sec:elastic_selection}. The background asymmetry term appearing on the RHS of Eq.~\eqref{MLestimators_bg} is defined as:
\begin{eqnarray}
  \lambda_{bg}^{(i)} &\equiv& f_{bg}^{(i)}h_i P_e A_y^{(i)}\left[ \left(S_{yt}^{(i)}\cos \varphi_i - S_{xt}^{(i)} \sin \varphi_i \right) P_{t}^{inel} \right. \nonumber \\
  & & \left. + \left(S_{y\ell}^{(i)} \cos \varphi_i - S_{x\ell}^{(i)} \sin \varphi_i \right)P_\ell^{inel} \right],
\end{eqnarray}
where $P_t^{inel}$ and $P_\ell^{inel}$ are the transferred polarization components of the inelastic background, which are measured using the inelastic events as described below.

The residual inelastic contamination of the final selection of elastic $ep$ events is estimated directly from the data using the procedure described in Sec.~\ref{sec:elastic_selection}. Averaged over the acceptance of the final cuts, the fractional contamination $f$ ranges from $0.16\%$ for $Q^2 = 2.5$ GeV$^2$, $\epsilon = 0.638$ to $4.89\%$ for $Q^2 = 8.5$ GeV$^2$ (see Tab.~\ref{tab:frac_bg}). The measured polarizations $P_{t,\ell}^{obs}$ are related to the signal and background polarizations by
\begin{eqnarray}
  P_{t,\ell}^{obs} &=& (1-f) P_{t,\ell}^{el} + f P_{t,\ell}^{inel}, \label{bgsubtract}
\end{eqnarray}
where $P_{t,\ell}^{el}$ and $P_{t,\ell}^{inel}$ are, respectively, the transferred polarizations of the elastic ``signal'' and the inelastic ``background''. 
\begin{figure}
  \begin{center}
    \includegraphics[width=0.85\columnwidth]{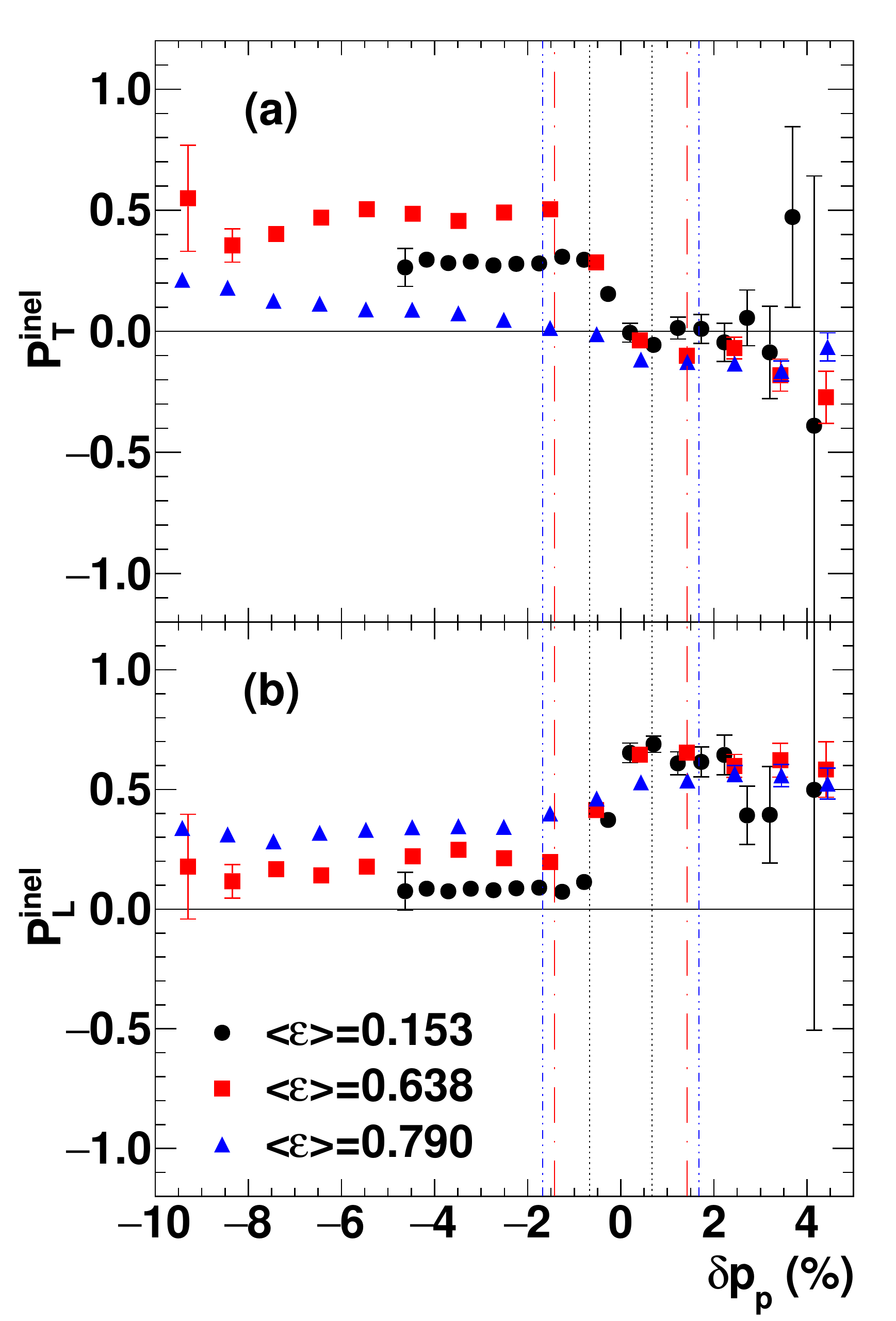}
  \end{center}
  \caption{\label{fig:Pinel_gep2g}  Transferred polarization components of the inelastic background vs. $\delta p_p$ for the GEp-2$\gamma$ kinematics. Panel (a) shows the transverse component $P_t^{inel}$, while panel (b) shows the longitudinal component $P_\ell^{inel}$. Vertical lines illustrate the approximate final cut regions for $\epsilon = 0.153$ (black dotted), $\epsilon = 0.638$ (red dot-dashed) and $\epsilon = 0.790$ (blue double dot-dashed). }
\end{figure}

\begin{figure}
  \begin{center}
    \includegraphics[width=0.85\columnwidth]{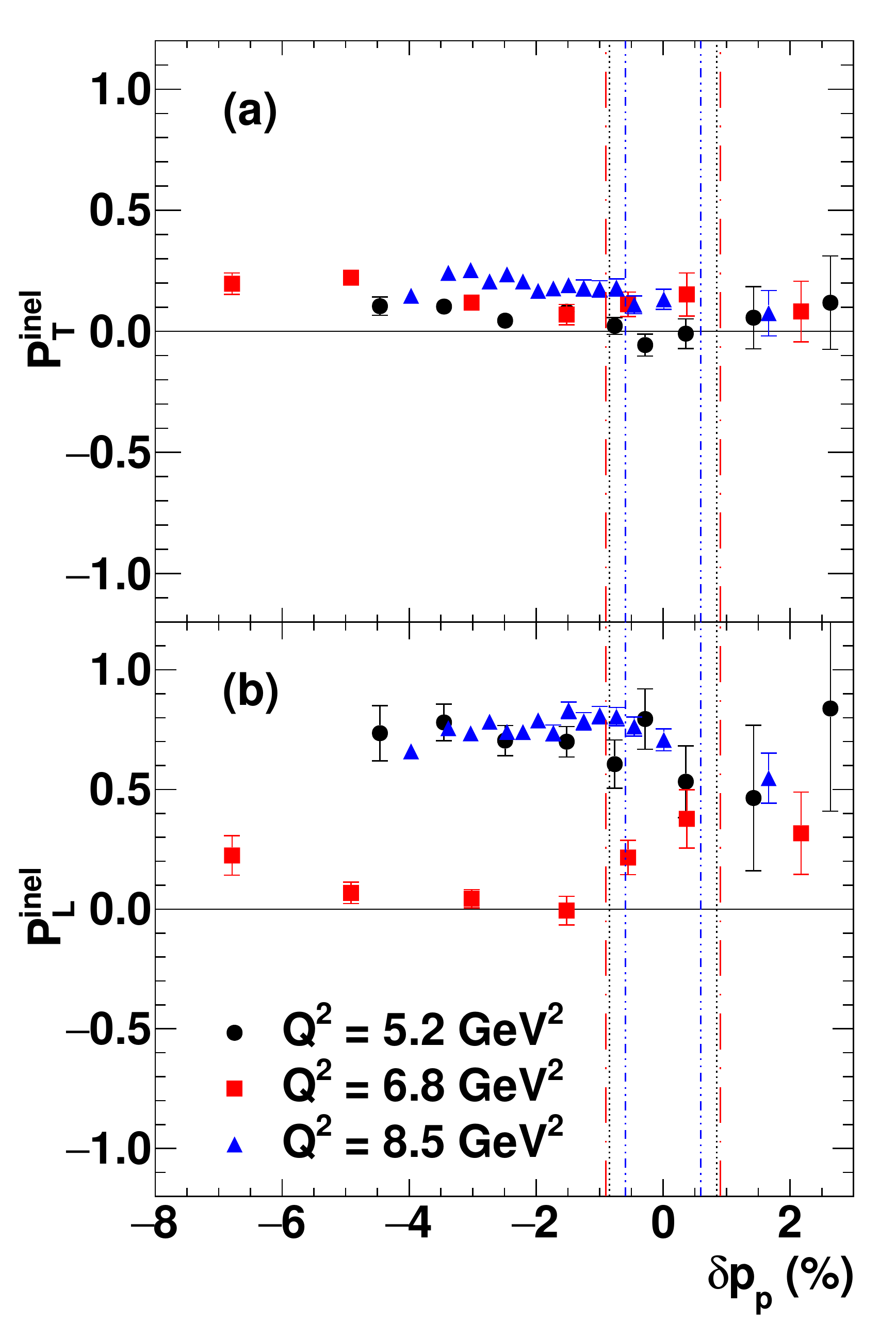}
  \end{center}
  \caption{\label{fig:Pinel_gep3}  Transferred polarization components of the inelastic background vs. $\delta p_p$ for the GEp-III kinematics. Panel (a) shows the transverse component $P_t^{inel}$, while panel (b) shows the longitudinal component $P_\ell^{inel}$. Vertical lines illustrate the approximate final cut regions for $Q^2 = 5.2$ GeV$^2$ (black dotted), $Q^2 = 6.8$ GeV$^2$ (red dot-dashed) and $Q^2 = 8.5$ GeV$^2$ (blue double dot-dashed). }
\end{figure}
Figures~\ref{fig:Pinel_gep2g} and \ref{fig:Pinel_gep3} show the $\delta p_p$ dependence of $P_{t,\ell}^{inel}$ for the GEp-2$\gamma$ and GEp-III kinematics, respectively. The background polarizations are extracted directly from the data by applying the maximum-likelihood method described above to the inelastic events, using the analyzing power resulting from the elastic events. Background events were selected by excluding a two-dimensional region of $(\delta p_e, \delta \phi)$ in which the elastic peak and radiative tail contributions are significant. The background polarizations exhibit a strong $\delta p_p$ dependence in the region of the elastic peak, a behavior explained by the different background processes involved and their relative contributions. In the inelastic region ($\delta p_p < 0$), which is dominated by $\pi^0p$ events, the background polarizations are approximately constant and differ strongly from the signal polarizations. The polarization transfer observables for $\vec{\gamma} p \rightarrow \pi^0 \vec{p}$, measured rather precisely as a byproduct of this experiment, are interesting in their own right, and were already the subject of a dedicated publication~\cite{Luo:2011uy}, which also addressed the induced polarization, which is non-negligible for the $\vec{\gamma} p \rightarrow \pi^0 \vec{p}$ process. The induced polarization of the $\pi^0p$ background is ignored here, as its effect on the extraction of the transferred polarization of the elastic signal is negligible. In the region of overlap with the elastic peak, the background polarizations evolve rapidly toward values that are similar (but not identical) to the signal polarizations. This transition reflects the sharp kinematic cutoff for $\pi^0p$ production and the transition to a regime in which the dominant background process is quasi-elastic Al$(e,e'p)$ scattering in the end windows of the cryotarget. The $\delta p_p$-dependences of the contamination $f$ and the background polarizations $P_{t,\ell}^{inel}$ are accounted for in the final, background-subtracted maximum-likelihood analysis. The total corrections to $R$, $P_t$ and $P_\ell$ are dominated by the lowest $\delta p_p$ bins within the final cut region, and are slightly smaller than would be implied by correcting the acceptance-averaged results using the acceptance-averaged values of $f$ and $P_{t,\ell}^{inel}$ using Eq.~\eqref{bgsubtract}.
\begin{table}
  \caption{\label{tab:bgcorrections} Inelastic background corrections to $P_t$, $P_\ell$, and $R$. Systematic uncertainties associated with the background correction are discussed in Ref.~\cite{Puckett:2017egz}.}
    \begin{ruledtabular}
      \begin{tabular}{ccccc} 
        $Q^2$ (GeV$^2$) & $\left<\epsilon\right>$ & $\Delta P_t$ & $\Delta P_\ell$ & $\Delta R$ \\ \hline 
        2.5 & 0.153 & -0.0013 & 0.0024 & 0.0043 \\
        2.5 & 0.638 & -0.0008 & 0.0005 & 0.0023 \\
        2.5 & 0.790 & -0.0002 & 0.0002 & 0.0007 \\
        5.2 & 0.382 & -0.0010 & 0.0015 & 0.0043 \\
        6.8 & 0.519 & -0.0009 & 0.0030 & 0.0036 \\
        8.5 & 0.243 & -0.0060 & 0.0096 & 0.0419 
      \end{tabular}
    \end{ruledtabular}
\end{table}
Table~\ref{tab:bgcorrections} shows the effect of the background subtraction on $P_t$, $P_\ell$ and $R$. The uncertainties associated with the background subtraction procedure are discussed in Ref.~\cite{Puckett:2017egz}. In all cases, the correction to $P_t$ ($P_\ell$) is negative (positive), and the resulting correction to $R$ is always positive. In general, the corrections to $R$ and $P_\ell$ are very small, except in the case of $Q^2 = 8.5$ GeV$^2$, for which the size of the correction to $R$ is comparable to the total systematic uncertainty. Despite the similar levels of inelastic contamination between $\left<\epsilon\right> = 0.638$ and $\left<\epsilon\right> = 0.790$ at 2.5 GeV$^2$, the corrections at $\left<\epsilon\right> = 0.790$ are significantly smaller, because of the smaller differences between the signal and background polarizations.

\subsection{Radiative Corrections}
The ``standard'', model-independent $\mathcal{O}(\alpha)$ radiative corrections (RC) to polarized elastic $\vec{e}p$ scattering have been discussed extensively in Refs.~\cite{Afanasev:2001jd,Afanasev:2001nn,Afanasev:2000aq,Akushevich:2011zy}, and include standard virtual RC such as the vacuum polarization and vertex corrections, and emission of real photons (Bremsstrahlung). Radiative corrections to double-polarization observables, such as the beam-target double-spin asymmetry in scattering on a polarized target, or polarization transfer as in this experiment, tend to be smaller than the RC to the unpolarized cross sections, because polarization asymmetries are ratios of polarized and unpolarized cross sections, for which the factorized, virtual parts of the RC tend to partially or wholly cancel in the expression for the relative RC to the asymmetry. Moreover, the effect of Bremsstrahlung corrections can be suppressed by the exclusivity cuts used to select elastic events. The ratio of transferred polarization components $P_t/P_\ell$, which is directly proportional to $G_E^p/G_M^p$ in the Born approximation, is a ratio of ratios of cross sections, and is subject to RC that are typically as small as or smaller than the RC to the individual asymmetries, depending on the kinematics and cuts involved.

\begin{table}
  \caption{\label{tab:RC_ratio} Estimated model-independent relative radiative corrections to $R = \mu_p G_E^p/G_M^p$ and the longitudinal transferred polarization component $P_\ell$, calculated using the approach described in Ref.~\cite{Afanasev:2001nn}. Note that a negative (positive) value for the radiative correction as presented below implies a positive (negative) correction to obtain the Born value from the measured value for the observable in question. These corrections have \emph{not} been applied to the final results shown in Tables~\ref{tab:FinalResultsGEP3} and \ref{tab:FinalResultsGEp2gamma}. See text for details.}
  \begin{ruledtabular}
    \begin{tabular}{lllrr}
      $Q^2$ (GeV$^2$) & $E_e$ (GeV) & $u_{max}$ (GeV$^2$) & $\frac{R_{obs}}{R_{Born}}-1$ & $\frac{P_\ell^{obs}}{P_\ell^{Born}}-1$\\ \hline
      2.5 & 1.87 & 0.03 & $-1.4 \times 10^{-3}$ & $1.2 \times 10^{-4}$ \\
      2.5 & 2.848 & 0.08 & $-2.8 \times 10^{-4}$ & $6.2 \times 10^{-4}$\\
      2.5 & 3.548 & 0.1 & $-1.6 \times 10^{-4}$ & $8.3 \times 10^{-4}$ \\
      2.5 & 3.680 & 0.1 & $-1.5 \times 10^{-4}$ & $8.4 \times 10^{-4}$ \\
      5.2 & 4.052 & 0.08 & $-5.0 \times 10^{-4}$ & $2.2 \times 10^{-4}$ \\
      6.8 & 5.710 & 0.12 & $-3.3 \times 10^{-4}$ & $3.2 \times 10^{-4}$ \\
      8.5 & 5.712 & 0.1 & $-8.0 \times 10^{-4}$ & $1.3 \times 10^{-4}$
    \end{tabular}
  \end{ruledtabular}
\end{table}
The model-independent RC to the ratio $R$ were estimated using the formulas described in Ref.~\cite{Afanasev:2001nn}. The results for the relative RC to $R$ and $P_\ell/P_\ell^{Born}$ are shown in Table~\ref{tab:RC_ratio}. The corrections are very small in all cases. For the ratio $R$, the correction is negative for every kinematic. The corrections to $P_\ell$ are also negligible in magnitude, and do not exceed $10^{-3}$ for any kinematic. The upper limit on the Lorentz-invariant ``inelasticity'' $u \equiv (k_1 + p_1 - p_2)^2$, with $k_1$, $p_1$, and $p_2$ denoting the four-momenta of incident electron, target proton, and recoil proton, respectively, was chosen according to the effective experimental resolution of $u$ by plotting the distribution of $u$ for events selected by the exclusivity cuts described in Sec.~\ref{sec:elastic_selection}. It is assumed in the calculations that only the outgoing proton is observed, and the kinematics of the unobserved scattered electron and/or the radiated hard Bremsstrahlung photon are integrated over. In reality, the tight exclusivity cuts applied to the kinematics of both the electron and proton angles and the proton momentum are such that Bremsstrahlung corrections are even more strongly suppressed than in the case of a simple cut on $u$ reconstructed from the measured proton kinematics. The ``true'' model-independent RC to the ratio could be expected to be even smaller than those reported in Tab.~\ref{tab:RC_ratio}, which can be regarded as conservative upper limits. No radiative corrections have been applied to the final results for $R$ and $P_\ell/P_\ell^{Born}$ reported in Sec.~\ref{sec:results} below, as the estimated values of the RC are essentially negligible compared to the statistical and systematic uncertainties of the data. Note also that no hard TPEX corrections are applied to the results, as there is presently no model-independent theoretical prescription for these corrections. Existing calculations give a wide variety of results, varying both in sign and magnitude, but are in general agreement that these corrections are small.


\section{Results}
\label{sec:results}
\subsection{Summary of the data}
\begin{figure}
  \begin{center}
    \includegraphics[height=0.85\columnwidth,angle=90]{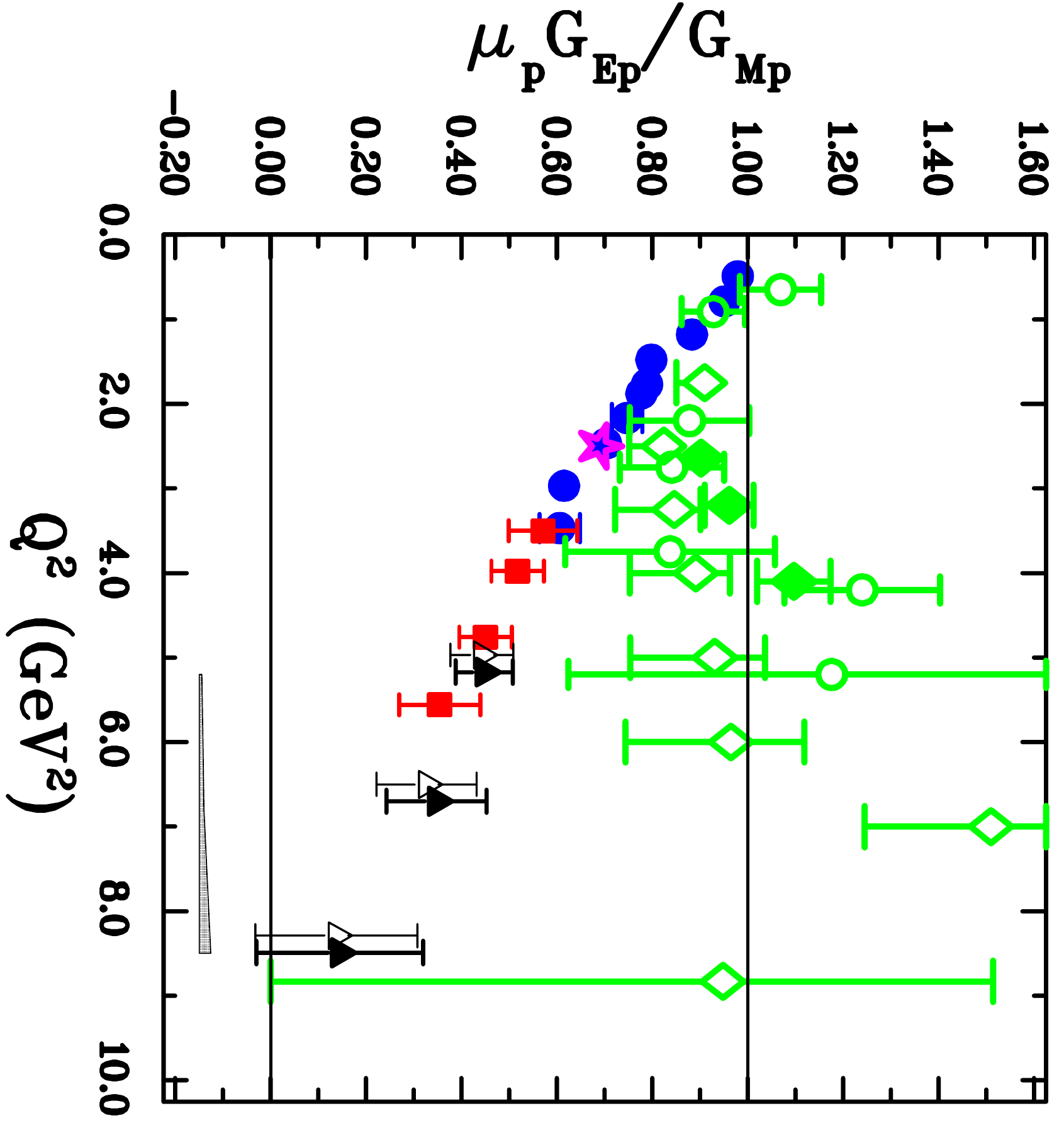}
  \end{center}
  \caption{\label{fig:FinalResultsGEP3}  Final results of GEp-III (black filled triangles) for $\mu_p G_E^p/G_M^p$, with selected existing data from cross section and polarization measurements. The error bars shown are statistical. The band below the data shows the final, one-sided systematic uncertainties for GEp-III. The originally published results~\cite{Puckett:2010ac} (black empty triangles) are shown for comparison, offset slightly in $Q^2$ for clarity. The final weighted-average result of GEp-2$\gamma$ for $R$ at $Q^2 = 2.5$ GeV$^2$ is shown as the pink empty star. Existing polarization transfer data are from Refs.~\cite{Punjabi:2005wq} (blue filled circles) and ~\cite{Puckett:2011xg,Gayou:2001qd} (red filled squares). Rosenbluth separation data are from Refs.~\cite{Christy:2004rc} (green empty circles), \cite{Andivahis:1994rq} (green empty diamonds), and \cite{Qattan:2004ht} (green filled diamonds).}
\end{figure}

\begin{table*}
  \caption{\label{tab:FinalResultsGEP3} Final results of the GEp-III experiment. These results supersede the originally published results from~\cite{Puckett:2010ac}. The central $Q^2$ value is defined by the HMS central momentum setting. The average beam energy $\left<E_{beam}\right>$ is the result of correcting the incident beam energy event-by-event for the mean energy loss in the target materials upstream of the reconstructed interaction vertex. The kinematics of each setting are described by the average, RMS deviation from the mean and total accepted range of $Q^2$ and $\epsilon$. The ratio $R = \mu_p G_E^p/G_M^p$ is quoted with its statistical and total systematic uncertainty. The polarization transfer components $P_t$ and $P_\ell$ are quoted with their statistical uncertainties to illustrate the relative statistical precision with which the two components are simultaneously measured\footnote{The difference between the absolute statistical errors $\Delta P_t$ and $\Delta P_\ell$ is entirely explained by spin precession.}. The quoted values of $P_t$ and $P_\ell$ are the maximum-likelihood estimators obtained after calibrating the analyzing power at each $Q^2$ as in Sec.~\ref{subsubsec:Ay}. The value of $P_\ell^{Born}$ is quoted with its statistical uncertainty, which is due solely to the uncertainty in $R$. $\rho(P_t,P_\ell)$ is the correlation coefficient between $P_t$ and $P_\ell$ resulting from the maximum-likelihood analysis. See text for details.}
  \begin{ruledtabular}
    \begin{tabular}{lrrr}
      Central $Q^2$ (GeV$^2$) & 5.200 & 6.800 & 8.537 \\ 
      $\left<E_{beam}\right>$ (GeV) & 4.049 & 5.708 & 5.710 \\ \hline 
      $\left<Q^2\right> \pm \Delta Q^2_{rms}$ (GeV$^2$)  & $5.17 \pm 0.12$ & $6.70 \pm 0.19$ & $8.49 \pm 0.17$ \\
      $(Q^2_{min}, Q^2_{max})$ (GeV$^2$) & $(4.90,5.47)$ & $(6.20,7.21)$ & $(8.14,8.87)$ \\
      $\left<\epsilon\right> \pm \Delta \epsilon_{rms}$ & $0.382 \pm 0.026$ & $0.519 \pm 0.027$ & $0.243 \pm 0.028$ \\
      $(\epsilon_{min}, \epsilon_{max})$ & $(0.32, 0.44)$ & $(0.45, 0.59)$ & $(0.18, 0.30)$ \\
      $R \pm \Delta R_{stat} \pm \Delta R_{syst}$ (final) & $0.448 \pm 0.060 \pm 0.006$ & $0.348 \pm 0.105 \pm 0.010$ & $0.145 \pm 0.175 \pm 0.024$ \\
      $P_t \pm \Delta_{stat} P_{t}$ & $-0.090 \pm 0.012$ & $-0.063 \pm 0.019$ & $-0.020 \pm 0.024$ \\
      $P_\ell \pm \Delta_{stat} P_{\ell}$ & $0.918 \pm 0.034$ & $0.842 \pm 0.027$ & $0.970 \pm 0.026$ \\
      $P_\ell^{Born} \pm \Delta_{stat} P_{\ell}^{Born}$ & $0.918 \pm 0.002$ & $0.851 \pm 0.002$ & $0.970 \pm 0.001$  \\
      $\rho(P_t, P_\ell)$ & -0.167 & -0.076 & 0.052
    \end{tabular}
  \end{ruledtabular}
\end{table*}

\begin{table*}
  \caption{\label{tab:FinalResultsGEp2gamma} Final results of the GEp-2$\gamma$ experiment. These results supersede the originally published results from~\cite{Meziane:2010xc}. Average kinematics and ranges are as in Tab.~\ref{tab:FinalResultsGEP3}. The central $\epsilon$ value corresponds to the average beam energy and the central $Q^2$ of 2.5 GeV$^2$. The results at $\left<\epsilon\right> = 0.790$ are obtained by combining the data collected at $E_e = 3.549$ GeV and $E_e = 3.680$ GeV (see Tab.~\ref{kintable}) and analyzing them together as a single setting, which is justified by the very similar acceptance-averaged values of $Q^2$ and $\epsilon$ at these two energies. The acceptance-averaged values of the ratio $R \equiv -\mu_p \frac{P_t}{P_\ell}\sqrt{\frac{\tau(1+\epsilon)}{2\epsilon}}$ and the longitudinal polarization transfer component $P_\ell$ are quoted with statistical and total systematic uncertainties. $R_{bcc}$ is the ``bin-centering-corrected'' value of $R$ at the central $Q^2$ of 2.5 GeV$^2$ (see Tab.~\ref{tab:bcc} and discussion in Sec.~\ref{subsec:bcc}). $P_t$ is quoted with its statistical uncertainty only\footnote{As in Tab.~\ref{tab:FinalResultsGEP3}, the quoted values of $P_t$ and $P_\ell$ correspond to the maximum-likelihood estimators obtained using the results of the analyzing power calibration of Sec.~\ref{subsubsec:Ay}, performed at $\left<\epsilon\right> = 0.153$ under the assumption $P_\ell = P_\ell^{Born}$ and applied to all three kinematic settings.}. The total systematic uncertainty in $P_\ell$ is dominated by the beam polarization measurement. The point-to-point systematic uncertainties are defined relative to $\epsilon = 0.790 (0.153)$ for $R (P_\ell/P_\ell^{Born})$. $\rho(P_t,P_\ell)$ is the correlation coefficient between $P_t$ and $P_\ell$ resulting from the maximum-likelihood analysis. See text for details.}
  \begin{ruledtabular}
    \begin{tabular}{lrrr}
      Central $Q^2$ (GeV$^2$) & 2.500 & 2.500 & 2.500 \\
      Central $\epsilon$ & 0.149 & 0.632 & 0.783 \\
      $\left<E_{beam}\right>$ (GeV) & 1.867 & 2.844 & 3.632 \\ \hline 
      $\left<Q^2\right> \pm \Delta Q^2_{rms}$ (GeV$^2$) & $2.491 \pm 0.032$ & $2.477 \pm 0.074$ & $2.449 \pm 0.105$ \\
      $(Q^2_{min}, Q^2_{max})$ (GeV$^2$) & $(2.42,2.58)$ & $(2.33,2.68)$ & $(2.18,2.75)$ \\
      $\left<\epsilon\right> \pm \Delta \epsilon_{rms}$ & $0.153 \pm 0.015$ & $0.638 \pm 0.018$ & $0.790 \pm 0.017$ \\
      $(\epsilon_{min}, \epsilon_{max})$ & $(0.11,0.19)$ & $(0.59,0.67)$ & $(0.73,0.83)$ \\
      $R \pm \Delta R_{stat} \pm \Delta R_{syst}^{total}$ (final) & $0.6953 \pm 0.0091 \pm 0.0079$ & $0.6809 \pm 0.0070 \pm 0.0040$ & $0.6915 \pm 0.0059 \pm 0.0039$ \\
      $\Delta R_{syst}^{ptp}$ (cf. $\left<\epsilon\right> = 0.790$) & $0.0043$ & $0.0002$ & $0.0001$ \\
      $R_{bcc} \pm \Delta_{stat} R_{bcc}$ & $0.6940 \pm 0.0091$ & $0.6776 \pm 0.0070$ & $0.6837 \pm 0.0059$ \\
      $P_t \pm \Delta_{stat} P_{t}$ & $-0.1481 \pm 0.0019$ & $-0.1881 \pm 0.0019$ & $-0.1618 \pm 0.0013$ \\
      $P_\ell \pm \Delta_{stat} P_{\ell} \pm \Delta_{syst}^{total} P_\ell$ & $0.9750 \pm 0.0020 \pm 0.0042$ & $0.7335 \pm 0.0020 \pm 0.0051$ & $0.5802 \pm 0.0014 \pm 0.0040$ \\
      $P_\ell^{Born} \pm \Delta_{stat} P_{\ell}^{Born}$ & $0.9753 \pm 0.0003$ & $0.7295 \pm 0.0008$ & $0.5720 \pm 0.0006$ \\
      $\frac{P_\ell}{P_\ell^{Born}} \pm \Delta_{stat}\left(\frac{P_\ell}{P_\ell^{Born}}\right) \pm \Delta_{syst}^{total} \left(\frac{P_\ell}{P_\ell^{Born}}\right) $ & N/A & $1.0055 \pm 0.0029 \pm 0.0070$ & $1.0143 \pm 0.0027 \pm 0.0071$ \\
      $\Delta_{syst}^{ptp} \left(\frac{P_\ell}{P_\ell^{Born}}\right)$ (cf. $\left<\epsilon\right> = 0.153$) & N/A & 0.0053 & 0.0061 \\
      $\rho(P_t, P_\ell)$ & 0.019 & 0.009 & 0.006 
    \end{tabular}
  \end{ruledtabular}
\end{table*}

The final results of the GEp-III and GEp-2$\gamma$ experiments are shown in Fig.~\ref{fig:FinalResultsGEP3} and reported in Tables~\ref{tab:FinalResultsGEP3} and \ref{tab:FinalResultsGEp2gamma}. The acceptance-averaged values of the relevant observables can be considered valid at the acceptance-averaged kinematics ($Q^2$ and $\epsilon$). The final results of the GEp-III experiment for $R = \mu_p G_E^p/G_M^p$ are essentially unchanged relative to the original publication~\cite{Puckett:2010ac}, showing small, statistically and systematically insignificant increases for all three $Q^2$ points, despite non-trivial modifications to event reconstruction and elastic event selection in the final analysis. The statistical uncertainties of the GEp-III data are also slightly modified, as it was discovered during the reanalysis of the data that the effect of the covariance term expressing the correlation between $P_t$ and $P_\ell$ was not included in the originally published statistical uncertainties, whereas it is included in this work. The effect of the covariance term on $\Delta R_{stat}$ is only significant for $Q^2 = 5.2$ GeV$^2$, for which the correlation coefficient is $\rho(P_t, P_\ell) \approx -0.17$. Because $P_t$ and $P_\ell$ are opposite in sign at this $Q^2$, a negative correlation coefficient tends to reduce the magnitude of the statistical error (see Eq.~\eqref{eq:dRstat}). The larger correlation coefficient observed at 5.2 GeV$^2$ compared to all the other kinematics is related to the unfavorable precession angle centered near 180 degrees and the reduced sensitivity of the measured asymmetry to $P_\ell$. The final systematic uncertainties of the GEp-III data are also smaller than those originally published, as a result of more thorough analysis of the data from the study of the non-dispersive plane optics of the HMS~\cite{Puckett:2017egz}, which reduced the uncertainty in the total bend angle in the non-dispersive plane to $\Delta_{syst} \phi_{bend} \approx 0.14$ mrad.

The values of $P_\ell^{Born}$ quoted in Tables~\ref{tab:FinalResultsGEP3} and ~\ref{tab:FinalResultsGEp2gamma} are acceptance-averaged values, computed event-by-event from Eq.~\eqref{eq:PtBornPlBorn} using the parametrized global $Q^2$ dependence of $R$ resulting from ``Global Fit II'' of appendix~\ref{sec:globalfit}, which includes the final results of GEp-III and GEp-2$\gamma$ reported in this work. The statistical uncertainty $\Delta P_\ell^{Born}$ is computed at each kinematic by propagating the statistical uncertainty in $R$ through Eq.~\eqref{eq:PtBornPlBorn}, and is basically negligible compared to the uncertainty in $P_\ell$ itself. The use of a global parametrization of $R(Q^2)$ to calculate $P_\ell^{Born}$ is necessary for a self-consistent extraction of the $\epsilon$ dependence of $P_\ell/P_\ell^{Born}$ at 2.5 GeV$^2$. For the GEp-III kinematics and the lowest $\epsilon$ measurement from GEp-2$\gamma$, which is used for the analyzing power calibration, the differences between $P_\ell^{Born}$ computed from the global parametrization of $R(Q^2)$ and $P_\ell^{Born}$ computed directly from the measurement result for $R$ are negligible.  

The results in Tables~\ref{tab:FinalResultsGEP3} and \ref{tab:FinalResultsGEp2gamma} are the product of a thorough reanalysis of the data, aimed at reducing the systematic and statistical uncertainties of the final results. The most significant difference between the analysis reported here and that of the original publications is that this work uses the full dataset of the GEp-2$\gamma$ experiment to achieve a significant reduction in the statistical uncertainties. The original analysis, published in Ref.~\cite{Meziane:2010xc}, applied acceptance-matching cuts to the data at $\left<\epsilon\right> = 0.638$ and $\left<\epsilon\right> = 0.790$ to match the envelope of events at the HMS focal plane populated by the $\left<\epsilon\right> = 0.153$ data, and further restricted the proton momentum to $\left|\delta\right| \le 2\%$ for all three settings. These cuts selected subsamples of the data with essentially the same average $Q^2$, and thus the same average analyzing power, and suppressed possible $\epsilon$-dependent systematic effects resulting from the different phase space regions populated by elastically scattered protons, including the momentum dependence of the analyzing power, ``bin centering'' effects, and the quality of the reconstruction of the proton kinematics and the calculation of the spin transport matrix elements. 

The acceptance-matching and $\delta$ cuts applied in the original analysis~\cite{Meziane:2010xc} reduced the total number of events by a factor of approximately 2.5(3.4) at $\epsilon = 0.638 (0.790)$ relative to the full-acceptance dataset. Subsequent analysis has shown that the momentum dependence of the analyzing power is adequately accounted for by the global $p_p^{-1}$ scaling of Eq.~\eqref{eq:Ayfppfactorized}, and that the HMS optics and spin transport are well-calibrated within the wider phase space regions populated by the two higher-$\epsilon$ settings (see Fig.~\ref{fig:R_PT_PL_Q2_gep2g} and additional discussion in Ref.~\cite{Puckett:2017egz}). As a result, the statistical uncertainties in $R$ and $P_\ell/P_\ell^{Born}$ are significantly reduced relative to Ref.~\cite{Meziane:2010xc}, without increasing the systematic uncertainty. Other changes in the final analysis common to both experiments are mainly related to event reconstruction and elastic event selection. Details of the improvements in event reconstruction and elastic event selection, and the final evaluation of systematic uncertainties, can be found in Ref.~\cite{Puckett:2017egz}.

\begin{figure}
  \begin{center}
    \includegraphics[width=0.85\columnwidth]{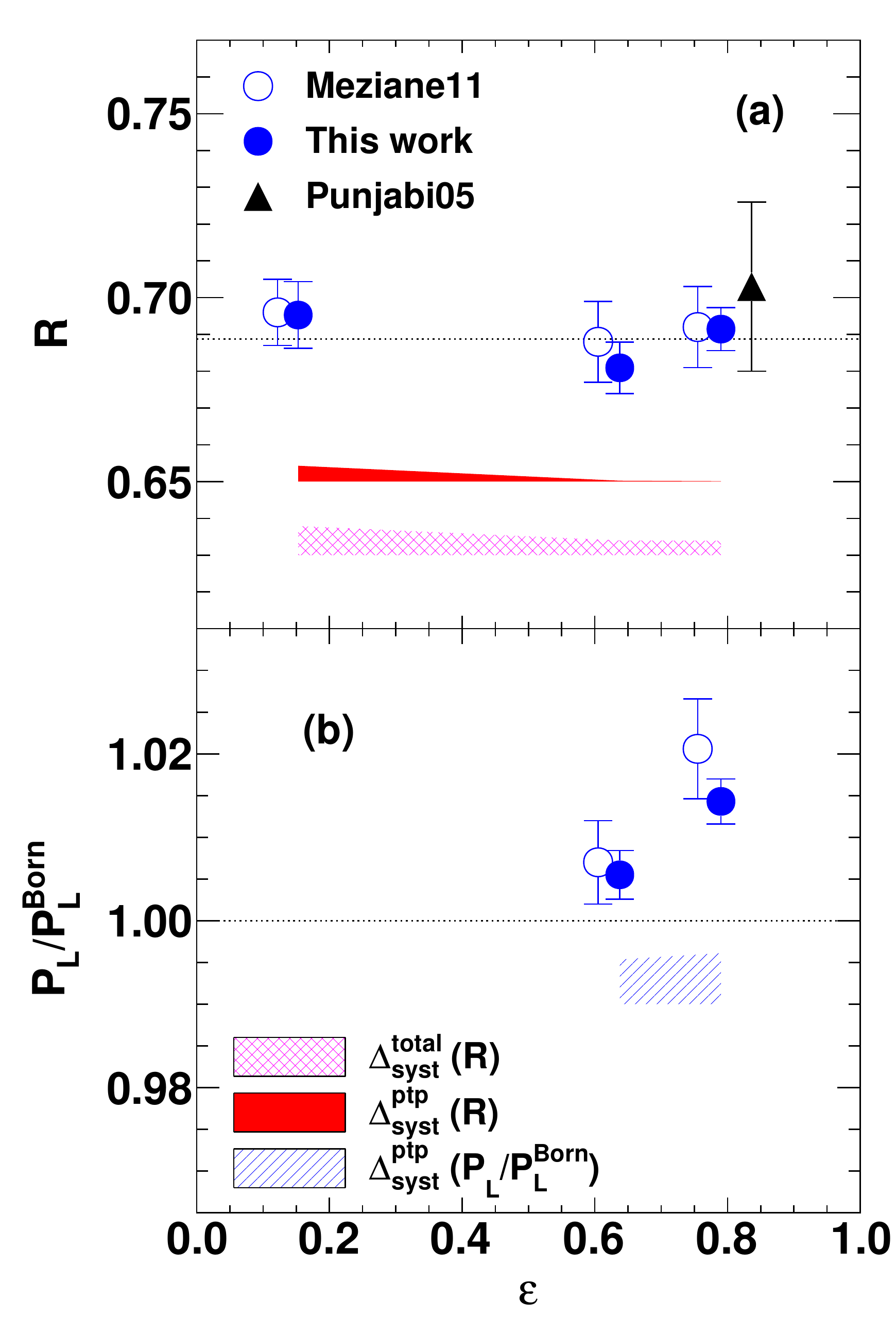}
  \end{center}
  \caption{\label{fig:twogammaresults}  Final, acceptance-averaged results of the GEp-2$\gamma$ experiment, without bin-centering corrections, as a function of $\epsilon$, for the ratio $R \equiv -\mu_p \frac{P_t}{P_\ell} \sqrt{\frac{\tau (1+\epsilon)}{2\epsilon}}$ (panel (a)), and the ratio $P_\ell/P_\ell^{Born}$ (panel (b)), compared to the originally published results~\cite{Meziane:2010xc} (Meziane11), and the GEp-I result~\cite{Punjabi:2005wq} (Punjabi05) at $Q^2 = 2.47$ GeV$^2$. Error bars on the data points are statistical only. For $R$, the (one-sided) total and point-to-point (relative to $\epsilon = 0.79$) systematic uncertainty bands are shown, while only the point-to-point (relative to $\left<\epsilon\right> = 0.153$) systematic errors are shown for $P_\ell/P_\ell^{Born}$ (also one-sided). The originally published points from Ref.~\cite{Meziane:2010xc} have been offset by -0.03 in $\epsilon$ for clarity. Note that $P_\ell/P_\ell^{Born} \equiv 1$ at $\left<\epsilon\right> = 0.153$.}
\end{figure}
Fig.~\ref{fig:twogammaresults} shows the final results for the $\epsilon$-dependence of $R$ and $P_\ell/P_\ell^{Born}$. The data collected at $E_e = 3.548$ GeV ($\left<\epsilon\right> = 0.779$) and $E_e = 3.680$ GeV ($\left<\epsilon\right> = 0.796$) were also analyzed separately and found to be consistent. The statistical compatibility of the separately analyzed results, the similarity of the average kinematics of the two settings, and the near-total overlap of their $Q^2$ and $\epsilon$ ranges justifies combining these two measurements into the single result reported in Tab.~\ref{tab:FinalResultsGEp2gamma} and shown in Fig.~\ref{fig:twogammaresults}. For both observables, the final results are consistent with the originally published results, but with significantly smaller statistical uncertainties at the two highest $\epsilon$ values. Notably, the enhancement of $P_\ell/P_\ell^{Born}$ at $\left<\epsilon\right> = 0.790$ relative to $\left<\epsilon\right> = 0.153$ persists in the full-acceptance analysis and is consistent with the $\sim 2\%$ enhancement seen in the original publication. The deviation from unity of the final result is 5.3 times the statistical uncertainty, 2.3 times the point-to-point systematic uncertainty, and 1.9 times the ``total'' uncertainty defined as the quadrature sum of the statistical and total systematic uncertainties. The $\sim 0.6\%$ enhancement at $\epsilon = 0.638$ is roughly a $2\sigma$ effect statistically, but also consistent with no enhancement within the point-to-point systematic uncertainty. The total and point-to-point systematic uncertainties in $P_\ell/P_\ell^{Born}$ are dominated by the point-to-point uncertainty $\Delta P_e/P_e = \pm 0.5\%$ in the beam polarization. It is worth noting that the global $\pm 1\%$ uncertainty of the M$\o$ller measurement of the beam polarization is irrelevant to the determination of the relative $\epsilon$ dependence of $P_\ell/P_\ell^{Born}$, because a global overestimation (underestimation) of the beam polarization is exactly compensated by an equal and opposite underestimation (overestimation) of the analyzing power at $\left<\epsilon\right> = 0.153$.

\subsection{``Bin centering'' effects in $R$ at $Q^2 = 2.5$ GeV$^2$}
\label{subsec:bcc}

In contrast with the original publication~\cite{Meziane:2010xc}, the acceptance-averaged results of the full-acceptance analysis of the GEp-2$\gamma$ data are quoted at significantly different average $Q^2$ values (see Tab.~\ref{tab:FinalResultsGEp2gamma}), such that the expected variation of $R$ with $Q^2$ can noticeably affect its apparent $\epsilon$-dependence, even in the absence of significant two-photon-exchange effects in this observable. The expected variation of $R$ with $Q^2$ within the acceptance of each point is much larger than its expected $\epsilon$ dependence, which is zero in the Born approximation and small in most model calculations of the hard TPEX corrections widely thought to be responsible for the cross section-polarization transfer discrepancy. For example, $R(Q^2)$ from the global fit described in appendix~\ref{sec:globalfit} varies by approximately seven times the statistical uncertainty of the acceptance-averaged result for $R$ within the $Q^2$ acceptance of the measurement at $\epsilon = 0.79$ (see Fig.~\ref{fig:R_PT_PL_Q2_gep2g}).

In order to correct the results for $R$ to a common central $Q^2$ of 2.5 GeV$^2$, a bin-centering correction to $R$ is computed for each kinematic under the assumption that $R$ depends only on $Q^2$, or, equivalently, under the weaker assumption that the global $Q^2$ dependence of $R$ factorizes from any potential $\epsilon$ dependence of $R$, at least within the acceptance of each kinematic. The corrected value of $R$ is obtained by multiplying the acceptance-averaged result, which corresponds to the average $Q^2$ and $\epsilon$, by the ratio $R(2.5$ GeV$^2)/R(\left<Q^2\right>)$, where $R(Q^2)$ is evaluated using the results of the global proton form factor fit\footnote{The corrections shown in Tab.~\ref{tab:bcc} are computed using the results of ``Global Fit II'' of appendix~\ref{sec:globalfit}. The corrections obtained using ``Global Fit I'' are indistinguishable.} described in appendix~\ref{sec:globalfit}. The corrected results are then plotted at the value of $\epsilon$ corresponding to the central $Q^2$, as opposed to the acceptance-averaged value of $\epsilon$. The bin-centering correction to $R$ is always negative, because the slope of $R(Q^2)$ is negative and the average $Q^2$ is less than the ``central'' $Q^2$ for all three settings (due to the $Q^2$ dependence of the acceptance-convoluted cross section). 
\begin{table}
  \caption{\label{tab:bcc} Summary of bin-centering corrections to $R$ at $Q^2 = 2.5$ GeV$^2$. $\left<Q^2\right>$ and $\left<\epsilon\right>$ are the acceptance-averaged kinematics. $\epsilon_c$ is the central $\epsilon$ value computed from the central $Q^2$ value and the average beam energy. $R_{bcc}$ is the bin-centering-corrected value of $R$ with statistical uncertainty. $R_{bcc} - R_{avg}$ is the bin-centering correction relative to the results for the average kinematics reported in Tab.~\ref{tab:FinalResultsGEp2gamma}.}
  \begin{ruledtabular} 
    \begin{tabular}{ccccc}
      $\left<Q^2\right>$ (GeV$^2$)  & $\left<\epsilon\right>$ & $\epsilon_c$ & $R_{bcc} \pm \Delta_{stat} R_{bcc}$ & $R_{bcc} - R_{avg}$\\ \hline 
      2.491 & 0.153 & 0.149 & $0.6940 \pm 0.0091$ & -0.0013 \\
      2.477 & 0.638 & 0.632 & $0.6776 \pm 0.0070$ & -0.0033 \\
      2.449 & 0.790 & 0.783 & $0.6837 \pm 0.0059$ & -0.0078
    \end{tabular}
  \end{ruledtabular}
\end{table}
Tab.~\ref{tab:bcc} shows the results for $R$ corrected to the ``central'' kinematics at $Q^2 = 2.5$ GeV$^2$. The magnitude of the correction is small but noticeable compared to the uncertainties for the two higher $\epsilon$ points, while being essentially negligible for $\epsilon = 0.153$. The differences between the average and central $\epsilon$ values are small. 
\begin{table}
  \caption{\label{tab:bcc2} Linear and constant fit results for the $\epsilon$ dependence of $R$, with and without bin-centering corrections. Quoted uncertainties in fit results are statistical only.}
  \begin{ruledtabular}
    \begin{tabular}{lcc}
      & No b.c.c. & b.c.c. \\ \hline 
      Slope $dR/d\epsilon$ & $-0.0076 \pm 0.0169$ & $-0.0173 \pm 0.0169$ \\
      Linear fit $\chi^2/ndf$ & 1.78/1 & 1.02/1 \\
      Linear fit ``$p$''-value & 0.18 & 0.31 \\ 
      Linear fit $R(\epsilon = 0)$ & $0.693 \pm 0.011$ & $0.694 \pm 0.011$ \\ 
      Constant fit $R$ & $0.6887 \pm 0.0040$ & $0.6837 \pm 0.0040$ \\
      Constant fit $\chi^2/ndf$ & 1.98/2 & 2.07/2 \\
      Constant fit ``$p$''-value & 0.37 & 0.36
    \end{tabular}
  \end{ruledtabular}
\end{table}
Tab.~\ref{tab:bcc2} shows the results of linear and constant fits to the $\epsilon$ dependence of $R$ for both the average and central kinematics. While the corrected and uncorrected data both favor a slightly negative slope for $R$ as a function of $\epsilon$, the slope is also compatible with zero in both cases. Indeed, the constant fits actually give higher ``$p$-values'' than the linear fits, although the comparison of these values is not particularly meaningful given the small number of degrees of freedom and the dramatically different shape of the theoretical $\chi^2$ distributions for $\nu = 1$ and $\nu = 2$. 

\begin{figure}
  \begin{center}
    \includegraphics[width=0.85\columnwidth]{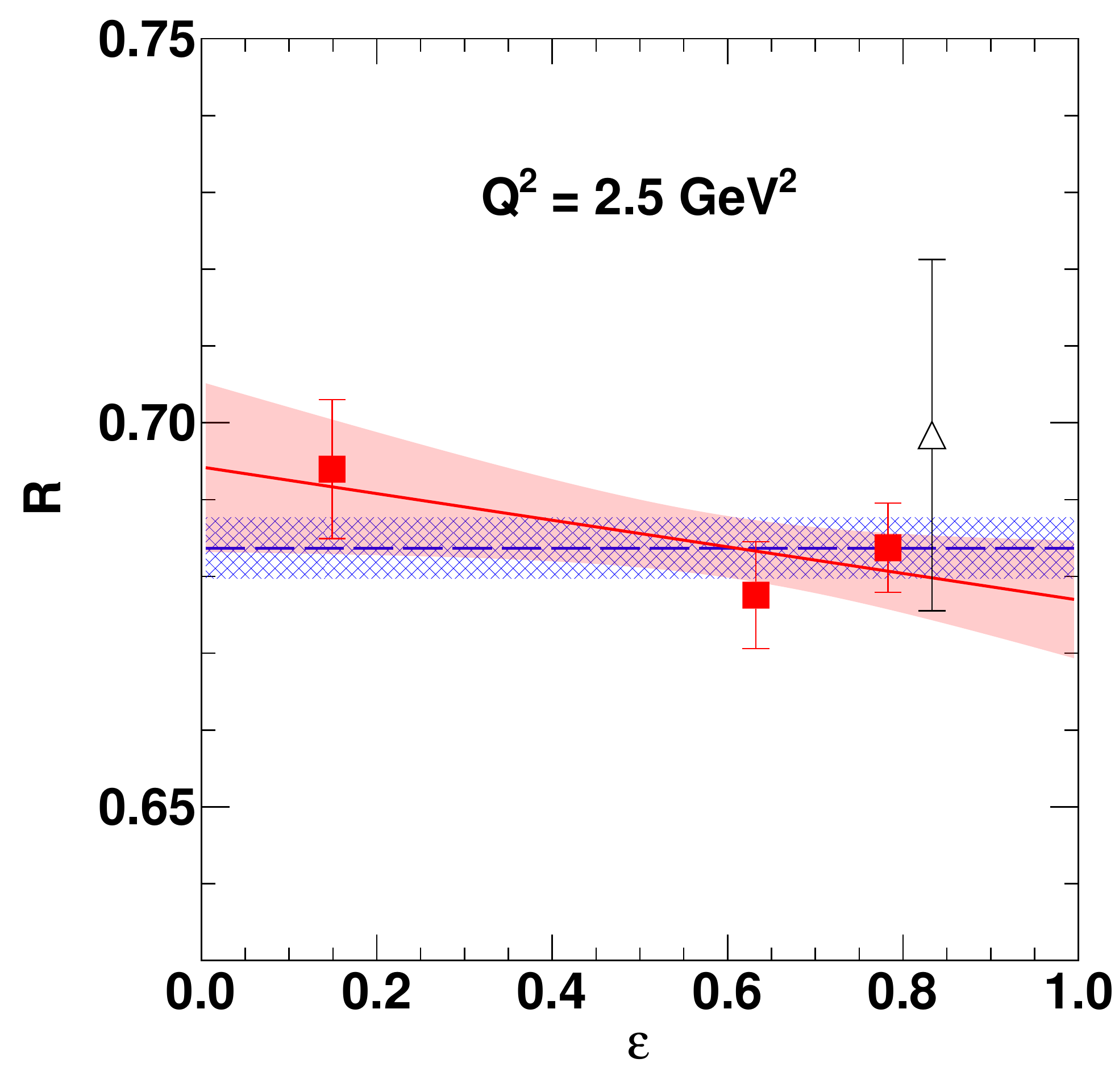}
  \end{center}
  \caption{\label{fig:Rbcctwogamma}  Bin-centering-corrected results for the $\epsilon$ dependence of the ratio $R$ at the common central $Q^2$ of 2.5 GeV$^2$ (red filled squares), with statistical uncertainties only. The red solid line is the linear fit to the corrected data reported in Tab.~\ref{tab:bcc2}. The red shaded region indicates the point-wise, $1\sigma$ uncertainty band of the linear fit (68\% confidence level). The blue dashed horizontal line is the weighted average of the three measurements assuming no $\epsilon$ dependence of $R$. The blue hatched region indicates the 68\% confidence interval ($1\sigma$) for the weighted average. The results of the constant fit are also quoted in Tab.~\ref{tab:bcc2}. The GEp-I result~\cite{Punjabi:2005wq} (empty triangle), corrected to 2.5 GeV$^2$ using the same approach as the GEp-2$\gamma$ data, is shown for comparison.}
\end{figure}
Fig.~\ref{fig:Rbcctwogamma} shows the final, bin-centering-corrected values of $R$ as a function of $\epsilon$ at 2.5 GeV$^2$. The linear fit quoted in Tab.~\ref{tab:bcc2} is also shown in Fig.~\ref{fig:Rbcctwogamma} with its 68\% confidence band. The full-acceptance data, which are significantly more precise at the two highest $\epsilon$ values than the originally published data~\cite{Meziane:2010xc}, slightly favor a small, negative slope $dR/d\epsilon = -0.017 \pm 0.017$ (see Tab.~\ref{tab:bcc2}), after correcting the data to the common central $Q^2$ of 2.5 GeV$^2$. The uncertainty in the slope $dR/d\epsilon$ is dominated by the statistical uncertainties of the data, as the point-to-point systematic uncertainties are small. The observed slope is consistent with zero, but is more likely to be negative than positive. No bin-centering corrections were necessary for the ratio $P_\ell/P_\ell^{Born}$, other than to quote the results at the central kinematics as opposed to the average kinematics. This is because the observed $Q^2$ dependence of $P_\ell$ closely follows the predicted $Q^2$ dependence of $P_\ell^{Born}$ (see Fig.~\ref{fig:R_PT_PL_Q2_gep2g}), such that the $Q^2$ dependence of the \emph{ratio} $P_\ell/P_\ell^{Born}$ is consistent with a constant within the acceptance of each kinematic.  

\section{Comparison to theoretical predictions}
\label{sec:theory}

\subsection{Theoretical interpretation of $G_E^p/G_M^p$ at large $Q^2$}
\label{subsec:theory_sub_a}


Among the primary motivations for measuring nucleon elastic electromagnetic form factors to larger $Q^2$ values is to observe the transition from strong coupling and confinement to the regime of perturbative QCD (pQCD) physics. However, the applicability of pQCD to hard exclusive processes such as elastic electron-nucleon scattering may require much larger momentum transfers than those currently accessible.
One fact that the new proton data have revealed beyond a doubt, is the importance of 
quark orbital angular momentum to the understanding of nucleon structure. The role of orbital 
angular momentum is also revealed in a global way, by the very fact that 
the nucleon magnetic moment is strongly anomalous, differing from 
the Dirac magnetic moment by $\sim\pm$ 2 units of the nuclear magneton, 
for the proton and neutron, respectively.
Solving the QCD equations from first principle for the nucleon is only 
possible on the lattice; until quite recently, the feasible $Q^2$ range for lattice calculations of nucleon FFs has been limited to $Q^2 \lesssim 3$ GeV$^2$ by computing power and other technical issues. The expectation, given increases in computational power and technical innovations in the methodology of the calculations, is that lattice QCD will be applicable 
up to 10 GeV$^2$ or higher in the near future. At the present time  
only phenomenological models which include some, but not all of the fundamental 
characteristics of QCD are possible. Some of the most 
successful models include Vector Meson Dominance (VMD), the relativistic Constituent 
Quark Models (RCQM), Generalized Parton Distributions (GPD), Dyson-Schwinger QCD, and others. We discuss a selection of these approaches in more detail here and compare them with the data.

\subsubsection{Vector Meson Dominance}
The earliest models explaining the global features of the nucleon form factors,
such as their apparent and approximate dipole behavior, were vector meson 
dominance (VMD) models. In this picture the photon couples to the nucleon 
through the exchange of vector mesons. A single vector meson exchange with simple couplings gives 
an $m_V^2/(m_V^2 - q^2)$ factor, from its propagator, for the falloff 
of the form factor. One can obtain a $Q^{-4}$ high momentum falloff, in accord with observation or with pQCD, from cancellations 
among two or more vector meson exchanges with different masses, or by giving the vector mesons themselves a 
form factor in their coupling to nucleons. 

An early example of a VMD fit to form factor data was given by Iachello, Jackson, and Lande~\cite{Iachello:1972nu} or IJL. They had 
several fits, but the one most cited is a 5-parameter fit with a more complicated $\rho$ propagator than the form noted above, to 
account for the large decay width of the $\rho$ meson.  (The $\omega$ and $\phi$ are narrow enough that modifying their propagators 
gives no numerical advantage.)  

The IJL work was improved by Gari and Kr\"umpelmann \cite{Gari:1984ia,Gari:1992qw} to better match the power law pQCD expectations at 
high $Q^2$, that $F_1 \sim Q^{-4}$ and $F_2 \sim Q^{-6}$, but also including some $\ln (Q^2)$ corrections to the falloffs based on the 
running behavior of the coupling $\alpha_s(Q^2)$.

Further improvement in VMD fits was made by Lomon \cite{Lomon:2001ga,*Lomon:2002jx,*Lomon:2006xb}, who included  a 
second $\rho$ as the $\rho'(1450)$, and 
later also a second $\omega$ as the $\omega'(1419)$, and obtained a good parameterization for all the nucleon form factors. The first 
of the polarization transfer  measurements~\cite{Jones:1999rz} became available in time for Lomon's 2001 work~\cite{Lomon:2001ga}. Lomon 
further tuned his fits~\cite{Lomon:2006xb} when the second set of polarization transfer data became available~\cite{Gayou:2001qd}.

\begin{figure}
\begin{center}
\includegraphics[angle=0,width =0.85\columnwidth]{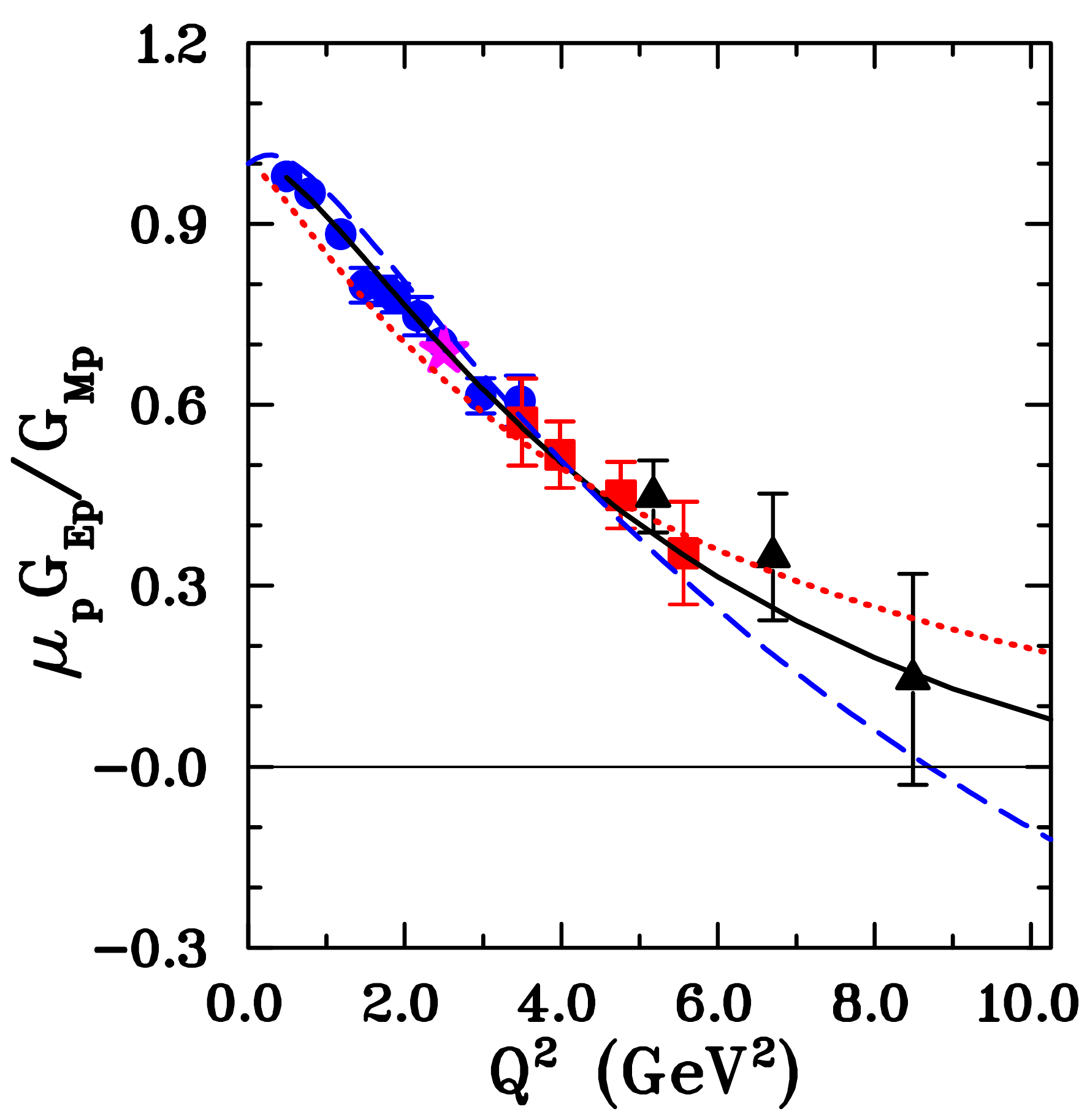}
\caption{ Several VMD fits compared to the JLab $G_E^p/G_M^p$ data.  The solid curve (black) is the fit of 
Lomon~\cite{Lomon:2006xb}, the 
dashed curve (blue) is that of Iachello, Jackson, and Lande~\cite{Iachello:1972nu}, and the dotted curve (red) is that of Bijker 
and Iachello~\cite{Bijker:2004yu}. Data are from Refs.~\cite{Punjabi:2005wq} (blue circles), ~\cite{Puckett:2011xg} (red squares), and the present work (black triangles for GEp-III and pink star for GEp-2$\gamma$). The GEp-2$\gamma$ result shown is the weighted average of the three $\epsilon$ points \emph{without} bin-centering corrections (see Tab.~\ref{tab:bcc2}). Figure adapted from Fig. 23 of Ref.~\cite{Punjabi:2015bba}. }
\label{fig:f2andf1}
\end{center}
\end{figure}

In addition, the original IJL fits~\cite{Iachello:1972nu} were not as good for the neutron as for the proton.  Both the spacelike neutron 
form factors and timelike nucleon form factors were addressed in what may be termed IJL updates, by Iachello 
and Wan~\cite{Iachello:2004aq} 
and Bijker and Iachello~\cite{Bijker:2004yu}, both in 2004.   Further, Lomon and Pacetti~\cite{Lomon:2012pn} have updated and analytically 
continued the earlier Lomon fits in order to also give a good account of data in both timelike and spacelike regions. The VMD models are of course fits to existing data, and they have been regularly updated as new data appeared. It will be interesting 
to check the ``predictions'' for the neutron form factors as new data appear. A plot of the existing situation for the proton is 
given in Fig.~\ref{fig:f2andf1}.

VMD models are a special case of the more general dispersion 
relation approach which relates the nucleon form factors in the space-like ($q^2 < 0$) region accessible in fixed-target electron scattering to the time-like ($q^2 > 0$) region accessible in annihilation experiments $e^+ e^- \rightarrow p\bar{p}$ (or $p\bar{p} \rightarrow e^+ e^-$). 
The analytic properties of FFs justify a common interpretation of scattering and annihilation experiments and the precision reachable at colliders requires a unified description of form factors for both space-like and time-like $q^2$. Although the separation of $G_E$ and $G_M$ has been challenging in the time-like region due to the low luminosities of $e^+ e^-$ colliders relative to fixed-target experiments, some data on the form factor ratio in the time-like region do exist, mainly from the study of the initial-state radiation (ISR) process $e^+ e^- \rightarrow p\bar{p} \gamma$. The most recent and precise data in the time-like region come from the BABAR collaboration~\cite{Aubert:2005cb,Lees:2013ebn}.

\subsubsection{Constituent Quark Models}

The early success of the non-relativistic constituent 
quark model was in explaining static properties, including magnetic moments and transition amplitudes. Examples are the 
models of De R\'ujula, Georgi, and Glashow~\cite{DeRujula:1972te} and of Isgur and Karl~\cite{Isgur:1978xj}. However, to 
describe the data presented here in terms of constituent quarks, 
it is necessary to include relativistic effects because the momentum transfers involved
are much larger than the constituent quark mass.

Constituent quark models (CQMs) have been used to understand the structure of nucleons, beginning when quarks were first hypothesized and 
predating the emergence of QCD as the theory of the strong interaction. In the CQM, ground state 
nucleons (and other baryons in the lowest-lying spin-1/2 octet and spin-3/2 decuplet) are composed of three valence quarks, selected from the three lightest flavors up ($u$), down ($d$) and strange ($s$), and described using 
$SU(6)$ spin-flavor wave functions and a completely antisymmetric color wave function.
\begin{figure}
\begin{center}
\includegraphics[angle=0,width=0.85\columnwidth]{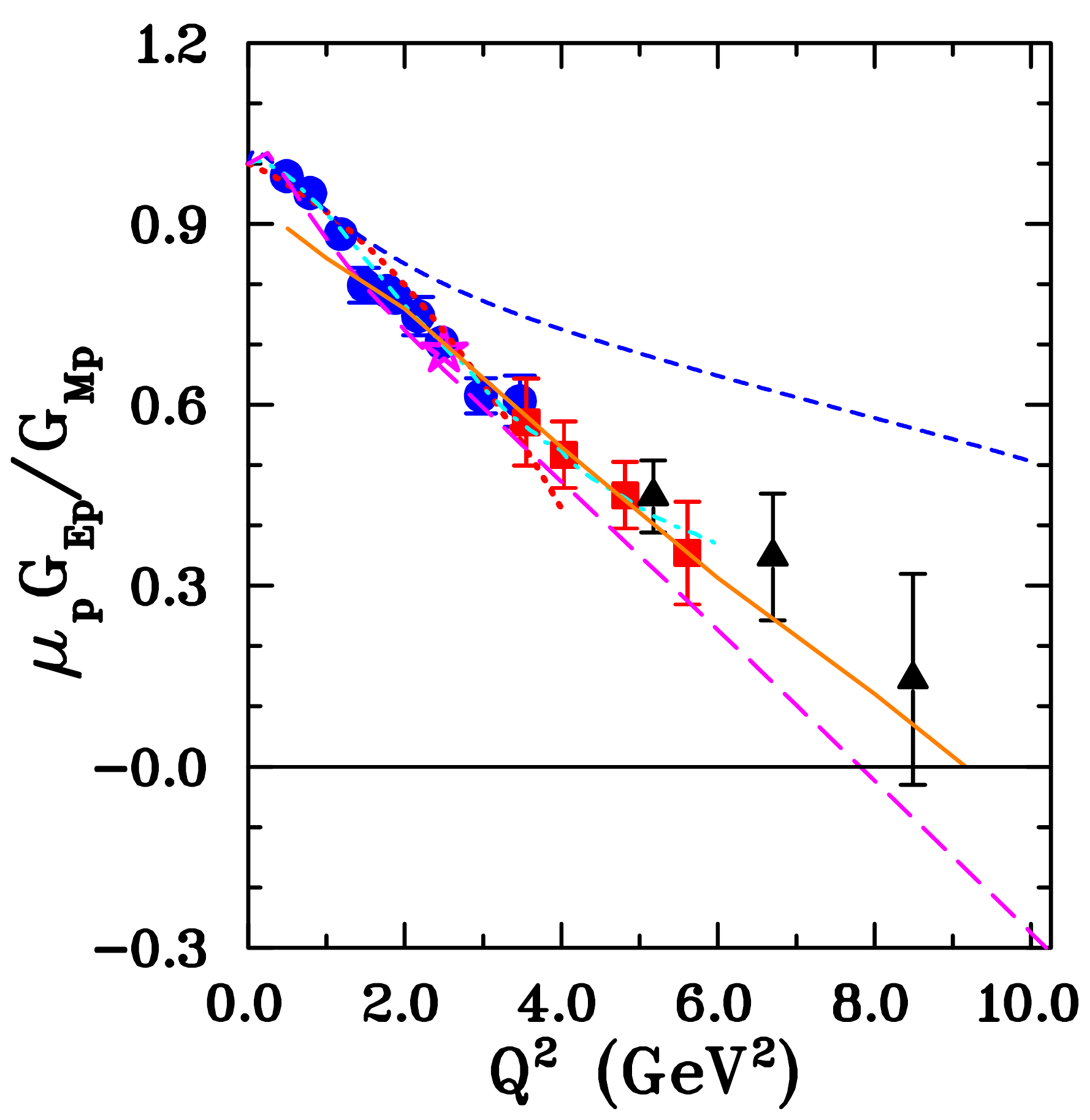}
\caption{ The JLab $G_E^p/G_M^p$ data compared to the results of a selection of constituent quark models.   The short dashed curve (blue) 
is from Boffi \textit{et al.}~\cite{Boffi:2001zb}, the solid (orange) from de Melo \textit{et al.}~\cite{deMelo:2008rj}, the long dash (magenta) 
from Gross \textit{et al.}~\cite{Gross:2006fg}, the dotted (red) from Chung and Coester~\cite{Chung:1991st}, and the dash-dot (cyan) from 
Cardarelli \textit{et al.}~\cite{Cardarelli:1995dc}. Data are the same as in Fig.~\ref{fig:f2andf1}. Figure adapted from Fig. 24 of Ref.~\cite{Punjabi:2015bba}.}
\label{fig:cqm}
\end{center}
\end{figure}
Figure~\ref{fig:cqm} compares a selection of CQM calculations to the polarization transfer data for $\mu_p G_E^p/G_M^p$ from the GEp-I, GEp-II, GEp-III and GEp-2$\gamma$ experiments. 

A crucial question for a form factor calculation, since the nucleon must be moving after or before the interaction or both, is how 
the wave function in the rest frame transforms to a moving frame. The relative ease of exactly transforming states from the frame 
where the wave functions are calculated or otherwise given, to any other frame, makes 
the light-front form attractive for form factor calculations.  The light-front form in this context was introduced by Berestetsky and 
Terentev~\cite{BerestetskyA,BerestetskyB}, and later developed by Chung and Coester~\cite{Chung:1991st}. The light-front form of the wave 
function is obtained by a Melosh or Wigner rotation of the Dirac spinors for each quark.  

Chung and Coester~\cite{Chung:1991st} used a Gaussian wave function. They did obtain a falling $G_E^p/G_M^p$ ratio.  This 
is apparently a feature shared by many relativistic calculations and is caused by the Melosh transformation~\cite{Miller:2002ig}.  
Frank, Jennings, and Miller~\cite{Frank:1995pv,Miller:2002qb} used the light-front nucleonic wave function of 
Schlumpf \cite{Schlumpf:1992vq,Schlumpf:1992pp} and found a decreasing $G_E^p/G_M^p$ ratio, obtaining a zero between  $Q^2$ of 5 and 6 GeV$^2$.
Cardarelli \textit{et al.}~\cite{Cardarelli:1995dc,Pace:1999as} also used the light-front formalism which used quark wave functions obtained 
from a potential of Capstick and Isgur~\cite{Capstick:1986bm}. They made the point that the one-gluon exchange is crucial to obtaining 
high momentum components in the wave function to explain the form factor data.

A comparable amount of high-momentum components in the nucleon wave function can be obtained in the Goldstone-boson-exchange (GBE) quark 
model \cite{Glozman:1997fs,Glozman:1997ag}. This model relies on constituent quarks and Goldstone bosons, which arise as effective degrees 
of freedom of low-energy QCD from the spontaneous breaking of the chiral symmetry. The GBE CQM was used by Boffi {\it et al.}
\cite{Boffi:2001zb} to calculate the nucleon electromagnetic form factors in the point-form. 
Relativistic CQM calculations by Wagenbrunn {\it et al.} \cite{Wagenbrunn:2005wk} compared using Goldstone-boson-exchange to one-gluon-exchange 
in the point-form and found little difference between the calculations  for proton form factors.

De Sanctis \textit{et al.}~\cite{DeSanctis:2007ft,Santopinto:2010zz} have calculated the ratio $G_E^p/G_M^p$ within the 
hypercentral constituent quark model including relativistic corrections. Parameters of the potential are fit to the baryon mass spectrum. With 
the inclusion of form factors for the constituent quarks, good fits are obtained for the nucleon form factors \cite{Santopinto:2010zz}, for 
the latest polarization transfer $G_E^p$ results~\cite{Puckett:2010ac}.

Another type of covariant CQM calculation was done by Gross, Ramalho, and Pe\~na~\cite{Gross:2006fg}, partly based on earlier work of 
Gross and Agbakpe~\cite{Gross:2004nt}, avoiding questions of dynamical forms by staying in momentum space. They performed  
CQM calculations using a covariant spectator model, where the photon interacts with one quark and the other two quarks are 
treated as an on-shell diquark with a definite mass.
They modeled the nucleon as a system of three valence constituent quarks with their own parameterized form factors, where 
the CQ form factors are obtained 
with parameters that they fit to the data. Their fit from the 9-parameter ``model IV'' achieves a rather good description of the existing data, including the recent higher-$Q^2$ data for $G_E^n$ from Ref.~\cite{Riordan:2010id}, which had not yet been published at the time.

\subsubsection{Perturbative QCD}
\label{subsubsec:pQCD}
\begin{figure}
\begin{center}
\includegraphics[width =0.85\columnwidth]{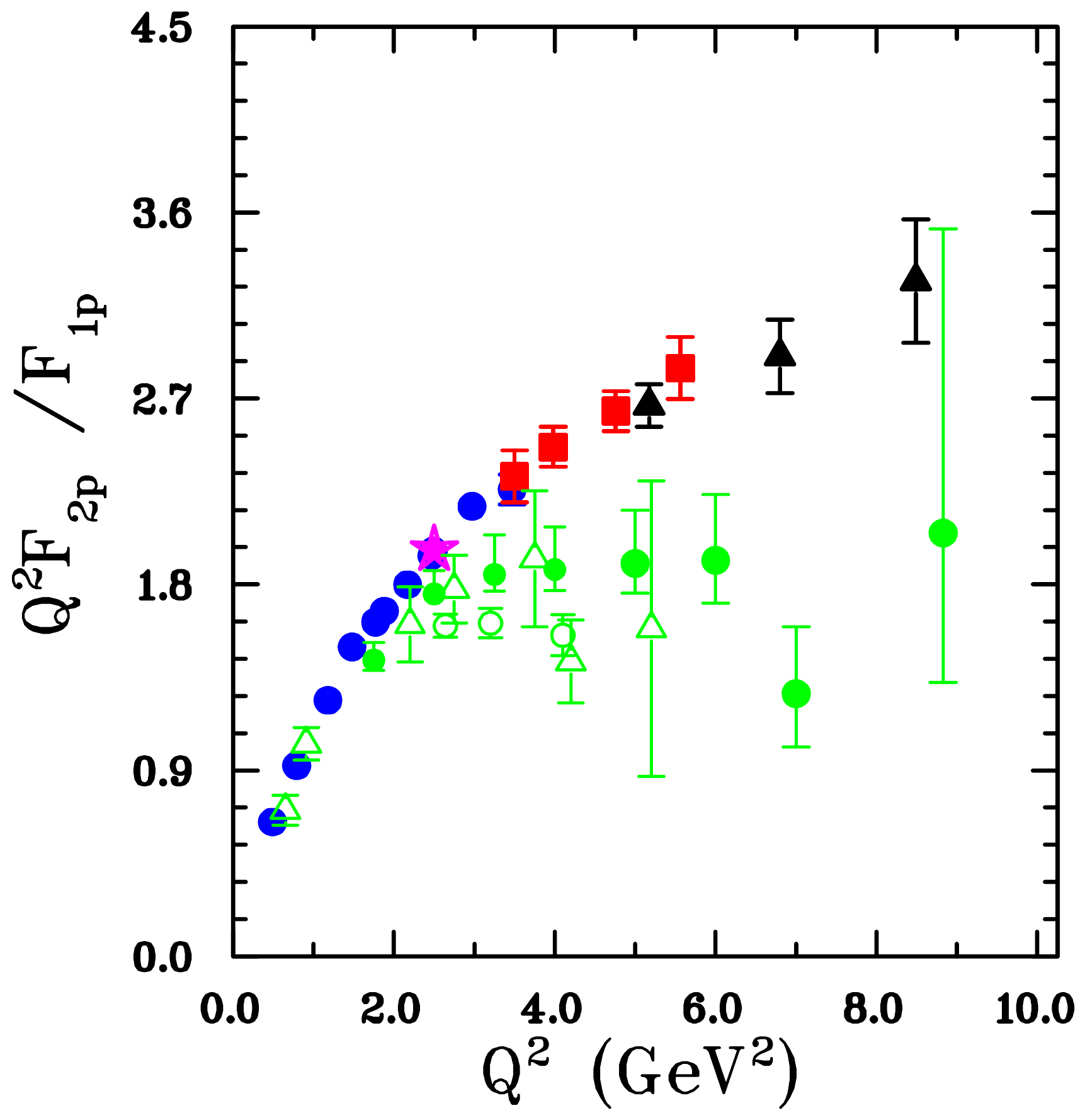}
\caption{  Selected data for $Q^2 F_{2}^p/F_{1}^p$ from cross section and polarization observables. Polarization transfer data and symbols are the same as in Fig.~\ref{fig:FinalResultsGEP3}. Cross section data are from Refs.~\cite{Andivahis:1994rq} (filled green circles), \cite{Qattan:2004ht} (empty green circles), and \cite{Christy:2004rc} (empty green triangles). The cross section data show flattening starting at $Q^2 \approx 3$ GeV$^2$. However, the polarization transfer data continue to rise up to $Q^2 = 8.5$ GeV$^2$.}
\label{fig:q2f2f1}
\end{center}
\end{figure}

In the context of elastic scattering and other hard exclusive processes, perturbative QCD (pQCD) is only expected to be applicable at very large momentum transfers~\cite{Isgur:1988iw,Isgur:1989cy}, perhaps one to several tens of GeV$^2$ in the most optimistic scenario. In this limit, the virtual photon makes
a hard collision with a single valence quark, which then shares the large momentum transfer with 
the other two, nearly collinear quarks through two hard gluon exchanges. 
pQCD predicts that $Q^4 F_1$ and $Q^2 F_{2}/F_{1}$ should
become constant for asymptotically large $Q^2$, where the extra power of $Q^2$ for $F_2$ relative to $F_1$ is a consequence of helicity conservation at high energies. 
The predictions were given by Brodsky and Farrar~\cite{Brodsky:1974vy,Brodsky:1973kr} and by Matveev {\it et al.} \cite{Matveev:1973ra}.  
By a simple rearrangement of Eq.~\eqref{eq:gepgmp}, the ratio of Dirac and Pauli FFs is given in terms of the Sachs ratio $r = G_E/G_M$ by $F_2/F_1 = (1-r )/(\tau+r)$. Figure~\ref{fig:q2f2f1} shows the JLab polarization data together with selected cross section data for $Q^2 F_{2}^p/F_{1}^p$. The cross section data (without TPEX corrections) show flattening for $Q^2 \gtrsim 3$ GeV$^2$. However, the GEp-I, GEp-II and GEp-III data do not yet show the pQCD scaling behavior.

\begin{figure}
\begin{center}
\includegraphics[width =0.85\columnwidth]{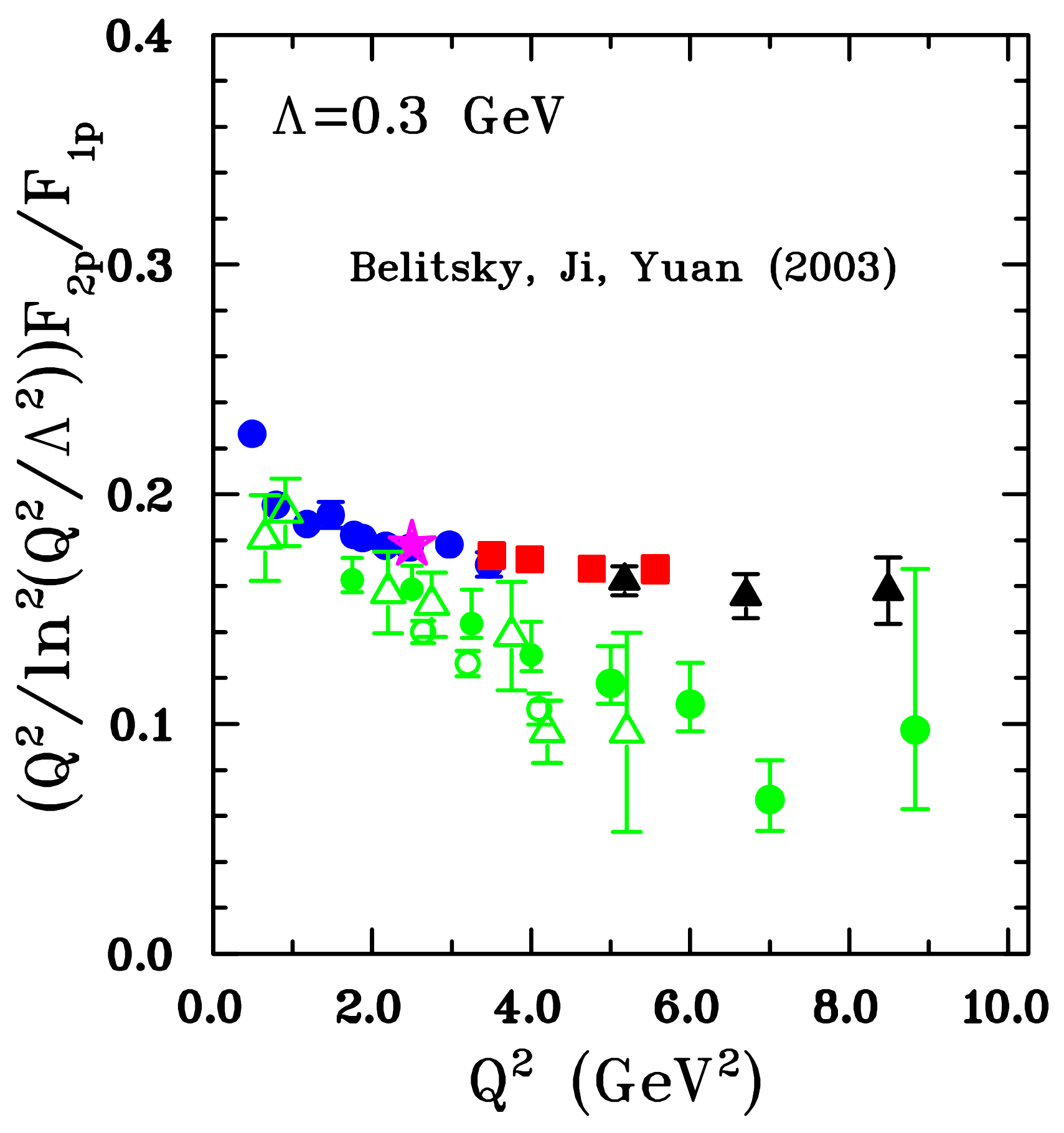}
\caption{  Same data as Fig.~\ref{fig:q2f2f1}, plotted as $\left(Q^2/\ln^2\left(Q^2/\Lambda^2\right)\right) F_{2}^p/F_{1}^p$ as proposed by Belitsky \textit{et al.}~\cite{Belitsky:2002kj}, for $\Lambda = 0.3$ GeV.}
\label{fig:bjy}
\end{center}
\end{figure}

In 2003 Belitsky {\it et al.} \cite{Belitsky:2002kj} 
investigated the assumption of quarks moving collinearly with the proton underlying the pQCD prediction. 
They reiterated the fact that the helicity of a massless (or very light) quark 
cannot be flipped by the virtual photon of the $ep$ reaction. 
Instead, the leading contribution to $F_2^p$ at large $Q^2$ requires one unit of orbital angular momentum in either the initial or final-state light-cone nucleon wave function, leading to a modified logarithmic scaling behavior $Q^2 F_2/F_1 \propto \ln^2 (Q^2 / \Lambda^2)$ at large $Q^2$, with
$\Lambda$ a non-perturbative mass scale. With $\Lambda = 0.3$~GeV, as shown in Fig.~\ref{fig:bjy}, the polarization data for $F_{2 p}/F_{1 p}$ agree qualitatively with such double-logarithmic enhancement\footnote{This observation is not particularly sensitive to the choice of $\Lambda$ within a range of values comparable to $\Lambda_{QCD}$ and/or $\Lambda \approx \tfrac{\hbar c}{r_p} \approx 0.235$ GeV}. 
Ralston~\cite{Ralston:2003mt} and Brodsky {\it et al.} \cite{Brodsky:2003pw} have also discussed the role of quark orbital angular momentum in producing a ratio $F_{2p}/F_{1p}$ which falls more slowly than $1/Q^2$. While the ``precocious'' scaling behavior observed in the proton's $F_2/F_1$ ratio is interesting, it is important to note that the neutron FF data up to 3.4 GeV$^2$~\cite{Riordan:2010id} do not support the logarithmic pQCD scaling behavior for a cutoff parameter $\Lambda$ similar to that which describes the proton data. The detailed flavor decomposition of the individual quark contributions to the nucleon form factors~\cite{Cates:2011pz,Qattan:2012zf} suggests that the pQCD-like scaling behavior observed for the proton's $F_2/F_1$ ratio is probably largely accidental, and a consequence of the delicate interplay between the $u$ and $d$ quark contributions to $F_1$ and $F_2$.

In 2006 Braun \textit{et al.}~\cite{Braun:2006hz} evaluated leading order contributions to the nucleon EMFFs within the light-cone sum rule (LCSR) approach, using both asymptotic distribution amplitudes (DAs) and DAs with QCD sum rule-based corrections. The LCSR approach with asymptotic DAs yields values of $G_{M}^p$ and $G_{M}^n$ which are close to the data in the range $Q^2 \sim 1$--$10$~GeV$^2$. The electric 
form factors were found to be much
more difficult to describe, with $G_{E}^n$ overestimated, and 
$G_{E}^p/G_{M}^p$ nearly constant. The ratio $G_E^p/G_M^p$ was found to be very sensitive to the details of the DAs. 
A qualitative description of the proton and neutron electric form factors was obtained by including twist-3 and twist-4 corrections to the nucleon DAs within a simple model. 
More recently, the LCSR approach was refined by Anikin \textit{et al.}~\cite{Anikin:2013aka} to include the next-to-leading-order pQCD corrections to the contributions of both twist-3 and twist-4 operators and a consistent treatment of nucleon mass corrections. In Ref.~\cite{Anikin:2013aka}, the DAs were extracted using the form factor data and compared to lattice QCD results, leading to a self-consistent description. The LCSR approach is, however, not yet able to describe all four nucleon EMFFs to a degree of accuracy comparable to that of the data. 

Kivel and Vanderhaeghen~\cite{Kivel:2010ns,Kivel:2012zz} investigated the soft
rescattering contribution to nucleon form factors using soft collinear effective theory (SCET).
They have been able to show that the soft or Feynman process 
can be factorized into three subprocesses with different scales:
 a hard rescattering , a hard-collinear scattering, and soft nonperturbative modes.
For the $Q^2$ range of the present data,  SCET qualitatively predicts that $Q^2 F_2/F_1$ should not be a constant, but exhibit 
a slow rise, as seen in the data.

\subsubsection{Generalized Parton Distributions}

The elementary hard scattering process in large-$Q^2$ electron-nucleon scattering is virtual photoabsorption by a single quark, embedded in the target nucleon as part of a complex, many-body, relativistic system of valence quarks, sea quark-antiquark pairs, and gluons, described by the Generalized Parton Distributions (GPDs). The GPDs provide a framework to describe the process of emission and re-absorption of a quark by a hadron in hard exclusive reactions via the ``handbag'' mechanism. The GPDs are universal non-perturbative objects arising in the QCD factorization of hard exclusive processes such as deeply virtual Compton scattering (DVCS) and deeply virtual meson production (DVMP). The form factors $F_1$ and $F_2$ are related to the vector ($H$) and tensor ($E$) GPDs by model-independent sum rules~\cite{Ji:1996ek}: 
\begin{eqnarray}
\int_{-1}^{+1}dx\, H^{q}(x,\xi ,Q^2)&=& F_{1}^{q}(Q^2)\, ,\nonumber\\
\int _{-1}^{+1}dx\, E^{q}(x,\xi ,Q^2)&=& F_{2}^{q}(Q^2)\, ,
\label{eq:ffsumrulehe}
\end{eqnarray}
where $F_1^q$ ($F_2^q$) represents the contribution of quark flavor $q$ to the Dirac (Pauli) FF of the nucleon. These relations allow us, if we have complete measurements or good models for the GPDs, to predict the electromagnetic form factors~\cite{Guidal:2004nd}. Alternatively, the measured form factors at high $Q^2$, when combined with the forward parton distributions measured in deep-inelastic scattering, provide fairly stringent constraints on the GPDs, particularly with respect to their behavior at large $x$ and/or $-t$ values~\cite{Diehl:2004cx,Diehl:2013xca}. Early theoretical developments in GPDs indicated that 
measurements of the separated elastic form factors of the nucleon to high $Q^2$ might also shed light on the nucleon spin decomposition, via Ji's angular momentum sum rule \cite{Ji:1996ek} for the total (spin and orbital) angular momentum $J_q$ carried by the parton flavor $q$:
\begin{eqnarray}
  2J_q &=& \int_{-1}^{1} \left[H_q(x, 0, 0) + E_q(x, 0, 0)\right]x dx.
\end{eqnarray} 
The model-independent extraction of GPDs from observables of hard exclusive processes is an area of high current activity and interest. Some recent and less-recent reviews of the subject can be found in Refs.~\cite{Ji:1998pc,Goeke:2001tz,Diehl:2003ny,Belitsky:2005qn,Ji:2004gf,Guidal:2013rya}. 
The GPDs can be represented in impact-parameter space via two-dimensional Fourier transforms of the $t$-dependence of GPDs at zero skewness~\cite{Burkardt:2002hr}, allowing a three-dimensional ``tomography'' of the nucleon in two transverse spatial dimensions and one longitudinal momentum dimension. By forming the charge-squared-weighted sum over quark flavors and integrating the impact-parameter-space GPDs over longitudinal momentum fractions $x$, Miller~\cite{Miller:2007uy,Miller:2010nz} derived model-independent expressions for the impact-parameter-space charge and magnetization densities of the nucleon in terms of two-dimensional Fourier-Bessel transforms of $F_1$ and $F_2$:
\begin{eqnarray}
  \rho_{ch}(b) &=& \int_0^\infty \frac{Q}{2\pi} J_0(Qb) F_1(Q^2) dQ \\
  \tilde{\rho}_M(b) &=& \frac{b}{2\pi} \sin^2 \phi \int_0^{\infty} \frac{Q^2}{2\pi} J_1(Qb) F_2(Q^2) dQ,
\end{eqnarray} 
in which $b$ is the magnitude of the transverse displacement from the center of the nucleon, and $\phi$ is the angle between the direction of $\mathbf{b}$ and the direction of the transverse magnetic field or, equivalently, the transverse nucleon polarization. Venkat \textit{et al.}~\cite{Venkat:2010by} performed a first extraction with realistic uncertainty estimation of $\rho_{ch}(b)$ and $\tilde{\rho}_M(b)$ for the proton.

\subsubsection{Lattice QCD}

Lattice gauge theory is presently the only known method for calculating static and dynamic properties of strongly interacting systems from first-principles, non-perturbative QCD in the regime of strong coupling and confinement. Practical computations in lattice gauge theory involve numerical solutions of QCD on a finite-volume lattice of discrete space-time points. In the recent past, these calculations have often been performed for unphysically large quark masses due to computational limitations, whereas modern calculations often work at or near the physical pion mass. Calculations are typically performed for several lattice volumes, spacings and quark masses and then extrapolated to the infinite-volume, continuum limit and to the physical pion mass. Early calculations of nucleon electromagnetic form factors in lattice QCD emphasized the isovector ($p - n$) form factors, which are simpler to calculate since contributions from disconnected diagrams are suppressed~\cite{Alexandrou:2011db}. 
Until quite recently, most calculations of nucleon form factors in lattice QCD~\cite{Alexandrou:2011db,Collins:2011mk,Shanahan:2014cga,Shanahan:2014uka,Capitani:2015sba} have been restricted to relatively low momentum transfers $Q^2 \lesssim 3$ GeV$^2$, because the rapid falloff with $Q^2$ of the form factors leads to very small signal-to-noise ratios in the extraction of hadronic three-point correlators, and related systematic uncertainties due to excited-state contamination, among other issues. Lin \textit{et al.}~\cite{Lin:2010fv} employed a novel technique using anisotropic lattices with both quenched and dynamical ensembles with $m_\pi \ge 450$ MeV to reach $Q^2 \approx 6$ GeV$^2$. 

The prospects for lattice QCD form factor calculations to reach high $Q^2$ have recently been improved by a novel application of the Feynman-Hellman theorem~\cite{Chambers:2017tuf}, through which hadronic matrix elements can be related to energy shifts. In the context of nucleon form factor calculations on the lattice, the Feynman-Hellman method allows access to the matrix elements relevant to form factor calculations via two-point correlators as opposed to more complicated three-point functions, and exploits strong correlations in the gauge ensembles to enhance the signal-to-noise ratios for high-momentum states. 
\begin{figure}
  \begin{center}
    \includegraphics[angle=90,width=0.85\columnwidth]{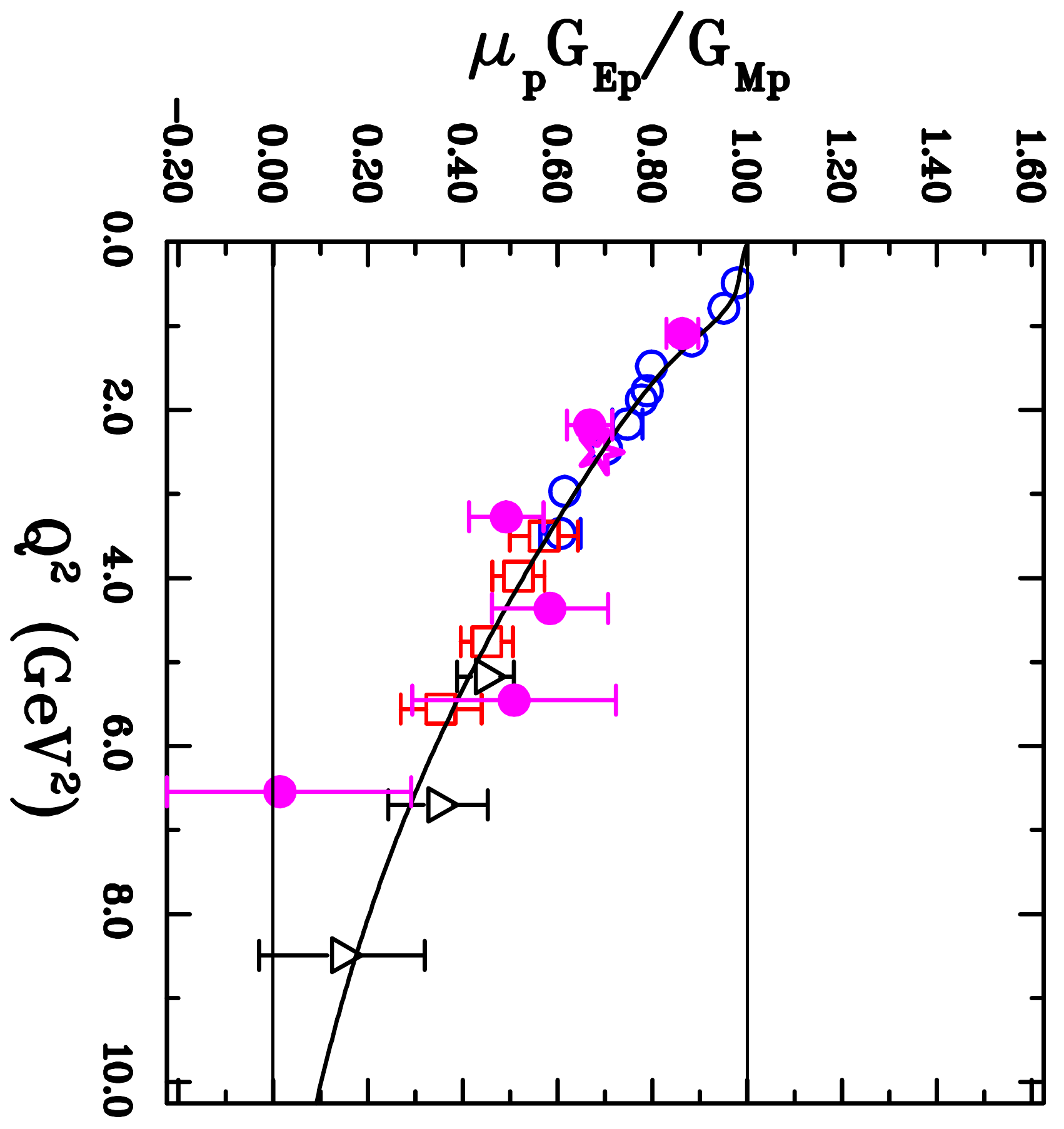}
  \end{center}
  \caption{\label{fig:LatticeQCD}  Lattice QCD results for $\mu_p G_E^p/G_M^p$ obtained using a novel method based on the Feynman-Hellman theorem~\cite{Chambers:2017tuf} (pink filled circles), compared to polarization transfer data from Refs.~\cite{Jones:1999rz,Punjabi:2005wq} (blue empty circles), \cite{Gayou:2001qd,Puckett:2011xg} (red empty squares), the final GEp-III data (black empty triangles), and the weighted-average of the final GEp-2$\gamma$ data (pink empty star). The solid curve is the fit to the data using Eqn. 44 from Ref. \cite{Punjabi:2015bba}, and has not been re-fitted using the final results reported in this work.}
\end{figure}
Figure~\ref{fig:LatticeQCD} shows an initial result from the QCDSF/UKQCD/CSSM collaborations~\cite{Chambers:2017tuf} for $\mu_pG_E^p/G_M^p$ reaching $Q^2 \approx 6.5$ GeV$^2$ with uncertainties approaching the precision of the experimental data. 

\subsubsection{Dyson-Schwinger Equations}

In recent years, significant progress has also been realized in the explanation and prediction of static and dynamic properties of ``simple'' hadronic systems such as the pion, the nucleon and the $\Delta(1232)$ in continuum non-perturbative QCD, within the framework of QCD's Dyson-Schwinger Equations (DSEs)~\cite{Cloet:2008re}. Where the calculation of nucleon electromagnetic form factors is concerned, the DSE approach requires the solution of a Poincar\'{e} covariant Faddeev equation. One analytically tractable, symmetry-preserving truncation scheme that has achieved considerable success in describing the observed behavior of the nucleon FFs involves dressed quarks and non-pointlike scalar and axial vector diquarks as the dominant degrees of freedom. 

\begin{figure}
  \begin{center}
    \includegraphics[angle=90,width=0.85\columnwidth]{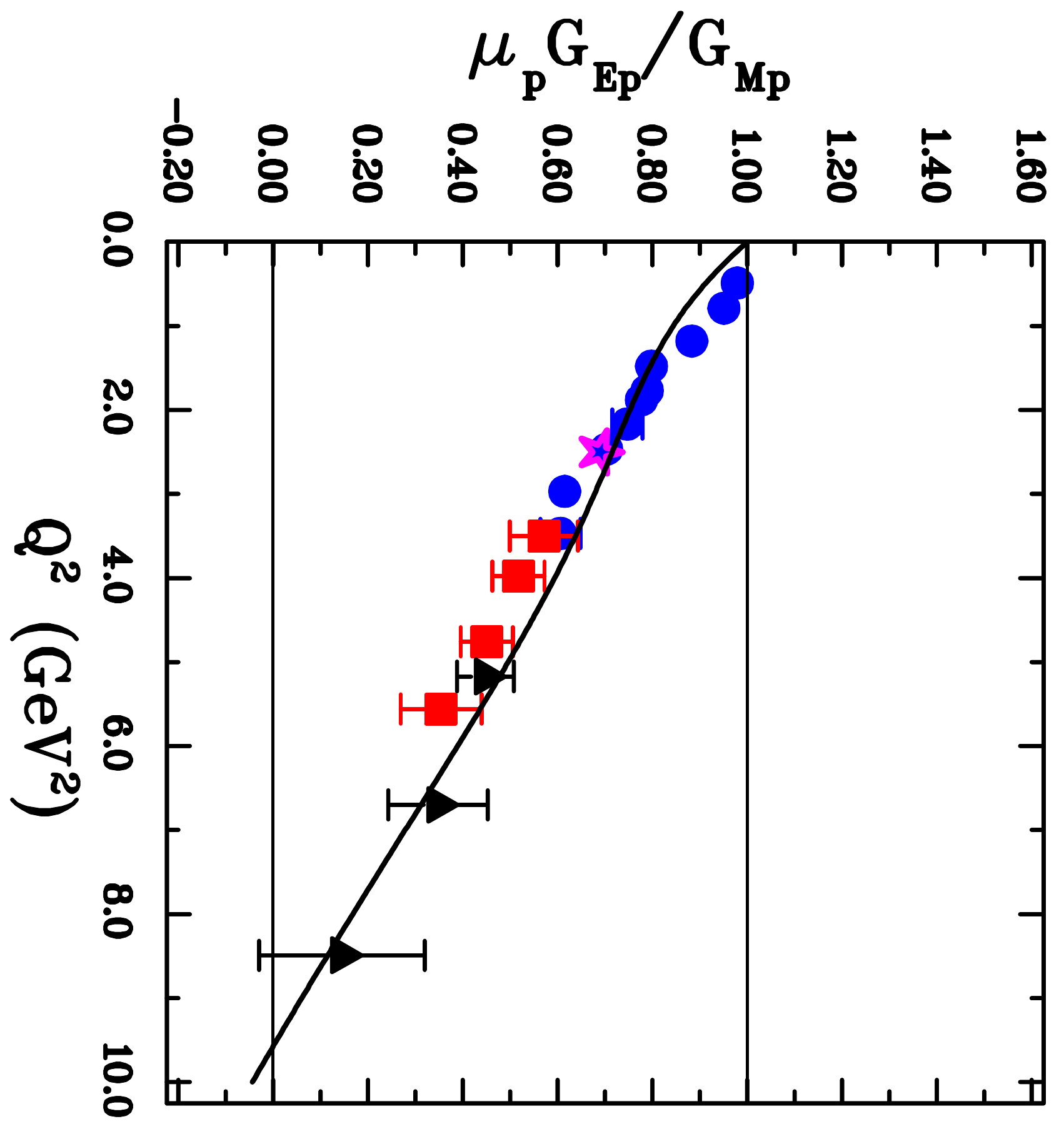}
  \end{center}
\caption{\label{fig:DSE} Comparison of polarization transfer data for $\mu_p G_E^p/G_M^p$ with the DSE based calculation of Ref.~\cite{Segovia:2014aza}.}
\end{figure}
In the DSE framework, the nucleon EMFFs at large $Q^2$ values are sensitive to the momentum dependence of the running masses and couplings in the strong interaction sector of the Standard Model~\cite{Cloet:2013jya}. In a recent study, Segovia \textit{et al.}~\cite{Segovia:2014aza} achieved simultaneously good descriptions of the nucleon and $\Delta(1232)$ elastic and transition form factors using identical propagators and interaction vertices for the relevant dressed quark and diquark degrees of freedom. One notable prediction is a zero crossing in the ratio $G_E^p/G_M^p$ at $Q^2 = 9.5$ GeV$^2$ and in the neutron FF ratio $G_E^n/G_M^n$ at $Q^2 \approx 12$ GeV$^2$. In this framework, any change in the quark-quark interaction that shifts the location of the zero in $G_E^p$ to larger $Q^2$ implies a corresponding shift in the location of a zero in $G_E^n$ to smaller $Q^2$. The location of the zero in $G_E^p$ is particularly sensitive to the rate of transition of the dressed quark mass function between the non-perturbative and perturbative regimes, with a slower fall-off of $G_E^p/G_M^p$ corresponding to a faster transition to the perturbative regime, consistent with the ``dimensional scaling'' expectation discussed in Sec.~\ref{subsubsec:pQCD}. This prediction will be severely tested by planned near-future precision measurements of $G_E^p$ ($G_E^n$) to $Q^2 \approx 12$ (10) GeV$^2$ at JLab. Fig.~\ref{fig:DSE} shows the calculation of Segovia \textit{et al.}~\cite{Segovia:2014aza} for $\mu_p G_E^p/G_M^p$, compared to the polarization transfer data from Halls A and C. 

\subsection{Implications of GEp-2$\gamma$ for TPEX}

\label{subsec:theory_sub_b}
Shortly after the publication of GEp-I and GEp-II, two groups independently suggested that the difference between cross section and double polarization results might be attributable to
previously neglected hard TPEX processes; these were Guichon and Vanderhaeghen \cite{Guichon:2003qm}, 
and Blunden $et~al.$ \cite{Blunden:2003sp}. Notably, some of the earliest polarization experiments for elastic $ep$ were done to assess the contribution of the two photon exchange process~\cite{bizot63,bizot65,Powell:1970qt,DeRujula:1972te}. In general, cross section data
require large radiative corrections, whereas
double-polarization ratios do not~\cite{Afanasev:2001jd,Afanasev:2001nn}. Several calculations and/or extractions of the hard TPEX contribution involving various models, assumptions and approximations have been published over the last decade. A partial list of these efforts includes Refs.~\cite{Afanasev:2005mp,Arrington:2003ck,Kondratyuk:2005kk,Bystritskiy:2006ju,Vanderhaeghen:2000ws,Carlson:2007sp}. Many of the calculations partially resolve the discrepancy, but a model-independent theoretical prescription for TPEX corrections constrained directly by data remains elusive. Recent reviews of the subject can be found in Refs.~\cite{Arrington:2011dn,Afanasev:2017gsk}.

\begin{figure}
  \begin{center}
    \includegraphics[width=0.85\columnwidth]{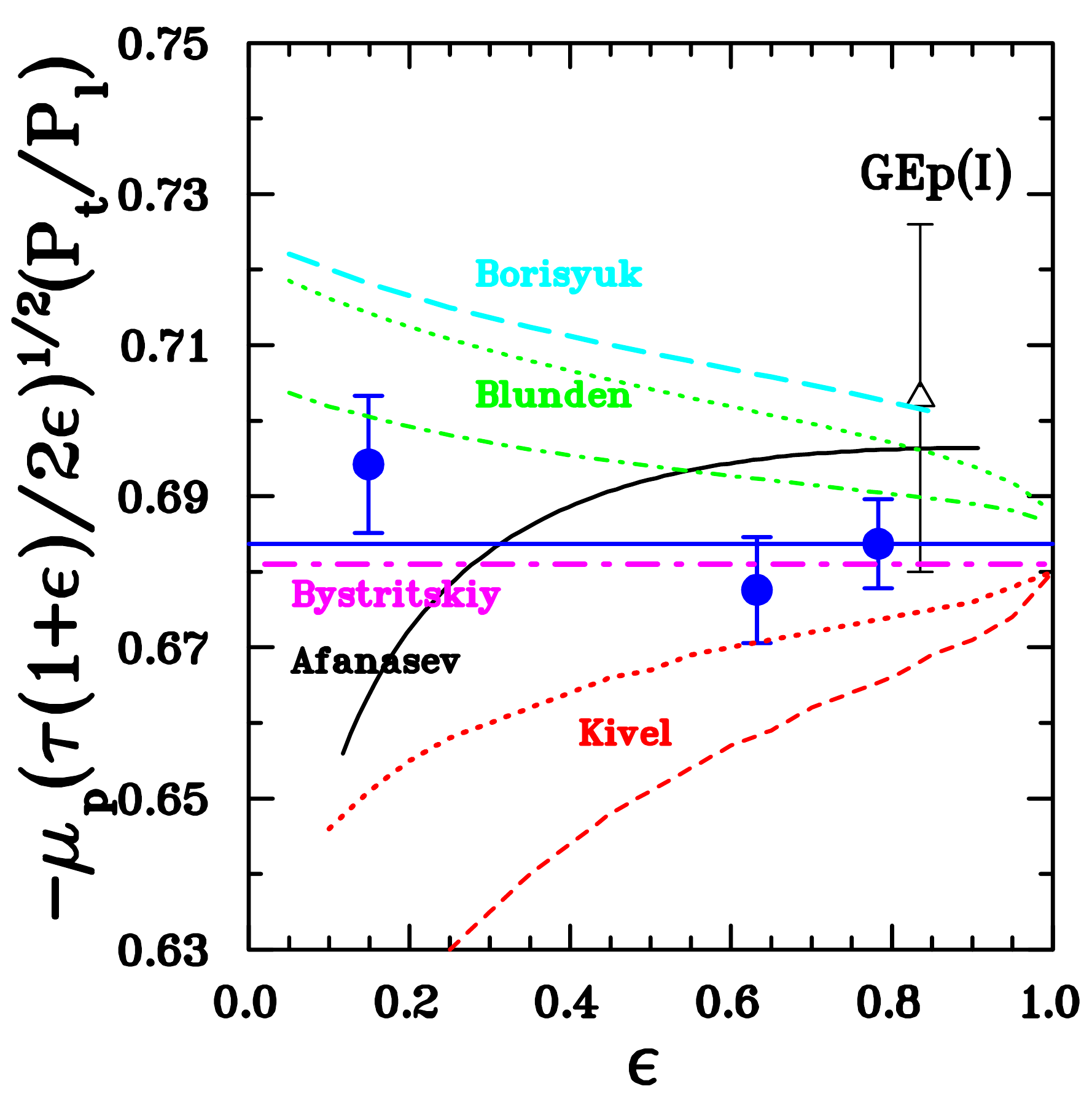}
  \end{center}
  \caption{\label{fig:Rptwogamma} Final, bin-centering-corrected results of GEp-2$\gamma$ for the ratio $R$, compared to several theoretical predictions for the $\epsilon$ dependence of $R$ at $Q^2 = 2.5$ GeV$^2$ due to TPEX corrections. The blue solid horizontal line is the weighted average of the corrected data (see Tab.~\ref{tab:bcc2} and Fig.~\ref{fig:Rbcctwogamma}). Curves are: Borisyuk \textit{et al.}~\cite{Borisyuk:2013hja} (cyan dashed), Blunden \textit{et al.}~\cite{Blunden:2017nby} (green dot-dashed ($N$ only) and green dotted ($N+\Delta$)), Bystritskiy \textit{et al.}~\cite{Bystritskiy:2006ju} (pink dot-long dashed), Afanasev \textit{et al.}~\cite{Afanasev:2005mp} (black solid), and Kivel \textit{et al.}~\cite{Kivel:2009eg} (red dotted (BLW) and red dashed (COZ)). Note that because the ratio $R$ is proportional to the Born value of $\mu_p G_E^p/G_M^p$, each curve can be renormalized, in principle, by an overall multiplicative factor. See text for details.}
\end{figure}
In addition to the significant theoretical work to resolve the discrepancy, major experimental efforts were carried out over the last decade to search for experimental signatures of significant TPEX contributions to elastic $eN$ scattering. These signatures include possible non-linearities of the Rosenbluth plot~\cite{Qattan:2004ht,Tvaskis:2005ex}, a non-zero target-normal single-spin asymmetry~\cite{Zhang:2015kna} or induced normal recoil polarization, and deviations from the Born approximation in polarization transfer observables, as in this work and Ref.~\cite{Meziane:2010xc}. The beam-normal single spin asymmetries in elastic $eN$ scattering have also been precisely measured as byproducts of a large number of parity violation experiments~\cite{Wells:2000rx,Maas:2004pd,Armstrong:2007vm,Androic:2011rh,Abrahamyan:2012cg,Waidyawansa:2016znm,Rios:2017vsw}, albeit in a $Q^2$ range well below the region of the discrepancy. These beam-spin asymmetries are typically at the few-ppm level, and are also sensitive to the imaginary part of the TPEX amplitudes. The most direct observable to access the real part of the TPEX amplitude is a deviation of the $e^+p/e^-p$ cross section ratio from unity~\cite{Arrington:2003ck}, as the real part of the interference term between the Born and TPEX diagrams changes sign with the charge of the lepton beam. Three major experiments with very different and complementary technical approaches have recently measured the $e^+p/e^-p$ cross section ratio~\cite{Rachek:2014fam,Adikaram:2014ykv,Rimal:2016toz,Henderson:2016dea}.


Figure~\ref{fig:Rptwogamma} shows the final, bin-centering-corrected results of GEp-2$\gamma$ for the $\epsilon$ dependence of $R$, compared to several theoretical predictions for the hard TPEX corrections to this observable. Blunden \textit{et al.}~\cite{Blunden:2017nby} recently evaluated the hard TPEX corrections to elastic $ep$ scattering within a dispersive approach, which avoids off-shell uncertainties inherent in the direct evaluation of loop diagrams~\cite{Blunden:2005ew}. The box and crossed diagrams for TPEX corrections involving nucleon and $\Delta$ intermediate hadronic states were evaluated both algebraically and numerically within the dispersive approach using empirical parametrizations of the nucleon elastic and $N \rightarrow \Delta$ transition form factors. The result including the contributions of both $N$ and $\Delta$ intermediate states is consistent in slope with the final GEp-2$\gamma$ data, and also achieves a reasonable description of the $e^+p/e^-p$ cross section ratios which, however, are only measured for $Q^2 \lesssim 2.1$ GeV$^2$. At $Q^2 = 2.5$ GeV$^2$, it appears that a description in terms of hadronic degrees of freedom with only the nucleon elastic and $\Delta$ intermediate states is adequate. At higher $Q^2$ values where the discrepancy between cross section and polarization data is more severe, the effects of higher-mass resonances, inelastic nonresonant intermediate states including the $\pi N$ continuum, and the finite widths of resonances are expected to increase in importance. Borisyuk \textit{et al.}~\cite{Borisyuk:2013hja} also used the dispersive approach to compute the contribution of the $P_{33}$ partial wave of the $\pi N$ channel to the TPEX amplitude, which effectively includes the $\Delta$ contribution with realistic shape, width, and nonresonant background ``automatically''. The prediction of Ref.~\cite{Borisyuk:2013hja} for the ratio $R$ exhibits similar behavior to the calculation of Blunden \textit{et al.}, which is not surprising, given its similar physics content. Bystritskiy \textit{et al.}~\cite{Bystritskiy:2006ju} used the electron structure function method to compute the higher-order radiative corrections to all orders in perturbative QED in the leading logarithm approximation. Their method predicts no noticeable $\epsilon$ dependence at the level of precision of the GEp-2$\gamma$ data, consistent with our results. 

Afanasev \textit{et al.}~\cite{Afanasev:2005mp} approached the TPEX corrections to elastic $ep$ scattering in a parton-model approach assuming dominance of the ``handbag'' mechanism, in which both hard virtual photons are exchanged with a single quark, embedded in the nucleon via GPDs. This approach is expected to be valid for simultaneously large values of $s$, $-u$, and $Q^2$. The parton-model evaluation of TPEX corrections predicts a strong, non-linear $\epsilon$ dependence for $R$ that is not observed the in GEp-2$\gamma$ data. Kivel \textit{et al.}~\cite{Kivel:2009eg} computed the hard TPEX correction to elastic $ep$ in a perturbative QCD approach in which the leading contribution for asymptotically large $Q^2$ involves two hard photon exchanges occuring on different valence quarks, and a single hard gluon exchange occurring on the third valence quark. In the pQCD approach, the TPEX amplitude can be expressed in a model-independent way in terms of leading-twist nucleon distribution amplitudes (DAs). In Fig. 27, the calculation of Ref.~\cite{Kivel:2009eg} is shown for two different models for the DAs: that of Braun \textit{et al.} (BLW~\cite{Braun:2006hz}), and that of Chernyak \textit{et al.} (COZ~\cite{Chernyak:1987nu}). The GPD and pQCD models for the hard TPEX correction predict a significant positive slope $dR/d\epsilon$ which is disfavored by the data. It must be noted, however, that the GEp-2$\gamma$ measurement at $\left<\epsilon\right> = 0.153$ in particular lies outside the expected kinematic range of applicability of a partonic description. 

The deviation from unity of $P_\ell/P_\ell^{Born}$ at large $\epsilon$, given the absence of significant $\epsilon$ dependence of the ratio $R$, implies a similar deviation from the Born approximation in $P_t$ that cancels in the ratio $P_t/P_\ell$. This deviation was not predicted by any of the TPEX calculations available at the time of the original publication~\cite{Meziane:2010xc}, which generally expected small TPEX corrections to this observable. A deviation from unity in $P_\ell/P_\ell^{Born}$ was subsequently predicted within the SCET approach by Kivel \textit{et al.}~\cite{Kivel:2012vs}. Guttmann \textit{et al.}~\cite{Guttmann:2010au} performed an extraction of the TPEX amplitudes from a global analysis of elastic $ep$ scattering data including the original GEp-2$\gamma$ results~\cite{Meziane:2010xc} and the Hall A ``Super-Rosenbluth'' data at the similar $Q^2$ of 2.64 GeV$^2$~\cite{Qattan:2004ht}, using the formalism of Eqs.~\eqref{eq:siggen}-\eqref{eq:rgen}. Under the assumptions used in their analysis, the observed deviation from unity of $P_\ell/P_\ell^{Born}$ and the constant value of $R$ imply that the TPEX amplitudes $\mathcal{Y}_E \equiv \Re \left(\delta \tilde{G}_E/G_M\right)$ and $\mathcal{Y}_3 \equiv \left(\nu / M^2 \right)\Re\left(\tilde{F}_3/G_M\right)$ (see Eqs.~\eqref{eq:regm}-\eqref{eq:nu2gamma}), which are mainly driven by the original GEp-2$\gamma$ data, are roughly equal in magnitude and opposite in sign, and approach the 2-3\% level at $\epsilon \approx 0.8$ and $Q^2 = 2.5$ GeV$^2$.

\section{Conclusions}

\label{sec:conclusions}
This article has described two proton form factor experiments, GEp-III and GEp-2$\gamma$, which utilized the recoil polarization method in Hall C at Jefferson Lab to
 measure the ratio of the proton’s electric and magnetic form factors, $R \equiv \mu_p G_E^p/G_M^p$. The results of these experiments were previously published in two separate articles~\cite{Puckett:2010ac,Meziane:2010xc}. The purpose of this article was to provide an expanded description of the apparatus and analysis method common to both experiments and report the results of a full reanalysis of the data with significant improvements in detector calibration, event reconstruction, elastic event selection, and the evaluation of systematic uncertainties.
The final results of GEp-III are essentially unchanged relative to the originally published results~\cite{Puckett:2010ac}. The new analysis has resulted in a significant reduction in the systematic uncertainty, due to a more thorough evaluation of the systematic uncertainty in the total bend angle of the proton trajectory in the non-dispersive plane of the HMS. The high-$Q^2$ points confirmed the results of the GEp-I and GEp-II experiments from Hall A, namely that $R$ continues to decrease toward zero, but with clear indication that the rate of this decrease is slowing down. The impressive agreement of the measurements of $R$ in GEp-III and GEp-2$\gamma$ with the previous Hall A measurements at the same or similar $Q^2$ (but not necessarily the same $\epsilon$) demonstrates that the systematic uncertainties of the recoil polarization method are well understood, and that deviations from the Born approximation in the extraction of $G_E^p/G_M^p$ from polarization transfer observables are not large within the $Q^2$ range presently accessible to experiment. 

The GEp-2$\gamma$ data, originally published in Ref.~\cite{Meziane:2010xc}, consist of measurements for three different $\epsilon$ values at a fixed $Q^2$ of 2.5 GeV$^2$, obtained by changing the electron beam energy and the detector angles. The relative $\epsilon$ dependence of the ratio ${P_\ell}/{P_\ell^{Born}}$ was also extracted from the GEp-2$\gamma$ data with small uncertainties by exploiting the fact that the polarimeter analyzing power, the proton momentum, and the HMS magnetic field were the same for all three $\epsilon$ values. The lowest $\epsilon$ point was used to calibrate the polarimeter analyzing power, given the large value of $P_\ell$ and its negligible sensitivity to $R$ at this $\epsilon$. The results of the reanalysis of the GEp-2$\gamma$ data reported in this work include the previously unpublished full-acceptance data for the two highest $\epsilon$ points, increasing the statistics by a factor of 2.5 (3.4) at $\left<\epsilon\right> = 0.638 (0.790)$.
 
The GEp-2$\gamma$ experiment serves as a precise test of the validity of the polarization transfer method. Indeed, as expected from the Born approximation, the GEp-2$\gamma$ data demonstrate that $R$ is compatible with a
constant for a wide range of $\epsilon$ between 0.15 and 0.79. The only deviation from the Born approximation is observed
in the longitudinal polarization at $\epsilon = 0.79$: ${P_\ell}/{P_\ell^{Born}} = 1.0143 \pm 0.0027 \pm 0.0071$. This deviation is largely compensated by a similar relative deviation in $P_t$, such that the form factor ratio remains constant.
In addition, the statistically improved, simultaneous measurements of the independent observables $P_\ell/P_\ell^{Born}$ and $R$ at the same kinematics provide important tools for
testing TPEX models and constraining the extraction of TPEX form factors.

The accelerator at Jefferson Lab has recently been upgraded to a maximum beam energy of 12 GeV. 
There are approved experiments at Jefferson Lab that will extend the knowledge of $G_E^p/G_M^p$ to $Q^2 = 12 $ GeV$^2$, $G_E^n/G_M^n$ to $Q^2 = 10 $ GeV$^2$, and $G_M^n$ to 14 GeV$^2$. Dedicated measurements of the elastic $ep$ unpolarized differential cross section over a wide $Q^2$ range from 2-16 GeV$^2$ with $\lesssim 2\%$ total uncertainties have already been completed in Hall A in 2016 and are currently being analyzed. These measurements will significantly improve upon the existing knowledge of $G_M^p$ within the entire $Q^2$ range accessible with JLab's upgraded electron beam. The program of high-$Q^2$ form factor measurements using the upgraded JLab electron beam will enable the detailed flavor decomposition of the nucleon EMFFs to $Q^2 = 10 $ GeV$^2$, providing significant constraints on the predictions of theoretical models, and insight 
into the important degrees of freedom in understanding nucleon structure across a broad range of $Q^2$.

\section{Acknowledgments}

The collaboration thanks the Hall C technical staff and the Jefferson
Lab Accelerator Division for their outstanding support during the
experiment. This material is based upon work supported by the U.S. Department of Energy, Office of Science, Office of Nuclear Physics, under Award Number DE-SC-0014230 and contract Number(s) DE-AC02-06CH11357 and DE-AC05-06OR23177, the U.S.
National Science Foundation, the
Italian Institute for Nuclear Research, the French Commissariat
\`a l'Energie Atomique and Centre National de la Recherche Scientifique
(CNRS), and the Natural Sciences and Engineering
Research Council of Canada.

\appendix

\section{Global Proton Form Factor Fit(s) Using Kelly Parametrization}

\label{sec:globalfit}

\begin{figure}
  \begin{center}
    \includegraphics[width=0.85\columnwidth]{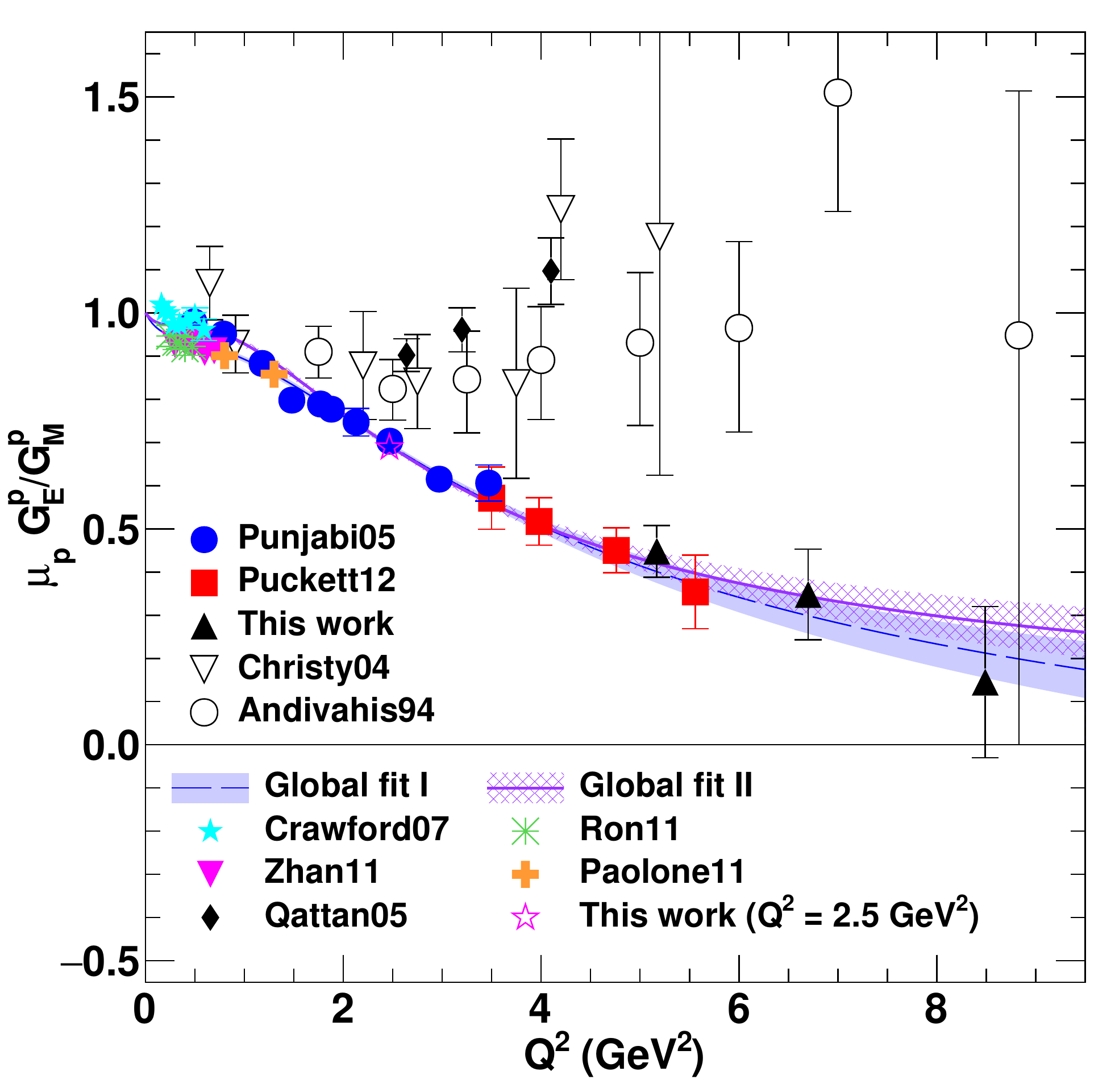}
  \end{center}
  \caption{\label{fig:gep3finalresults} Global fit results for the proton form factor ratio $\mu_p G_E^p/G_M^p$, compared to selected data from measurements of cross sections and polarization observables, including the final results of GEp-III (black solid triangles) and GEp-2$\gamma$ (pink empty star, weighted average). Other polarization data are from Refs.~\cite{Punjabi:2005wq,Jones:1999rz} (Punjabi05),~\cite{Puckett:2011xg,Gayou:2001qd} (Puckett12),~\cite{Crawford:2006rz} (Crawford07),~\cite{Ron:2011rd} (Ron11),~\cite{Zhan:2011ji} (Zhan11), and~\cite{Paolone:2010qc} (Paolone11). Rosenbluth separation data are from~\cite{Andivahis:1994rq} (Andivahis94),~\cite{Christy:2004rc} (Christy04), and~\cite{Qattan:2004ht} (Qattan05). Global fit I includes the data from Refs.~\cite{Ron:2011rd,Zhan:2011ji,Paolone:2010qc}, while excluding the data from Ref.~\cite{Crawford:2006rz} and the two lowest $Q^2$ points from Ref.~\cite{Punjabi:2005wq}. Global fit II excludes the data from Refs.~\cite{Ron:2011rd,Zhan:2011ji,Paolone:2010qc}. Shaded regions indicate $1\sigma$, pointwise uncertainty bands.}
\end{figure}

Several global fits of the proton form factors to measurements of differential cross sections and polarization observables in elastic $ep$ scattering were performed for this analysis using a procedure similar to that described in Ref.~\cite{Puckett:2010kn}. The results were used for the GEp-2$\gamma$ analysis to estimate the bin centering effects in the ratio $R$ and to calculate the event-by-event and acceptance-averaged values of $P_\ell^{Born}$ in the maximum-likelihood analysis. As in Ref.~\cite{Puckett:2010kn}, the ``first-order'' Kelly~\cite{Kelly:2004hm} parametrization was used in which $G_E^p$ and $G_M^p/\mu_p$ are described as ratios of a polynomial of degree $n$ and a polynomial of degree $n+2$ in $\tau = Q^2/4M_p^2$ (with $n = 1$). The Kelly parametrization enforces $G_E^p(0) = G_M^p(0)/\mu_p = 1$ and also enforces the ``dimensional scaling'' behavior at asymptotically large $Q^2$ predicted by perturbative QCD: $Q^4 F_1 \propto Q^6 F_2 \propto $ constant.  

Compared to Ref.~\cite{Puckett:2010kn}, the fits presented here differ in a few key respects. The data selection for differential cross section measurements is largely the same as before, and includes representative results from twelve different experiments spanning approximately $0.005$ GeV$^2 \le Q^2 \le 31$ GeV$^2$ (Refs.~\cite{Janssens:1965kd,Berger:1971kr,Price:1971zk,Kirk:1972xm,Bartel:1973rf,Borkowski:1974tm,Borkowski:1974mb,Simon:1980hu,Sill:1992qw,Walker:1993vj,Andivahis:1994rq,Christy:2004rc,Qattan:2004ht}). However, the database of polarization observables is modified substantially. First, the final results of GEp-III and GEp-2$\gamma$ reported in this work are now included in the fit, whereas in the original fit, the GEp-III results from Ref.~\cite{Puckett:2010ac} were used and the GEp-2$\gamma$ results were not included at all, as they were not yet published at the time. The three highest $Q^2$ points from the original GEp-II data~\cite{Gayou:2001qd} have been replaced by the results of the reanalysis of these data published in Ref.~\cite{Puckett:2011xg}. The data from Ref.~\cite{Ron:2007vr} have also been replaced by the reanalysis results published in Ref.~\cite{Ron:2011rd}. Finally, the high-precision data from Refs.~\cite{Paolone:2010qc,Zhan:2011ji} have been added. Given the apparent inconsistency of the various polarization experiments at low $Q^2$, an inconsistency which is not yet explained, two different fits were performed. In the first fit, hereafter referred to as ``Global fit I'', the recent precise data from Refs.~\cite{Paolone:2010qc,Zhan:2011ji, Ron:2011rd} were included, while the polarized target asymmetry data from Ref.~\cite{Crawford:2006rz} and the two lowest $Q^2$ points from GEp-I~\cite{Punjabi:2005wq} were excluded from the fit. In the second fit, referred to as ``Global fit II'', the data from Refs.~\cite{Paolone:2010qc,Zhan:2011ji, Ron:2011rd} were excluded, while all other $R_p$ data from polarization observables were included. 

The prescription for treating the cross section data, particularly in the high-$Q^2$ region where the inconsistency with the polarization transfer data exists, is also slightly modified here compared to Ref.~\cite{Puckett:2010kn}. As before, three iterations of the fit are performed, using the resulting parameters and their uncertainties and correlations from the previous fit as the starting point for the subsequent fits. In Ref.~\cite{Puckett:2010kn}, the value of $G_E^p(Q^2)$ was fixed for $Q^2 \ge 1$ GeV$^2$ using the result of the previous fit, or, on the first iteration, Kelly's 2004 result~\cite{Kelly:2004hm}, when computing the $\chi^2$ contribution of individual cross section data, effectively forcing $G_E^p$ to be entirely determined by polarization data for $Q^2 \ge 1$ GeV$^2$. In the fits reported here, $G_E$ ($G_M$) was fixed in the same way when the fractional contribution of the $\epsilon G_E^2$ ($\tau G_M^2$) term in the reduced cross section was less than 10\%, regardless of $Q^2$. This prescription removes the influence of individual cross section measurements on the determination of $G_E$ ($G_M$) at high (low) $Q^2$ when said measurements have very low sensitivity to the respective form factors. In particular, a cutoff of 10\% of the reduced cross section excludes \emph{all} cross section data for $Q^2 \gtrsim 2.2$ GeV$^2$ from participating in the determination of $G_E$, and some lower-$Q^2$ data, depending on $\epsilon$. 
\begin{table}
  \caption{\label{tab:fitresults} Summary of global proton FF fit results. Form factor parametrization is $G(Q^2) = \frac{1+a_1 \tau}{1+b_1 \tau + b_2\tau^2 + b_3 \tau^3}$, where $G(Q^2) = G_E(Q^2)$ or $G_M(Q^2)/\mu_p$. The uncertainty bands shown in Fig.~\ref{fig:gep3finalresults} represent the pointwise, $1\sigma$ errors computed from the full covariance matrix of the fit result. The asymptotic values of the form factors shown below are normalized to a dipole form $G_D = \left(1+Q^2/\Lambda^2\right)^{-2}$ with scale parameter $\Lambda^2 = 0.66$ GeV$^2$ corresponding to an RMS radius $r_p = 0.84$ fm. The total $\chi^2$ and degrees of freedom are shown along with the breakdown of $\chi^2$ contributions among cross section ($\sigma_R$) and polarization ($R_p^{pol}$) data. The $\chi^2$ contributions of cross section measurements are also separated into ``low'' ($Q^2 \le 1$ GeV$^2$) and ``high'' ($Q^2 > 1$ GeV$^2$) data. The best-fit normalization constants of the cross section experiments are omitted for brevity.}
  \begin{ruledtabular}
    \begin{tabular}{ccc}
      Fit & Global fit I & Global fit II \\ \hline
      $a_1^E$ & $-0.21 \pm 0.09$ & $-0.01 \pm 0.14$ \\
      $b_1^E$ & $12.21 \pm 0.18$ & $12.16 \pm 0.25$ \\
      $b_2^E$ & $12.6 \pm 1.1$ & $9.7 \pm 1.3$ \\
      $b_3^E$ & $23 \pm 4$ & $37 \pm 7$ \\
      $a_1^M$ & $0.058 \pm 0.022$ & $0.093 \pm 0.025$ \\
      $b_1^M$ & $10.85 \pm 0.073$ & $11.07 \pm 0.08$ \\
      $b_2^M$ & $19.9 \pm 0.2$ & $19.1 \pm 0.2$ \\
      $b_3^M$ & $4.4 \pm 0.6$ & $5.6 \pm 0.7$ \\
      $\lim_{Q^2\to \infty} \frac{G_{E}^p}{G_D(r_p = 0.84\text{ fm})} $ & $-0.26 \pm 0.15$ & $-0.01 \pm 0.11$ \\ 
      $\lim_{Q^2 \to \infty} \frac{G_{M}^p}{\mu_p G_D(r_p = 0.84\text{ fm})} $ & $0.38 \pm 0.09$ & $0.47 \pm 0.07$ \\
      $\chi^2/ndf$ (all data) & 706/460 & 696/455 \\
      $\chi^2/n_{data}$ ($\sigma_R$) & 672/427 & 653/427 \\
      $\chi^2/n_{data}$ ($R_p^{pol}$) & 34/53 & 44/48 \\
      $\chi^2/n_{data}$ ($\sigma_R, Q^2 \le 1$ GeV$^2$) & 337.7/275 & 308.4/275 \\
      $\chi^2/n_{data}$ ($\sigma_R, Q^2 > 1$ GeV$^2$) & 334.5/152 & 344.1/152 
    \end{tabular}
  \end{ruledtabular}
\end{table}
The other significant difference between the fits reported here and those of Ref.~\cite{Puckett:2010kn} is that in Ref.~\cite{Puckett:2010kn}, the overall normalization uncertainties in the absolute cross section data were essentially ignored in the $\chi^2$ calculation, whereas in the fits presented here, the overall normalization of each of the twelve experiments included in the global fit was allowed to float within a range of $\pm 2.5$ times the quoted normalization uncertainty. All of the best-fit normalization constants were well within their allowed ranges in both fits. In ``Global Fit II'', no experiment was renormalized by more than 3\%, whereas in ``Global fit I'' several experiments were renormalized downward by up to 5\%. This result reflects a subtle interplay between the tension with existing data of the precise polarization measurements of $R_p$ from Refs.~\cite{Paolone:2010qc,Zhan:2011ji, Ron:2011rd} in the 0.1-1 GeV$^2$ region on the one hand, and the discrepancy between cross section and polarization data at large $Q^2$ on the other. Allowing the cross section normalizations to float leads to a reduction of the $\chi^2$ per degree-of-freedom from 1.78 in Ref.~\cite{Puckett:2010kn} to approximately 1.54 in the fits reported here. 

Table~\ref{tab:fitresults} summarizes the global fit results. The best-fit values of the parameters describing $G_E^p$ and $G_M^p$ and their ($1\sigma$) uncertainties are presented together with the implied asymptotic values of $G_E^p$ and $G_M^p$, normalized to a dipole form factor with a scale parameter $\Lambda^2= 0.66$ GeV$^2$, corresponding to an RMS radius of 0.84 fm, consistent with the proton charge radius extracted from measurements of the Lamb shift in muonic hydrogen~\cite{Pohl:2010zza}. As pointed out in Ref.~\cite{Higinbotham:2015rja}, a dipole form factor with $r_p = 0.84$ fm describes the low-$Q^2$ $G_E^p$ data better than the ``standard'' dipole form factor with $\Lambda^2 = 0.71$ GeV$^2$ (corresponding to $r_p = 0.81$ fm). As measured by $\chi^2$, the overall quality of both fits is relatively good, except for the cross section data in the high $Q^2$ region, for which the $\chi^2$ per datum exceeds two. No attempt was made to correct the high-$Q^2$ cross section data for the effects of two-photon-exchange thought to be responsible for the discrepancy, as these effects are presently only poorly constrained experimentally and incompletely understood theoretically~\cite{doi:10.1063/PT.3.3541}. Instead, the ``excess'' $\epsilon$-dependence of the reduced cross sections observed in the high-$Q^2$ data (i.e., the ``excess'' slope in the Rosenbluth plot relative to the expectation from polarization transfer data) is simply averaged over in determining $G_M$, with the ratio $G_E/G_M$ fixed by the polarization data. While this procedure may bias the determination of $G_M$ in principle, the potential size of the effect on $G_M$ in the high-$Q^2$ region is mitigated by the smallness of the fractional contribution of $G_E^2$ to the reduced cross section. The inconsistency among polarization experiments in the low-$Q^2$ region is another issue that awaits resolution. While the fits reported here provide an adequate representation of the proton FFs in the $Q^2$ region in which they are directly constrained by data, the values and uncertainties in the extrapolation of these fits to larger $Q^2$ should not be taken too seriously. The high-precision polarization data for $R$ in both the 0.1-1 GeV$^2$ region~\cite{Punjabi:2005wq,Crawford:2006rz,Paolone:2010qc,Zhan:2011ji, Ron:2011rd} and at 2.5 GeV$^2$ as reported in this work, combine to exert significant influence on the extrapolation of $G_E$ and $G_M$ to $Q^2$ values beyond the reach of existing data, as is evident from the noticeably different asymptotic behaviors of the two fits, which differ only in the choice of low-$Q^2$ polarization data. This is a consequence of fitting a smooth, relatively inflexible parametrization of the form factors, with no specific theoretical justification other than its asymptotic behavior, to high-precision data at significantly different $Q^2$ values.

\bibliography{gep3_prc_master_references}

\end{document}